\documentclass[superscriptaddress,aps,prd,reprint,nofootinbib,longbibliography, amsmath,amssymb]{revtex4-1}

\usepackage{mathrsfs}
\usepackage{blindtext}
\usepackage{mathtools}
\usepackage[usenames,dvipsnames]{color}
\usepackage{adjustbox}
\usepackage{graphicx}
\usepackage[ISO]{diffcoeff}
\usepackage{bm}
\normalsize
\usepackage{booktabs}
\usepackage{siunitx}
\usepackage{tabularx}
\newcolumntype{Y}{>{\centering\arraybackslash}X}
\usepackage{soul}
\usepackage[utf8]{inputenc}
\usepackage[english]{babel}
\usepackage{rotating}
\usepackage{hyperref}
\hypersetup{
    colorlinks=true,
    linkcolor=blue,
    citecolor=red,
    filecolor=magenta,      
    urlcolor=magenta,
}
\usepackage[all]{hypcap}
\usepackage[capitalise]{cleveref}
\newcommand{\Ovec}{{\bm \Omega}}
\newcommand{\rvec}{{\bm r}}
\newcommand{\xvec}{{\bm x}}

\newcommand{\yvec}{{\bm y}}

\newcommand{\zvec}{{\bm z}}

\newcommand{\kvec}{{\bm k}}

\newcommand{\pvec}{{\bm p}}

\newcommand{\mvec}{{\bm m}}

\newcommand{\dvec}{{\bm \nabla}}
\newcommand{\Avec}{{\bm A}}
\newcommand{\Evec}{{\bm E}}
\newcommand{\Bvec}{{\bm B}}
\newcommand{\Dvec}{{\bm D}}
\newcommand{\Hvec}{{\bm H}}
\newcommand{\jvec}{{\bm j}}
\newcommand{\evec}{{\bm e}}
\newcommand{\nvec}{{\bm n}}
\newcommand{\psivecB}{{\bm \psi}_B}
\newcommand{\psiomega}{\psi_\omega}

\newcommand{\bara}{\bar{a}}
\newcommand{\dotbara}{\dot{\bara}}
\newcommand{\barAvec}{\bar{\Avec}}
\newcommand{\barEvec}{\bar{\Evec}}
\newcommand{\barBvec}{\bar{\Bvec}}
\newcommand{\barDvec}{\bar{\Dvec}}
\newcommand{\barHvec}{\bar{\Hvec}}
\newcommand{\barB}{\bar{B}}

\newcommand{\dotbarBvec}{\dot{\barBvec}}

\newcommand{\barE}{\bar{E}}

\newcommand{\da}{\delta a}
\newcommand{\dotda}{\delta \dot{a}}
\newcommand{\dotdotda}{\delta \ddot{a}}
\newcommand{\dphi}{\delta \phi}

\newcommand{\dAvec}{\delta \hspace{-0.04cm} \Avec}
\newcommand{\dotdAvec}{\delta \hspace{-0.04cm} \dot{\Avec}}

\newcommand{\dEvec}{\delta \hspace{-0.02cm} \Evec}
\newcommand{\dotdEvec}{\delta \hspace{-0.02cm} \dot{\Evec}}

\newcommand{\dBvec}{\delta \hspace{-0.02cm} \Bvec}
\newcommand{\dotdBvec}{\delta \hspace{-0.02cm} \dot{\Bvec}}

\newcommand{\dDvec}{\delta \hspace{-0.02cm} \Dvec}
\newcommand{\dotdDvec}{\delta \hspace{-0.02cm} \dot{\Dvec}}

\newcommand{\dHvec}{\delta \hspace{-0.02cm} \Hvec}

\newcommand{\Ohat}{\hat{\Ovec}}
\newcommand{\rhat}{\hat{\rvec}}
\newcommand{\xhat}{\hat{\xvec}}
\newcommand{\yhat}{\hat{\yvec}}
\newcommand{\zhat}{\hat{\zvec}}
\newcommand{\mhat}{\hat{\mvec}}
\newcommand{\ehat}{\hat{\evec}}

\newcommand{\barBhat}{\hat{\barBvec}}
\newcommand{\nhat}{\hat{\nvec}}

\newcommand{\tF}{\tilde{F}}

\newcommand{\per}{\;.}
\newcommand{\com}{\;,}

\newcommand{\MMR}{\text{MMR}}
\newcommand{\EHR}{\text{EHR}}

\newcommand{\RNS}{R_\text{\sc ns}}
\newcommand{\RLC}{R_\text{\sc lc}}
\newcommand{\wpl}{\omega_\mathrm{pl}}

\newcommand{\wB}{\omega_B}
\newcommand{\Lscr}{\mathscr{L}}
\newcommand{\ii}{\mathrm{i}}
\newcommand{\ee}{\mathrm{e}}
\newcommand{\dd}{\mathrm{d}}

\newcommand{\Lcal}{\mathcal{L}}
\newcommand{\Fcal}{\mathcal{F}}

\newcommand{\Mcal}{\mathcal{M}}
\newcommand{\Pbb}{\mathbb{P}}
\newcommand{\Pag}{\Pbb_{a\to\gamma}}
\newcommand{\rad}{r}
\newcommand{\radres}{\rad_\mathrm{res}}
\newcommand{\dres}{d_\mathrm{res}}
\newcommand{\ka}{k_a}
\newcommand{\kavec}{\kvec_a}

\newcommand{\pprime}{{\prime\prime}}

\newcommand{\gagg}{g_{a\gamma\gamma}}
\newcommand{\gaNN}{g_{aNN}}
\newcommand{\ggggg}{g_{\gamma\gamma\gamma\gamma}}

\newcommand{\neV}{\;\mathrm{neV}}
\newcommand{\ueV}{\;\mu\mathrm{eV}}
\newcommand{\meV}{\;\mathrm{meV}}
\newcommand{\eV}{\;\mathrm{eV}}
\newcommand{\keV}{\;\mathrm{keV}}
\newcommand{\MeV}{\;\mathrm{MeV}}
\newcommand{\GeV}{\;\mathrm{GeV}}

\newcommand{\THz}{\;\mathrm{THz}}
\newcommand{\Gauss}{\;\mathrm{G}}
\renewcommand{\sec}{\;\mathrm{sec}}

\newcommand{\cm}{\;\mathrm{cm}}
\newcommand{\km}{\;\mathrm{km}}
\newcommand{\erg}{\;\mathrm{erg}}

\newcommand{\Jy}{\;\mathrm{Jy}}
\newcommand{\pc}{\;\mathrm{pc}}
\newcommand{\kpc}{\;\mathrm{kpc}}

\newcommand{\eref}[1]{Eq.~(\ref{#1})} 
\newcommand{\erefs}[2]{Eqs.~(\ref{#1})~and~(\ref{#2})}

\newcommand{\fref}[1]{Fig.~\ref{#1}}
\newcommand{\Fref}[1]{Figure~\ref{#1}}
\newcommand{\sref}[1]{Sec.~\ref{#1}}

\newcommand{\aref}[1]{Appendix~\ref{#1}}

\newcommand{\rref}[1]{\textcite{#1}}

\newcommand{\ba}[1]{\begin{align} #1 \end{align}}
\newcommand{\bes}[1]{\begin{equation}\begin{split} #1 \end{split}\end{equation}}
\newcommand{\bsa}[2]{\begin{subequations}\label{#1}\begin{align} #2 \end{align}\end{subequations}}
\newcommand{\nn}{\nonumber \\}

\begin{document}
\title{Resonant conversion of axion dark radiation \\ into terahertz electromagnetic radiation \\ in a neutron star magnetosphere}
\author{Andrew J. Long}
\affiliation{Department of Physics and Astronomy, Rice University, Houston, Texas 77005, U.S.A}
\author{Enrico D. Schiappacasse}
\affiliation{Vicerrector\'ia de Investigaci\'on y Doctorados, Universidad San Sebasti\'an, Sede Santiago, Avenida del C\'ondor 720, 8580704 Huechuraba, Regi\'on Metropolitana, Chile}
\begin{abstract}
In the strong magnetic field of a neutron star's magnetosphere, axions coupled to electromagnetism develop a nonzero probability to convert into photons.  Past studies have revealed that the axion-photon conversion can be resonantly enhanced.  We recognize that the axion-photon resonance admits two parametrically distinct resonant solutions, which we call the mass-matched resonance and the Euler-Heisenberg assisted resonance.  The mass-matched resonance occurs at a point in the magnetosphere where the radially-varying plasma frequency crosses the axion mass $\omega_\mathrm{pl} \approx m_a$.  The Euler-Heisenberg assisted resonance occurs where the axion energy satisfies $\omega \approx (2 \omega_\mathrm{pl}^2 / 7 g_{\gamma\gamma\gamma\gamma} \bar{B}^2 )^{1/2}$.  This second resonance is made possible though the strong background magnetic field $\bar{B}$ as well as the nonzero Euler-Heisenberg four-photon self interaction, which has the coupling $g_{\gamma\gamma\gamma\gamma} = 8 \alpha^2 / 45 m_e^4$.  We study the resonant conversion of relativistic axion dark radiation into photons via the Euler-Heisenberg assisted resonance, and we calculate the expected electromagnetic radiation assuming different values for the axion-photon coupling $g_{a\gamma\gamma}$ and different amplitudes for the axion flux onto the neutron star $\Phi_a$.  We briefly discuss several possible sources of axion dark radiation.  Achieving a sufficiently strong axion flux to induce a detectable electromagnetic signal seems unlikely.
\end{abstract}

\maketitle
\tableofcontents

\section{Introduction}
\label{sec:introduction}

Perhaps a QCD axion~\cite{Peccei:1977hh, Peccei:1977ur,Weinberg:1977ma} or an axion-like particle~\cite{Marsh:2015xka,Irastorza:2018dyq,Choi:2020rgn} exists in nature.  
If so, the dense and highly magnetized environment of a neutron star provides a natural laboratory in which to probe these axions through their couplings to the Standard Model~\cite{Raffelt:1996wa}.  
If the axion couples to electromagnetism though the interaction $\Lscr_\mathrm{int} = - \gagg a F_{\mu\nu} \tF^{\mu\nu} / 4$, then axions develop a nonzero probability to convert into photons when propagating through a strong magnetic field~\cite{Maiani:1986md,Raffelt:1987im}.  
This axion-photon interconversion is the basis of many experimental searches for axions on Earth~\cite{Carosi:2013rla,Graham:2015ouw,Adams:2022pbo}.  

The axion-photon interconversion may be resonantly enhanced, depending on the model parameters and the system under consideration. 
In a neutron star magnetosphere, it is necessary to account for the presence of an electron-proton plasma, which modifies the electric constitutive relation in a way that depends on the plasma frequency $\wpl$.  
Additionally, due to the star's strong magnetic field, it is necessary to account for the Euler-Heisenberg four-photon self-interaction, which arises in the low-energy limit of quantum electrodynamics.  
Both effects play a role in the axion-photon resonance~\cite{Yoshimura:1987ma,Lai:2006af,Bondarenko:2022ngb,Song:2024rru}. 

Here we identify that the axion-photon resonance admits two solution branches.  
On the more familiar branch, which we call the mass-matched resonance (\MMR{}), the resonance condition is satisfied when the plasma frequency is approximately equal to the axion mass $\wpl \approx m_a$. 
The role of the \MMR{} has been studied extensively in astrophysical and cosmological axion-photon interconversion; a few examples are Refs.~\cite{Yanagida:1987nf,Higaki:2013qka,Tashiro:2013yea,Angus:2013sua,Kraljic:2014yta,Evoli:2016zhj,Fortin:2018ehg,Harris:2020qim,Buen-Abad:2020zbd,Fortin:2021cog,Schiavone:2021imu,Kar:2022ngx,Carenza:2023nck,Fortin:2023jlg,Sun:2023wqq,Mondino:2024rif,Beadle:2024jlr,Ferreira:2024ktd}. 
Importantly, the \MMR{} may be accessible for either relativistic or nonrelativistic axions. 
For instance, it could lead to a radio emission line when axion dark matter is incident on a neutron star~\cite{Pshirkov:2007st,Huang:2018lxq,Hook:2018iia,Safdi:2018oeu,Foster:2020pgt,Buckley:2020fmh,Edwards:2020afl,Nurmi:2021xds,Witte:2021arp,Millar:2021gzs,Choi:2022btl,Witte:2022cjj,McDonald:2023ohd,McDonald:2023shx,Tjemsland:2023vvc,Gines:2024ekm,McDonald:2024uuh}. 

In this work we focus on the second solution branch, which we call the Euler-Heisenberg assisted resonance (\EHR{}), because it has not been as extensively studied.  
Provided that the ambient magnetic field $\barB$ is sufficiently strong, the resonance condition is satisfied where the axion energy $\omega$ has $\omega = (2 \wpl^2 / 7 \ggggg \barB^2)^{1/2}$ where $\ggggg = 8 \alpha^2 / 45 m_e^4$ is the Euler-Heisenberg coupling.  
For typical parameters, this condition is only satisfied for relativistic axions.  
We study the \EHR{} and estimate the associated electromagnetic signal that is expected to arise if a flux of relativistic axions are incident on a neutron star.  
We discuss the role that this emission could play in searches for axion dark radiation.  

\section{Resonant conversion of relativistic axions}
\label{sec:conversion}

In this section we study the resonant conversion of relativistic axions into photons.  
The resonance is made possible by the presence of a plasma, \textit{i.e.} nonzero plasma frequency $\omega_\mathrm{pl}$, and a strong background magnetic field.  
Systems that exhibit both of these properties include the early universe and the magnetosphere of a neutron star.  
In this article, we focus on the neutron star system.

\subsection{Modeling a neutron star magnetosphere}
\label{sub:magnetosphere}

We adopt the Goldreich-Julian (GJ) Model~\cite{Goldreich:1969sb} of the neutron star magnetosphere.  
This simple model is expected to provide a reliable approximation for quiet and stable neutron stars \cite{Haensel:2007yy}, and it has been adopted in previous studies of axion dark matter such as Ref.~\cite{Pshirkov:2007st}. 
For older neutron stars that have crossed their death line, a combination of analytical modeling and numerical simulation \cite{1985MNRAS.213P..43K,Spitkovsky:2002wg,Cerutti:2016ttn} reveal that the magnetosphere is more reliably modeled as a vacuum spinning dipole (electrosphere), which we do not consider in this work. 

The neutron star itself is modeled as a ball with radius $\RNS$ that rotates with angular velocity $\Ovec$ about an axis through its center.  
We also define the angular speed $\Omega = |\Ovec|$, the unit vector $\Ohat = \Ovec / \Omega$, and the rotation period $P = 2\pi/\Omega$. 
Outside of the neutron star, the magnetic field $\barBvec(\rvec,t)$ is modeled as a dipole, with magnetic dipole moment $\mvec(t)$ having constant magnitude $m = |\mvec(t)|$ and varying orientation $\mhat(t) = \mvec(t) / m$ in a frame where $\Ovec$ is fixed.  
The field outside the star is written as 
\bes{\label{eq:B_dipole}
    & \barBvec(\rvec,t) = B_0 \, \psivecB(\rhat,t) \, \biggl( \frac{\rad}{\RNS} \biggr)^{\! \! -3} 
    \\ & 
    \quad \text{where} \quad 
    \psivecB(\rhat,t) = \tfrac{3}{2} \bigl( \mhat(t) \cdot \rhat \bigr) \, \rhat - \tfrac{1}{2} \mhat(t) 
    \com
}
and $\rad = |\rvec|$ is the distance from the center of the star to a point with position $\rvec = \rad \rhat$.  
At the north magnetic pole ($\rvec = \RNS \mhat$) the magnetic field is $\barBvec = 2 \mvec / \RNS^3$; along the magnetic equator ($\rvec \perp \mhat$) it is $\barBvec = - \mvec / \RNS^3$; and at the south magnetic pole ($\rvec = -\RNS \mhat$) it is again $\barBvec = 2 \mvec / \RNS^3$.  
We define $B_0 = 2 m / \RNS^3$ to denote the polar surface magnetic field strength.  
For a typical pulsar, $\RNS \approx 10 \km$, $P \approx 0.1$--$10 \sec$, and $B_0 \approx 10^{11\text{--}13} \Gauss$ whereas a magnetar can reach $B_0 \approx \mathrm{few} \times 10^{14} \Gauss$~\cite{Faucher-Giguere:2005dxp,Harding:2006qn,Olausen:2013bpa,Kaspi:2017fwg}.  

The GJ model rests on the assumption that the magnetic field lines co-rotate with the star, and a consistent solution of Maxwell's equations is obtained by assuming a charged plasma to be present in the magnetosphere.  
The density of electric charge $n_c(\rvec,t)$ is required to be $n_c(\rvec,t) = [2 \Ovec \cdot \barBvec(\rvec,t)/e] / (1 - |\Ovec \times \rvec|^2)$ where $e \approx 0.303$ is the unit of electromagnetic charge and $\alpha = e^2 / 4\pi \approx 1/137$ is the electromagnetic fine structure constant.  
Assuming a nonrelativistic electron-proton plasma, the plasma frequency is approximately $\wpl(\rvec,t) \approx \sqrt{4 \pi \alpha n_e(\rvec,t) / m_e}$ where $m_e \approx 0.511 \MeV$ is the electron's mass, and the contribution from the heavier protons can be neglected. 
We take the density of charge as an estimate of the electron number density $n_e(\rvec,t) \approx |n_c(\rvec,t)|$ while noting that $n_e \gg |n_c|$ is possible in the polar regions~\cite{Eilek:2016hms}. 
The plasma frequency is 
\ba{\label{eq:barwpl}
    & \wpl(\rvec,t) 
    = \omega_{\mathrm{pl},0} \ 
    \psiomega(\rhat,t) \ 
    \biggl( \frac{r}{\RNS} \biggr)^{\! \! -3/2}
    \quad \text{where} \quad 
    \\ & \ 
    \psiomega(\rhat,t) = \bigl| 2 \Ohat \cdot \psivecB(\rhat,t) \bigr|^{1/2} 
    \quad \text{and where} \quad 
    \nn & \ 
    \omega_{\mathrm{pl},0} \approx 
    \bigl( 70 \ueV \bigr) 
    \biggl( \frac{B_0}{10^{14} \Gauss} \biggr)^{\! \! 1/2} 
    \biggl( \frac{P}{1 \sec} \biggr)^{\! \! -1/2} 
    \biggl( \frac{n_e}{|n_c|} \biggr)^{\! \! 1/2} 
    \per
    \nonumber
}
In general $\wpl \propto r^{-3/2}$ and the plasma frequency decreases further from the star's surface.  

We expect the GJ model to be a reliable description of the neutron star magnetosphere in the equatorial regions near the star.  
In the polar regions, acceleration of charged particles in the strong field leads to the formation of an electron-positron plasma \cite{Istomin:2007ge}, which is not captured by the GJ model.  
Far from the star's surface, the plasma cannot co-rotate with the star, since it would need to travel at greater than the speed of light.  
This boundary is known as the star's light cylinder, and it is located at a radius of approximately $\RLC = 1 / \Omega$.  
For typical fiducial parameters the light cylinder radius is 
\bes{\label{eq:RLS_approx}
    \RLC \approx 
    \bigl( 4.8 \times 10^3 \RNS \bigr) 
    \biggl( \frac{P}{1 \sec} \biggr) 
    \biggl( \frac{\RNS}{10 \km} \biggr)^{\! \! -1} 
    \per
}
We expect the GJ model to be reliable at radii $\rad$ satisfying $\RNS \leq \rad \ll \RLC$ and away from the polar regions. 

\subsection{Modeling axion and photon interactions}
\label{sub:axion}

To model the axion's interactions with electromagnetism, we assume the usual $aF\tF$ operator, which is expected to be present for the QCD axion in the KSVZ~\cite{Kim:1979if,Shifman:1979if} and DFSZ~\cite{Dine:1981rt,Zhitnitsky:1980tq} models as well as for axion-like particles that arise in the string axiverse~\cite{Gendler:2023kjt}. 
The Lagrangian for the axion-photon system in vacuum is taken to be 
\ba{\label{eq:Lagrangian}
    \Lscr & = 
    - \tfrac{1}{4} F_{\mu\nu} F^{\mu\nu} 
    + \tfrac{1}{2} \partial_\mu a \partial^\mu a 
    - \tfrac{1}{2} m_a^2 a^2 
    - \tfrac{1}{4} \gagg a F_{\mu\nu} \tF^{\mu\nu} 
    \nn & \qquad 
    + \tfrac{1}{16} \ggggg \Bigl[ \bigl( F_{\mu\nu} F^{\mu\nu} \bigr)^2 + \tfrac{7}{4} \bigl( F_{\mu\nu} \tF^{\mu\nu} \bigr)^2 \Bigr] 
}
where $a(x)$ is the axion field, $A_\mu(x)$ is the photon field, $F_{\mu\nu} = \partial_\mu A_\nu - \partial_\nu A_\mu$ is the electromagnetic field strength tensor, and $\tF^{\mu\nu} = \epsilon^{\mu\nu\rho\sigma} F_{\rho\sigma} / 2$ is the dual tensor.  
The axion-photon interaction is parameterized by the coupling $\gagg$.  
A strong axion-photon interaction would lead to undetected emission of axions from astrophysical bodies.  
Searches for axion emission from our Sun by the CAST helioscope lead to a constraint $|\gagg| < 6.6 \times 10^{-11} \GeV^{-1}$ at $95\%$ confidence for $m_a < 0.02 \eV$~\cite{CAST:2017uph}.  
For much lighter axions, X-ray and $\gamma$-ray observations of NGC 1275~\cite{Reynolds:2019uqt}, super star clusters~\cite{Dessert:2020lil}, and SN-1987A~\cite{Payez:2014xsa} constrain the axion-photon coupling at the level for $\gagg < \mathrm{few} \times 10^{-12} \GeV^{-1}$ for various mass ranges up to $m_a < 10^{-10} \eV$. 
See also the Particle Data Group's review on axions~\cite{ParticleDataGroup:2022pth} for a comprehensive summary of constraints. 

The $(FF)^2$ and $(F\tF)^2$ terms in \eref{eq:Lagrangian} are known as the Euler-Heisenberg four-photon self interaction~\cite{Heisenberg:1936nmg,Weisskopf:1936hya,Adler:1971wn}. 
These interactions arise in the low energy effective theory obtained from QED upon integrating out the electron/positron~\cite{Dunne:2004nc}, leading to $\ggggg = 8 \alpha^2 / 45 m_e^4 = 2 \alpha / (45 \pi B_\mathrm{crit}^2)$ where $B_\mathrm{crit} = m_e^2 / e \approx 4.4 \times 10^{13} \Gauss$.  
In formulas and plots we generally fiducialize the magnetic field strength to $B_0 = 10^{14} \Gauss$ although the higher-order terms in the Euler-Heisenberg effective action will become significant for $|\barBvec| \gtrsim B_\mathrm{crit}$; one is free to scale $B_0$ down in the formulas. 
For reference, note that 
\bes{\label{eq:ggggg_B0sq}
    \ggggg B_0^2 \approx \bigl( 5.3 \times 10^{-4} \bigr) \biggl( \frac{B_0}{10^{14} \Gauss} \biggr)^2 
    \com
}
and we can generally neglect $\ggggg B_0^2$ (or $\ggggg |\barBvec|^2$ or $\ggggg \barB^2$) when it appears in comparison with an order $1$ number.  
In the background of a strong magnetic field, these interactions modify the photon's dispersion relation and impact axion-photon mixing~\cite{Raffelt:1987im}. 

In a neutron star magnetosphere, the electromagnetic field obeys the in-medium form of Maxwell's equations (augmented by the axion-photon interaction and the Euler-Heisenberg self-interaction).  
We model the medium with a linear constitutive relation $\Dvec = \epsilon \Evec$ where $\epsilon(\rvec,t)$ is the dielectric tensor.  
Due to the strong magnetic field, electrons perform orbits around field lines, and consequently electromagnetic waves propagating either longitudinal or transverse to the field lines experience different indices of refraction.  
We model this effect by writing the dielectric tensor as $\epsilon_{ij} = \delta_{ij} - (\wpl^2 / \omega^2) \hat{B}_i \hat{B}_j$ \cite{Beskin:1993xx} where $\hat{\bm B}$ is the unit vector that locally points in the direction of the magnetic field and $\wpl$ is the plasma frequency.  

\subsection{Resonant axion-photon conversion}
\label{sub:probability}

Several previous studies \cite{Raffelt:1987im,Lai:2006af} contain derivations of the equations of motion for the axion-photon system, including effects of the axion-photon coupling, the Euler-Heisenberg photon self-interaction, and the ambient magnetized plasma.  
For the sake of completeness, we present a detailed derivation in \aref{app:derivation}. 
For convenience we make two simplifying assumptions.  
First we focus on aligned neutron stars for which the magnetic dipole moment aligns with the star's rotation axis, $\mhat(t) = \Ohat$.  
An aligned neutron star has a static magnetic field $\barBvec(\rvec) = B_0 \, \psivecB(\rhat) \, (\rad / \RNS)^3$ and plasma frequency $\wpl(\rvec) = \omega_{\mathrm{pl},0} \, \psiomega(\rhat) \, (\rad / \RNS)^{-3/2}$.  
Second we focus on axions that move along radial trajectories out from the center of the star.  
If the unit vector $\nhat$ points in the direction of the axion's motion, then this assumption corresponds to choosing $\nhat = \rhat$.  
We identify the magnetic field's longitudinal and transverse components as 
\ba{
    & \barB_L = \nhat \cdot \barBvec
    \quad \text{and} \quad 
    \barB_T = |\barBvec - \barB_L \, \nhat| 
    \\ & \text{with} \quad 
    \beta_L(\nhat) = \tfrac{\barB_L(\rvec,\nhat)}{\barB(\rvec)} 
    \quad \text{and} \quad 
    \beta_T(\nhat) = \tfrac{\barB_T(\rvec,\nhat)}{\barB(\rvec)} 
    \per 
    \nonumber 
}
Along a radial trajectory $\nhat = \rhat$, the ratios $\beta_L(\nhat)$ and $\beta_T(\nhat)$ are independent of $\rad$, and they satisfy $\beta_L^2 + \beta_T^2 = 1$.  

\subsubsection{Resonance condition}
\label{subsub:res_condit}

Subject to the simplifying assumptions above, the equations of motion for the axion-photon system reduce to a set of coupled first-order differential equations
\ba{
\label{eq:MatrixHE1}
    -\ii \frac{\dd}{\dd r}
\begin{pmatrix}
    a_{\omega,\nhat} \\
    \ii A_{\omega,\nhat}
\end{pmatrix}
    = 
    \begin{pmatrix}
    \Delta_a & \Delta_B \\ 
    \Delta_B & \Delta_\parallel 
    \end{pmatrix}
    \begin{pmatrix}
    a_{\omega,\nhat} \\
    \ii A_{\omega,\nhat}
    \end{pmatrix}
    \com
}
which are labeled by an angular frequency $\omega$ (equivalently, the axion or photon energy) and a unit vector $\nhat$ (corresponding to the direction of propagation).  
The entries in the mixing matrix are 
\bsa{eq:Delta_def}{
    \Delta_a(\rad) & = 0 \\ 
    \Delta_B(\rad) & = \frac{\omega}{2\ka} \frac{\gagg \, \beta_T \, \barB}{1 - \beta_L^2 \tfrac{\wpl^2}{\omega^2}} \\ 
    \Delta_\parallel(\rad) & = \frac{m_a^2}{2\ka} - \frac{\omega^2}{2\ka} \frac{\beta_T^2 \tfrac{\wpl^2}{\omega^2} - \tfrac{7}{2} \ggggg \, \beta_T^2 \, \barB^2}{1 - \beta_L^2 \tfrac{\wpl^2}{\omega^2}} 
    \com
}
where $\wpl = \wpl(\rad \nhat)$, $\barB = \barB(\rad \nhat)$, $\beta_L = \beta_L(\nhat)$, $\beta_T = \beta_T(\nhat)$, and $\ka = (\omega^2 - m_a^2)^{1/2}$.  
The complex amplitudes of the axion field $a_{\omega,\nhat}(\rad)$ and electromagnetic field $A_{\omega,\nhat}(\rad)$ vary as the waves propagate.  
In writing \eref{eq:Delta_def} we have dropped negligible terms suppressed by additional factors of $\gagg^2 \barB^2$ or $\ggggg \barB^2$; the more general expressions appear in \eref{eq:Delta_defs}.  

To study the conversion of axions into photons, one can solve \eref{eq:MatrixHE1} along with the initial condition that $a_{\omega,\nhat} = a_{\omega,\nhat,0}$ and $A_{\omega,\nhat} = 0$ at the surface of the neutron star where $r = \RNS$.  
Then the probability that an axion converts into a photon reaching spatial infinity is calculated as $\Pag(\omega,\nhat) = |A_{\omega,\nhat}(\infty) / a_{\omega,\nhat,0}|^2$.  

\begin{figure}[t]
\centering
\includegraphics[scale=0.49]{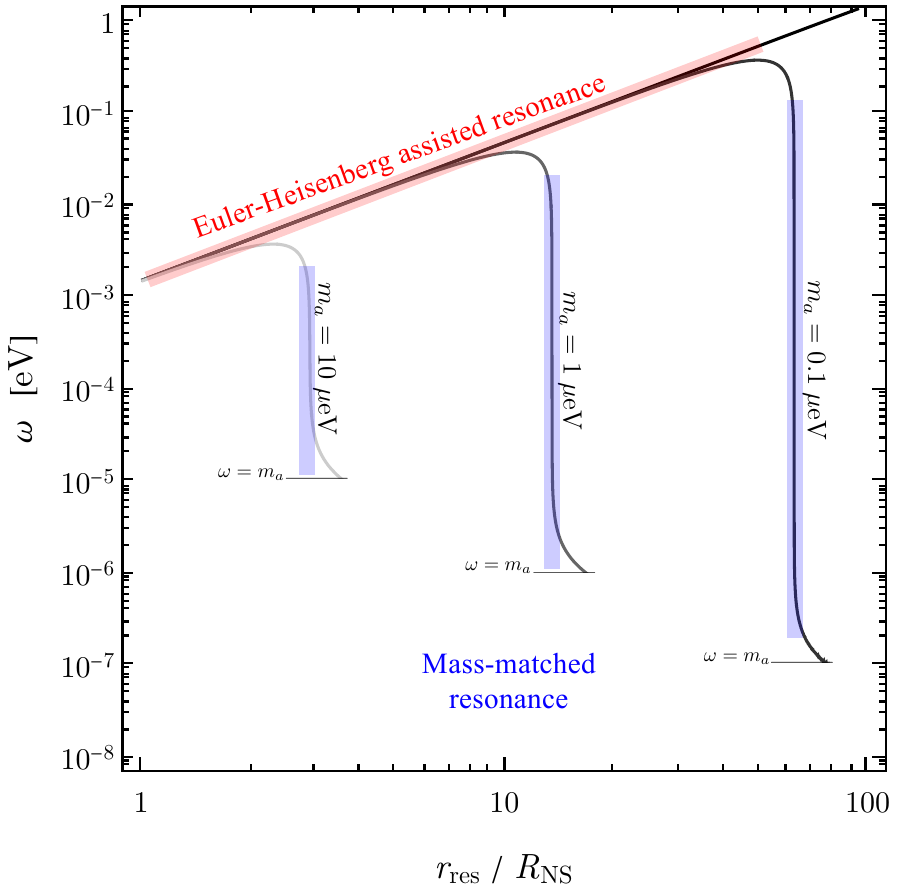}
\caption{\label{fig:res_condit}
A region of parameter space satisfying the resonance condition. We vary the axion energy $\omega$ and the resonance radius $\radres$, which is measured with respect to the neutron star radius $\RNS$. The black curves are solutions of \eref{eq:res_condit} for several values of the axion mass $m_a$.  The colored lines highlight the two solution branches, which we call the Euler-Heisenberg assisted resonance (red) and the mass-matched resonance (blue).  Toward large axion energy, the two solution branches merge (curved corner) and at higher $\omega$ there is no resonance.  Toward low axion energy, the curves are truncated at $\omega = m_a$ (short horizontal line segments).  To generate this figure we have chosen the parameters such that $B_0 = 10^{14} \Gauss$, $P = 1 \sec$, and $n_e = |n_c|$, which imply $\omega_{\mathrm{pl},0} \approx 70 \ueV$.  We neglect the orientation-dependent factors by setting $\beta_L(\nhat) = \beta_T(\nhat) = 1 / \sqrt{2}$ and $|\psivecB(\nhat)| = \psiomega(\nhat) = 1$.
}
\end{figure}

Often the axion-photon interconversion is resonantly enhanced, and when this occurs the probability can be approximated with a simple analytic expression.  
Resonant conversion occurs if there exists a point outside the neutron star ($\rad > \RNS$) at which $\Delta_a(\rad) = \Delta_\parallel(\rad)$~\cite{Raffelt:1987im}, and we call the solution the resonance radius $\radres$.  
The resonance condition is written equivalently as~\cite{Lai:2006af} 
\ba{\label{eq:res_condit}
    m_a^2 
    - m_a^2 \beta_L^2 \frac{\wpl^2}{\omega^2} 
    - \beta_T^2 \wpl^2 
    + \frac{7}{2} \ggggg \beta_T^2 \barB^2 \omega^2 
    \Bigr|_{\radres}
    = 0 
    \per 
}
Depending on the model ($m_a$ and $\ggggg$), on the neutron star ($\wpl$ and $\barB$), and on the axion ($\omega$ and $\nhat$) the resonance condition may admit zero, one, or two solutions.  
Here we identify that the solutions fall into two parametrically distinct classes.  
\Fref{fig:res_condit} shows solutions of the resonance condition on a slice of the parameter space.  
We see the there are two solution branches, which we study further in the next two subsections.  

\subsubsection{Mass-matched resonance}
\label{subsub:MM_res}

To identify the first resonance, we set $\ggggg = 0$ in the resonance condition \eqref{eq:res_condit}, which reduces to 
\ba{\label{eq:rc_MM}
    \wpl(\radres \, \nhat) = m_a \biggl( \beta_T^2(\nhat) + \frac{m_a^2}{\omega^2} \, \beta_L^2(\nhat) \biggr)^{\! \! -1/2} 
    \per
}
For nonrelativistic axions $\omega \approx m_a$ and the resonance occurs at a radius $\radres$ (assumed to be $> \RNS$) where $\wpl(\radres \, \nhat) \approx m_a$.  
For relativistic axions $\omega \gg m_a \, \beta_L / \beta_T$ and the resonance occurs where $\wpl(\radres \, \nhat) \approx m_a \beta_T^{-1}(\nhat)$.  
See \fref{fig:res_condit}. 
We call this the ``mass-matched'' resonance, since it occurs where the axion mass is approximately equal to the plasma frequency, which acts like a mass for electromagnetic waves in the plasma.  
If axion-photon interconversion proceeds via the \MMR{}, then the asymptotic conversion probability (for $r \gg \RNS$) is well-approximated by 
\ba{\label{eq:Pag_MM_main_text}
    & \Pag(\omega,\nhat) 
    = \frac{\pi}{3} \frac{\gagg^2 \barB^2 \omega^2 \radres}{\ka \wpl^2} \\ 
    & \approx 
    \bigl( 1.5 \times 10^{-3} \bigr) 
    \biggl( \frac{\gagg}{10^{-12} \GeV^{-1}} \biggr)^{\! \! 2} 
    \biggl( \frac{B_0}{10^{14} \Gauss} \biggr)^{\! \! 1/3} 
    \biggl( \frac{P}{1 \sec} \biggr)^{\! \! 5/3} 
    \nn & \quad \times 
    \biggl( \frac{\RNS}{10 \km} \biggr) 
    \biggl( \frac{n_e}{|n_c|} \biggr)^{\! \! -5/3} \, 
    \bigl| \psivecB(\nhat) \bigr|^{2} \, 
    \bigl| \psiomega(\nhat) \bigr|^{-10/3} 
    \nn & \quad \times 
    \begin{cases}
    \bigl( 0.1 \bigr) \, 
    \bigl( \frac{m_a}{1 \ueV} \bigr)^{10/3} \, 
    \bigl( \frac{\ka}{1 \neV} \bigr)^{-1} \\ 
    \bigl( \frac{m_a}{1 \ueV} \bigr)^{4/3} \, 
    \bigl( \frac{\omega}{10 \meV} \bigr)^{2} \, 
    \bigl( \frac{\ka}{10 \meV} \bigr)^{-1} \, 
    \bigl| \beta_T(\nhat) \bigr|^{-4/3} 
    \end{cases}
    \com
    \nonumber
}
where the top case corresponds to nonrelativistic axions with $\omega \approx m_a \gg \ka = (\omega^2 - m_a^2)^{1/2}$ and the bottom case corresponds to relativistic axions with $\omega \approx \ka \gg m_a \, \beta_L$.  
See \aref{app:mass_match_resonance} for additional details. 
The orientation-dependent factors $|\psivecB(\nhat)|$ and $\psiomega(\nhat)$ are discussed in \aref{app:numerical}.  

\subsubsection{Euler-Heisenberg assisted resonance}
\label{subsub:EH_res}

To identify the second resonance, we set $m_a = 0$ in the resonance condition \eqref{eq:res_condit}, which reduces to 
\bes{\label{eq:rc_EH}
    \omega 
    = \sqrt{ \frac{2}{7} \frac{\wpl^2(\radres \, \nhat)}{\ggggg \barB^2(\radres \, \nhat)} }  
    \per
}
For $\ggggg \barB^2 \ll 1$ as in \eref{eq:ggggg_B0sq}, it follows that $\omega \gg \wpl(\radres \, \nhat)$. 
Since this solution relies upon the Euler-Heisenberg photon self interaction ($\ggggg \neq 0$), we call it the ``Euler Heisenberg assisted'' resonance.  
It has also been called ``axion-photon resonance'' (collectively with the \MMR{}) by \rref{Lai:2006af}, ``double lens effect'' by \rref{Bondarenko:2022ngb}, and ``plasma-vacuum resonance'' by \rref{Song:2024rru}. 

For a given axion energy $\omega$ and neutron star profile functions, $\wpl(\rad)$ and $\barB(\rad)$, one can solve the resonance condition \eqref{eq:rc_EH} for $\rad = \radres$ to obtain the resonance radius 
\ba{\label{eq:rc}
    & \radres(\omega,\nhat) 
    = 
    \frac{7^{1/3}}{2^{1/3}} \ggggg^{1/3} B_0^{2/3} \frac{\omega^{2/3}}{\omega_{\mathrm{pl},0}^{2/3}} \frac{|\psivecB|^{2/3}}{|\psiomega|^{2/3}} \, \RNS
    \\ & \quad \approx 
    \bigl( 3.4 \RNS \bigr) 
    \biggl( \frac{B_0}{10^{14} \Gauss} \biggr)^{\! \! 1/3} 
    \biggl( \frac{P}{1 \sec} \biggr)^{\! \! 1/3} 
    \biggl( \frac{n_e}{|n_c|} \biggr)^{\! \! -1/3} 
    \nn & \qquad \times 
    \biggl( \frac{\omega}{10 \meV} \biggr)^{\! \! 2/3} \, 
    |\psivecB(\nhat)|^{2/3} \, 
    |\psiomega(\nhat)|^{-2/3} 
    \per
    \nonumber
}
We require $\radres > \RNS$ such that the conversion occurs outside the star, since otherwise the photon would likely scatter on the neutron star matter, rather than escaping from the system as radiation. 
The width of the resonance region, denoted by $\dres$, is related to how quickly $\Delta_a(\rad) - \Delta_\parallel(\rad)$ varies at the resonance radius.  
We find 
\ba{\label{eq:Pag_EH_main_text}
    & \dres(\omega,\nhat) 
    = 
    \sqrt{\frac{4\pi}{3}} \frac{\sqrt{\ka \radres}}{\wpl} \, 
    |\beta_T|^{-1} 
    \\ & \quad \approx 
    \bigl( 4.5 \times 10^{-2} \, \radres \bigr) 
    \biggl( \frac{B_0}{10^{14} \Gauss} \biggr)^{\! \! -1/6} 
    \biggl( \frac{P}{1 \sec} \biggr)^{\! \! 5/6} 
    \nn & \qquad \times 
    \biggl( \frac{\RNS}{10 \km} \biggr)^{\! \! -1/2} 
    \biggl( \frac{n_e}{|n_c|} \biggr)^{\! \! -5/6} \, 
    \biggl( \frac{\omega}{10 \meV} \biggr)^{\! \! 2/3} 
    \nn & \qquad \times 
    \biggl( \frac{\ka}{10 \meV} \biggr)^{\! \! 1/2} 
    | \beta_T(\nhat) |^{-1} \, 
    |\psivecB(\nhat)|^{2/3} \, 
    |\psiomega(\nhat)|^{-5/3}
    \com 
    \nonumber 
}
and one can find additional details in \aref{app:derivation}.  
Note that $\dres$ grows in relation to $\radres$ as the axion energy $\omega \approx \ka$ is increased.  
Provided that $\dres \ll \radres$, the axion-photon conversion probability may be calculated by employing the stationary phase approximation.  
To leading order in the axion-photon coupling $\gagg$, we find 
\ba{\label{eq:Pag_EH_main_text}
    & \Pag(\omega,\nhat) 
    = 
    \frac{2\pi}{21} \frac{\gagg^2 \radres}{\ggggg \ka} 
    = \frac{1}{4} \frac{\omega^2}{\ka^2} \gagg^2 \barB_T^2 \dres^2   
    \\ & \quad \approx 
    \bigl( 3.7 \times 10^{-2} \bigr) 
    \biggl( \frac{\gagg}{10^{-12} \GeV^{-1}} \biggr)^{\! \! 2} 
    \nn & \qquad \times 
    \biggl( \frac{B_0}{10^{14} \Gauss} \biggr)^{\! \! 1/3} 
    \biggl( \frac{P}{1 \sec} \biggr)^{\! \! 1/3} 
    \biggl( \frac{\RNS}{10 \km} \biggr)^{} 
    \biggl( \frac{n_e}{|n_c|} \biggr)^{\! \! -1/3} \, 
    \nn & \qquad \times 
    \biggl( \frac{\omega}{10 \meV} \biggr)^{\! \! 2/3} 
    \biggl( \frac{\ka}{10 \meV} \biggr)^{\! \! -1} 
    \nn & \qquad \times 
    |\psivecB(\nhat)|^{2/3} \, 
    |\psiomega(\nhat)|^{-2/3} 
    \per
    \nonumber
}
For relativistic axions $\ka \approx \omega$ and $\Pag \propto \omega^{2/3} \ka^{-1} \approx \omega^{-1/3}$, implying that conversion via the \EHR{} is less likely for higher energy axions.  
For comparison, the \MMR{} has instead $\Pag \propto \omega^2 \ka^{-1} \approx \omega^1$ such that higher-energy axions are more likely to convert into photons. 

\begin{figure}[t]
\centering
\includegraphics[width=0.45\textwidth]{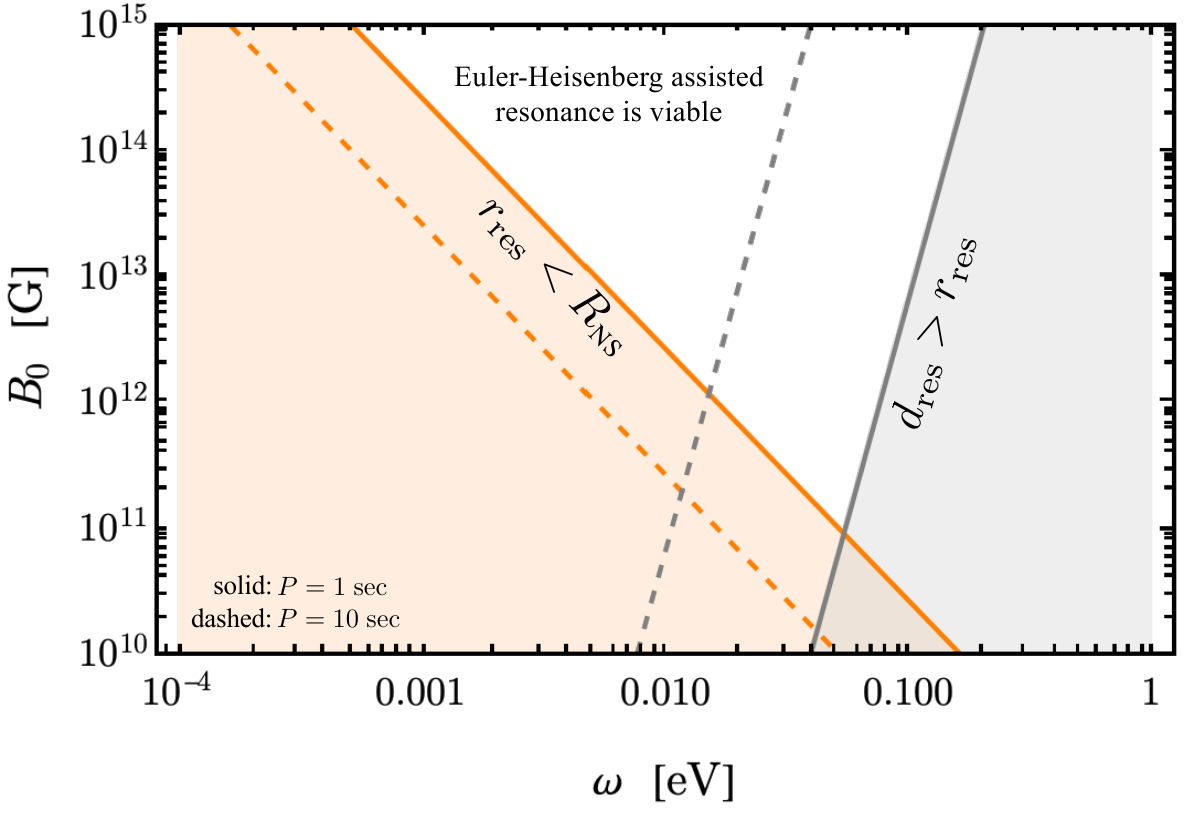}
\caption{\label{fig:omega_range}
An illustration of the parameter space over which the Euler-Heisenberg assisted resonance is viable.  We show the axion energy $\omega$ and the polar magnetic field strength $B_0$.  In the orange shaded region the resonance radius would be inside the neutron star, $\radres < \RNS$.  In the gray shaded region, the resonance width exceeds the resonance radius, $\dres > \radres$.   We neglect the orientation-dependent factors by setting $|\psivecB(\nhat)| = \psiomega(\nhat) = \beta_T(\nhat) = 1$, and we take $\RNS = 10 \km$ and $n_e = |n_c|$.  The solid and dashed lines correspond to the boundaries for $P = 1 \sec$ and $10 \sec$, respectively. For example, if $B_0 = 10^{14} \Gauss$ then the Euler-Heisenberg assisted resonance is accessible for $\omega$ from about $1 \meV$ to $100 \meV$.   If the electron density were larger, $n_e \gg |n_c|$, the resonance would shift to higher $\omega$.
}
\end{figure}

In order for this calculation to be self-consistent it is necessary to impose two conditions.  
First, the resonance must occur outside of the neutron star; this condition requires $\RNS < \radres$.  
Second, the resonance region must be narrow (since we calculate $\Pbb_{a\to\gamma}$ using the stationary phase approximation); this condition requires $\dres < \radres$.  
Together these two conditions bracket a range of axion energies $\omega_\mathrm{min} < \omega < \omega_\mathrm{max}$ where 
\ba{\label{eq:EHres_omega_range}
    & \omega_\mathrm{min}(\nhat)  
    = \sqrt{\frac{2}{7}} \frac{\omega_{\mathrm{pl},0}}{\sqrt{\ggggg} B_0} \frac{|\psiomega|}{|\psivecB|} 
    \\ & \quad \approx 
    \bigl( 1.6 \meV \bigr) 
    \biggl( \frac{B_0}{10^{14} \Gauss} \biggr)^{\! \! -1/2} 
    \biggl( \frac{P}{1 \sec} \biggr)^{\! \! -1/2} 
    \biggl( \frac{n_e}{|n_c|} \biggr)^{\! \! 1/2} \, 
    \nn & \qquad \times 
    |\psiomega(\nhat)|^{} \, 
    |\psivecB(\nhat)|^{-1} 
    \nn 
    & \omega_\mathrm{max}(\nhat) 
    = \frac{3^{3/7}}{2^{4/7} 7^{2/7} \pi^{3/7}} \frac{\RNS^{3/7} \omega_{\mathrm{pl},0}^{10/7}}{\ggggg^{2/7} B_0^{4/7}} \frac{|\beta_T|^{6/7} \, |\psiomega|^{10/7}}{|\psivecB|^{4/7}} 
    \nn & \quad \approx 
    \bigl( 140 \meV \bigr) 
    \biggl( \frac{B_0}{10^{14} \Gauss} \biggr)^{\! \! 1/7} 
    \biggl( \frac{\RNS}{10 \km} \biggr)^{\! \! 3/7} 
    \biggl( \frac{P}{1 \sec} \biggr)^{\! \! -5/7} 
    \biggl( \frac{n_e}{|n_c|} \biggr)^{\! \! 5/7} \, 
    \nn & \qquad \times 
    |\beta_T(\nhat)|^{6/7} \, 
    |\psiomega(\nhat)|^{10/7} \, 
    |\psivecB(\nhat)|^{-4/7} 
    \per 
    \nonumber
}
For typical parameters, the \EHR{} is relevant for axion energies from $1$ to $100 \meV$.  
This is illustrated in \fref{fig:omega_range}.  
The window widens for larger $B_0$, and it closes off entirely for $B_0 \lesssim (10^{11} \Gauss) (P / 1 \sec)^{1/3}$.  
Consequently, the \EHR{} is only expected to occur in the strong-field environment of a neutron star magnetosphere.  
As the neutron star period decreases, the window of axion energies shifts to smaller values.  
For a millisecond pulsar ($P \sim 1-10 \; \mathrm{ms}$) \cite{Petri:2016tqe} the window would be in the microwave band; however, the typical magnetic field strength ($B \lesssim 10^{10} \Gauss$) would make the \EHR{} inaccessible for millisecond pulsars.  

\subsubsection{Nonresonant conversion}
\label{sub:nonres}

It is also possible that the resonance condition \eqref{eq:res_condit} does not admit any real solutions with $\radres > \RNS$.  
For instance if the axion energy $\omega$ is large compared to the axion mass $m_a$ and the plasma frequency $\wpl$, such that $\ggggg \barB^2 \omega^2 \gg m_a^2$ and $\wpl^2$, then the first three terms in \eref{eq:res_condit} can be neglected with respect to the fourth, and the resonance condition has no real solution.  
In this situation, the axion-photon interconversion is said to be nonresonant.  
Several studies~\cite{Morris:1984iz,Raffelt:1987im,Lai:2001di,Gill:2011yp,Fortin:2018aom,Dessert:2019sgw,Buschmann:2019pfp,Dessert:2021bkv,Fortin:2021sst,Dessert:2022yqq} have noted that nonresonant axion-photon conversion can play an important role in generating X-ray signals of axion emission from neutrons stars (including magnetars) and white dwarf stars, while other work~\cite{Dobrynina:2014qba} has studied the conversion of TeV gamma rays into axions as they propagate across the galaxy.  
If the axion mass and plasma frequency can be neglected, then a simple formula is available for the nonresonant conversion probability~\cite{Fortin:2018aom}
\ba{\label{eq:Pag_NR_main_text}
    & \Pag(\omega,\nhat) 
    \approx \frac{\Gamma(\tfrac{2}{5})^2}{2^{2/5} 5^{6/5} 7^{4/5}} \frac{\gagg^2}{\ggggg^{4/5}} \barB_{T,0}^{2/5} \RNS^{6/5} \frac{\omega^{2/5}}{\ka^{6/5}} \\ 
    & \quad 
    \approx 
    \bigl( 5.1 \times 10^{-6} \bigr) 
    \biggl( \frac{\gagg}{10^{-12} \GeV^{-1}} \biggr)^{\! \! 2} 
    \biggl( \frac{B_0}{10^{14} \Gauss} \biggr)^{\! \! 2/5} 
    \nn & \qquad \times 
    \biggl( \frac{\RNS}{10 \km} \biggr)^{\! \! 6/5} 
    \biggl( \frac{\omega}{1 \keV} \biggr)^{\! \! 2/5} 
    \biggl( \frac{\ka}{1 \keV} \biggr)^{\! \! -6/5} 
    \nn & \qquad \times 
    |\beta_T(\nhat)|^{2/5} \, 
    |\psivecB(\nhat)|^{2/5} 
    \com 
    \nonumber 
}
where $\barB_{T,0} = B_0 \, |\beta_T| \, |\psivecB|$. 
For relativistic axions $\ka = (\omega^2 - m_a^2)^{1/2} \approx \omega$ and $\Pag \propto \omega^{2/5} \ka^{-6/5} \approx \omega^{-4/5}$.  
For comparison the \EHR{} has $\Pag \propto \omega^{-1/3}$, which decreases less quickly toward large axion energy.  
See \aref{app:nonres} for additional details and generalizations. 

\subsubsection{Comparison of the two resonances}
\label{sub:comparison}

\begin{figure}[t]
\centering
\includegraphics[width=0.49\textwidth]{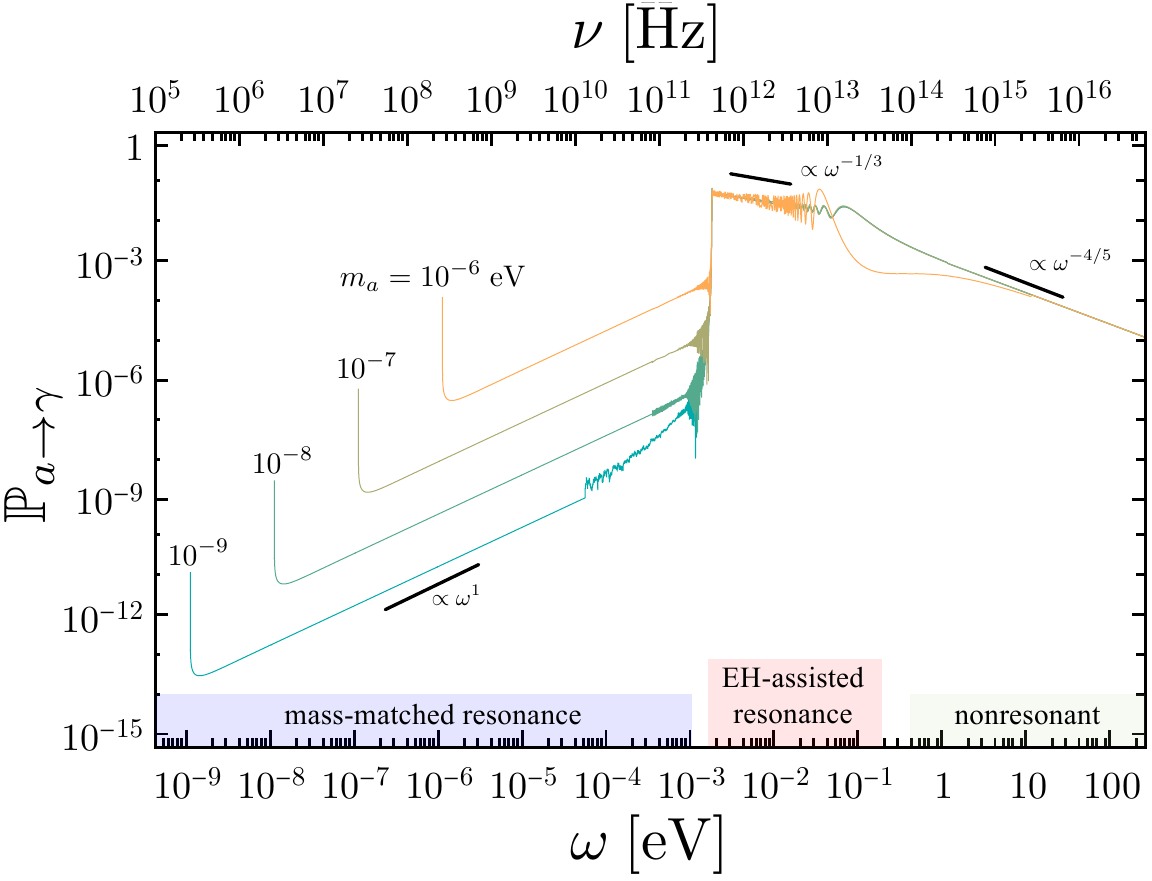}
\caption{\label{fig:Pag_v_omega}
Axion-photon conversion probability $\Pag$ as a function of axion energy $\omega$ (or equivalent frequency $\nu = \omega / 2\pi$) for several values of the axion mass $m_a$. Toward small $\omega$ the curves are truncated at $\omega \approx m_a$ corresponding to $k_a = (\omega^2 - m_a^2)^{1/2} = 10^{-3} m_a$.  The power-law dependence on $\omega$ is very well approximated by the analytical formulas in the text:  at small $\omega$ the mass-matched resonance \eqref{eq:Pag_MM_main_text} has $\Pag \propto \omega^{2} \ka^{-1}$, at intermediate $\omega$ the Euler-Heisenberg assisted resonance \eqref{eq:Pag_EH_main_text} has $\Pag \propto \omega^{2/3} \ka^{-1}$, and at large $\omega$ the nonresonant conversion \eqref{eq:Pag_NR_main_text} has $\Pag \propto \omega^{2/5} \ka^{-6/5}$.  The mass-matched resonance is suppressed toward small axion mass $m_a$, even if the axion is relativistic $\omega \gg m_a$; whereas for these parameters the Euler-Heisenberg assisted resonance is insensitive to the axion mass for $m_a \lesssim \mu\mathrm{eV}$.  We take the same fiducial parameters used in the main text: $\gagg = 10^{-12} \GeV^{-1}$, $B_0 = 10^{14} \Gauss$, $P = 1 \sec$, $\RNS = 10 \km$, and $n_e = |n_c|$. We neglect the orientation-dependent factors by setting $|\psivecB(\nhat)| = \psiomega(\nhat) = 1$ and $\beta_L(\nhat) = \beta_T(\nhat) = 1/\sqrt{2}$.
}
\end{figure}

\Fref{fig:Pag_v_omega} shows the predicted axion-photon conversion probability $\Pag$ as a function of the axion energy $\omega$ for a fiducial parameter set (see caption).  
To produce this figure, we solved the coupled axion-photon system \eqref{eq:MatrixHE1} numerically, and we evaluated the analytic formulas from the previous subsections.  
The curves shown in the figure combine numerical results at intermediate $\omega$ with analytical results at small and large $\omega$.  
The figure illustrates how the \MMR{} dominates at small $\omega$, the \EHR{} dominates at intermediate $\omega$, and the nonresonant conversion dominates at large $\omega$.  

Toward small axion energy, the axion-photon interconversion is controlled by the \MMR{}, and the parametric behavior can be understood from \eref{eq:Pag_MM_main_text} in which $\Pag \propto m_a^{4/3} \omega^2 \ka^{-1}$.  
For a given $m_a$ we truncate the curves at $\ka = 10^{-3} m_a$ corresponding to $\omega \approx m_a$.  
For nonrelativistic axions, which have $\omega \approx m_a \gg \ka$, the probability spikes, since $\Pag \propto \ka^{-1}$.  
For relativistic axions, which have $\omega \approx \ka \gg m_a$, the probability grows linearly, since $\Pag \propto \omega^2 \ka^{-1} \approx \omega$.  
This behavior is partly because the probability increases as the width of the resonance region increases: $\Pag \propto \dres^2 \omega^2 / \ka^2$ and $\dres \propto \ka^{1/2}$.  
For the curve with $m_a = 10^{-9} \eV$, as $\omega$ is increased the width of the resonance region $\dres$ becomes comparable to the resonance radius $\radres$ before the \EHR{} is reached, and this explains why the curve departs from the $\omega^1$ scaling.  
The dependence on axion mass follows the expectation for the \MMR{} in which $\Pag \propto m_a^{4/3} \omega^2$.  
Lowering the axion mass reduces the probability, because the resonance radius obeys $\wpl(\radres) \approx m_a$, and a smaller $m_a$ implies a larger $\radres$ where the magnetic field is weaker, leading to a smaller probability. 

At intermediate axion energy, the axion-photon conversion is controlled by the \EHR{}, and the parametric behavior can be understood from \eref{eq:Pag_EH_main_text} in which $\Pag \propto m_a^{0} \omega^{2/3} \ka^{-1}$.  
For the parameters that were chosen to generate this figure, the probability is insensitive to the axion mass as long as $m_a \lesssim \mu\mathrm{eV}$.  
Since $m_a \ll \omega \approx \ka$, axions in the \EHR{} window are relativistic.  
The sharp jump at $\omega \approx 10^{-3} \eV$ corresponds to the point where the \EHR{} resonance radius is comparable to the neutron star radius $\radres \approx \RNS$; see also \fref{fig:omega_range} and \eref{eq:EHres_omega_range}.  
For smaller $\omega$ we have $\radres < \RNS$ and the \EHR{} is inaccessible; whereas for larger $\omega$ we have $\radres > \RNS$ and the \EHR{} dominates over the \MMR{}.  
As the axion energy increases further, the axion-photon conversion probability tends to decrease as $\Pag \propto \omega^{-1/3}$, which is the expected behavior for the \EHR{}.  
This is mostly because $\radres \propto \omega^{2/3}$ so a higher energy axion converts further from the star, where the magnetic field is weaker, leading to a smaller probability.  
For $\omega \gtrsim 0.1 \eV$ the resonance width exceeds the resonance radius $\dres > \radres$ and the \EHR{} gives way to nonresonant conversion; see also \fref{fig:omega_range} and \eref{eq:EHres_omega_range}.  
If the axion mass were larger than $m_a \sim \mu\mathrm{eV}$ then the \EHR{} and \MMR{} would merge, see \fref{fig:res_condit}, and the probability would be suppressed at high axion energy.  

At high axion energy, the axion-photon conversion is nonresonant, and the parametric behavior can be understood from \eref{eq:Pag_NR_main_text} in which $\Pag \propto m_a^{0} \omega^{2/5} \ka^{-6/5}$.  
For relativistic axions this implies $\Pag \propto \omega^{-4/5}$, which agrees with the behavior seen in the figure.  

Comparing the two resonances against one another, we find that the \EHR{} tends to yield a larger axion-photon conversion probability than the \MMR{} for low-mass and high-energy relativistic axions.  
This conclusion is robust against changing the magnetic field strength $B_0$, the axion-photon coupling $\gagg$, and the neutron star radius $\RNS$ since both resonances have $\Pag \propto \gagg^2 B_0^{1/3} \RNS^{}$.  
As for the neutron star rotation period $P$, the \MMR{} has $\Pag \propto P^{5/3}$ while the \EHR{} has $\Pag \propto P^{1/3}$, but most (non-millisecond) pulsars have similar rotation periods $P \approx 1-10 \sec$. 
The conversion probability also depends on the orientation of the axion's trajectory $\nhat = \rhat$ via the factors $|\psivecB(\nhat)|$, $\psiomega(\nhat)$, and $\beta_T(\nhat)$, which we have neglected in generating \fref{fig:Pag_v_omega}.  
We explore the angular dependence further in \aref{app:numerical}. 

\subsubsection{EFT considerations}
\label{sub:EFT}

Since the Wilson coefficient $\ggggg$ of the dimension-8 Euler-Heisenberg operator appears in the denominator of the resonance condition \eqref{eq:rc_EH}, one might worry that the effective field theory (EFT) breaks down at the \EHR{}.  
However, through the following power-counting argument, we argue that this is not the case.  

At the Lagrangian level, we compare the various terms that impact the electromagnetic field.  
There is a dim-2 effective mass term due to medium effects, schematically written as $\Lscr_2 \sim \wpl^2 A^2$, and there is the dim-8 EH term, schematically written as $\Lscr_8 \sim \ggggg F^4 \sim (e^4 / m_e^4) \omega^2 \barB^2 A^2$ where $\ggggg \sim e^4 / m_e^4$.  
We focus on only the terms that contain two photon field operators (\textit{i.e.}, $\propto A^2$); there are also interaction operators (\textit{i.e.}, $\propto A^4$), but these do not impact axion-photon mixing.  
Going beyond mass dimension-8 operators, one would next encounter terms like $\Lscr_{12} \sim (e^6 / m_e^8) F^6 \sim (e^6 / m_e^8) \omega^6 \barB^2 A^2$ at mass dimension-12~\cite{Dunne:2004nc}.  
At the \EHR{}, the dim-2 and dim-8 operators are comparable by definition $\Lscr_8 / \Lscr_2 = O(1)$; note that \eref{eq:rc_EH} gives $\wpl^2 \sim (e^4 / m_e^4) \omega^2 \barB^2$.  
The dim-12 operator goes as $\Lscr_{12} / \Lscr_2 \propto \rad^6$, which is small at the star's surface and grows with increasing distance from the star.  
At the resonance radius $\radres$ it reaches a value of $\Lscr_{12} / \Lscr_2 \sim e^2 \omega^4 / m_e^4 \sim 10^{-30} (\omega / 10 \meV)^4$.  
However, provided that $\omega \ll m_e / \sqrt{e} \approx 1 \MeV$, the dim-12 operator is tiny in comparison to the dim-2 and dim-8 operators at the resonance radius.  
In short, although the resonance condition \eqref{eq:rc_EH} needs $\omega$ to be much larger than the plasma frequency, it still allows $\omega$ to be much smaller than the electron mass, which sets the UV cutoff of the EFT.  
Therefore, for the parameters of interest, we conclude that the EFT is under control at the \EHR{}.  

\section{Terahertz emission due to axion dark radiation}
\label{sec:emission}

Here we consider the electromagnetic radiation that may arise when axion dark radiation, \textit{i.e.} a population of relativistic axions, is incident on a neutron star's magnetosphere.  
We assess the conditions under which resonant axion-photon conversion is expected to occur, and we calculate the resultant electromagnetic radiation spectrum. 

\subsection{Modeling axion dark radiation}
\label{sub:axion_radiation}

We assume that relativistic axions are isotropically incident on the magnetosphere.  
This is a reasonable expectation for all of the sources of axion dark radiation discussed in \sref{sec:sources}.  
To take advantage of the system's approximate spherical symmetry, it is convenient to imagine that the axions are being emitted from the center of the neutron star at $\rvec = 0$.  
We model the incident axion dark radiation with an energy flux spectrum, $\Phi_a(\omega) = \dd E_a / \dd^2 \nhat \dd t \dd \omega$, which has the units of axion energy per unit solid angle per unit time per unit angular frequency.  
We assume the emission is isotropic and static, meaning that $\Phi_a$ doesn't depend on either $\nhat$ or $t$.  
For the sake of simplicity, we assume that the flux of axion radiation onto the neutron star can be modeled as a power-law across the observable frequency band.  
This assumption lets us write 
\bes{
    \Phi_a(\omega) & = \Phi_{a,0} \, \bigl( \omega / \omega_0 \bigr)^n 
    \com
}
where the exponent $n$ is a (possibly non-integer) index.  

\subsection{Predicted electromagnetic radiation}
\label{sub:EM_radiation}

If an axion with angular frequency $\omega$ that propagates in the direction $\nhat$ (such that its momentum is $\kavec = (\omega^2 - m_a^2)^{1/2} \nhat$) experiences resonant axion-photon conversion with probability $\Pag(\omega,\nhat)$, then the average resultant energy flux spectrum of the emitted electromagnetic radiation is given by 
\ba{
    \Phi_\gamma(\omega,\nhat) = \Phi_a(\omega) \, \Pag(\omega,\nhat) 
    \com
}
where $\Phi_\gamma(\omega,\nhat) = \dd E_\gamma / \dd^2 \nhat \dd t \dd \omega$ has the units of photon energy per unit solid angle per unit time per unit angular frequency. 
Note that the probability depends on the orientation of the axion's momentum $\nhat$, because of the anisotropy of the dipolar magnetic field and plasma frequency.  
Here we've also used the fact that an axion with energy $\omega$ converts into a photon with the same energy. 
The electromagnetic luminosity spectrum is obtained by integrating the flux over solid angle: 
\ba{
    \Lcal_\gamma(\omega) = \int_{4\pi} \! \dd^2 \nhat \ \Phi_\gamma(\omega,\nhat)
    \com
}
where $\Lcal_\gamma(\omega) = \dd E_\gamma / \dd t \dd \omega$ has units of photon energy emitted per unit time per unit angular frequency.  
The total electromagnetic luminosity is obtained by further integrating over angular frequency: 
\ba{
    L_\gamma 
    = \int_0^\infty \! \dd \omega \ \Lcal_\gamma(\omega) 
    = \int_{4\pi} \! \dd^2 \nhat \int_0^\infty \! \dd \omega \ \Phi_\gamma(\omega,\nhat) 
    \com
}
and it has units of photon energy emitted per unit time.  

We calculate the flux that's incident at Earth as 
\ba{
    \Fcal_\gamma(\omega) = \Lcal_\gamma(\omega) / 4 \pi d^2 
    \quad \text{or} \quad 
    F_\gamma = L_\gamma / 4 \pi d^2 
    \com
}
where $\Fcal_\gamma(\omega) = \dd E_\gamma / \dd^2 A \dd t \dd \omega$ has the units of photon energy received per unit area per unit time per unit angular frequency and $F_\gamma = \dd E_\gamma / \dd^2 A \dd t$ has the units of photon energy received per unit area per unit time.  
This calculation assumes that the neutron star is a distance $d$ away from Earth.  
Additionally, this calculation neglects absorption of the radiation along the path of propagation, which effectively sets the optical depth $\tau$ to zero. 

\subsection{Numerical estimates}
\label{sub:estimates}

We are interested in assessing whether the electromagnetic radiation that arises from resonant conversion of axion dark radiation in a neutron star's magnetosphere could be detected from Earth.  
Combining formulas from the previous subsections gives 
\bes{
    \omega \Fcal_\gamma(\omega) 
    & = \frac{\Phi_{a,0} \, \omega^{n+1}}{4 \pi d^2 \omega_0^n} \int_{4\pi} \! \dd^2 \nhat \, \Pbb_{a\to\gamma}(\omega, \nhat) 
    \per
}
Assuming that the \EHR{} is dominant, so $\Pbb_{a\to\gamma}$ is given by \eref{eq:Pag_EH_main_text}, we evaluate the integral to obtain 
\ba{\label{eq:omega_F_gamma_estimate}
    \omega \Fcal_\gamma(\omega) & \approx 
    \bigl( 6 \times 10^{-11} \erg / \mathrm{cm}^2 / \mathrm{sec} \bigr) \, 
    \\ & \quad \times 
    \biggl( \frac{\gagg}{10^{-12} \GeV^{-1}} \biggr)^{\! \! 2} 
    \biggl( \frac{B_0}{10^{14} \Gauss} \biggr)^{\! \! 1/3} 
    \nn & \quad \times 
    \biggl( \frac{P}{1 \sec} \biggr)^{\! \! 1/3} 
    \biggl( \frac{\RNS}{10 \km} \biggr)^{} 
    \biggl( \frac{n_e}{|n_c|} \biggr)^{\! \! -1/3} \, 
    \nn & \quad \times 
    \biggl( \frac{\omega}{10 \meV} \biggr)^{n+5/3} 
    \biggl( \frac{\ka}{10 \meV} \biggr)^{\! \! -1} 
    \biggl( \frac{\omega_0}{10 \meV} \biggr)^{\! \! -n} 
    \nn & \quad \times 
    \biggl( \frac{d}{100 \pc} \biggr)^{\! \! -2}  
    \biggl( \frac{\Phi_{a,0}}{10^{30} \erg / \mathrm{sr} / \mathrm{sec} / \mathrm{meV}} \biggr) 
    \per
    \nonumber
}
For reference $1 \Jy \approx 1.5 \times 10^{-11} \erg / \mathrm{cm}^2 / \mathrm{sec} / \mathrm{meV}$.  
We have fiducialized the distance to $100 \pc$, and for reference the nearby isolated neutron star RX J1856.5-3754 is located at $123_{-15}^{+11} \pc$ \cite{Walter:2010xxx,Mignani:2016fwz}.  
We have fiducialized the axion flux spectrum to a value that gives a potentially detectable signal. 
In \sref{sec:sources}, we estimate $\Phi_{a,0}$ for possible sources of axion dark radiation. 

For reference, if we model the Sun as a blackbody with surface temperature $T \approx 6000 \; \mathrm{K} \approx 500 \meV$, then at $\omega = 10 \meV$ the flux reaching the Earth's upper atmosphere is $\omega \, \Fcal_\gamma(\omega) \approx 1.4 \times 10^{5} \erg / \mathrm{cm}^2 / \mathrm{sec}$.  
As another reference, the diffuse extragalactic background at these energies straddles the cosmic microwave background (CMB) and the cosmic infrared background (CIB)~\cite{Hauser:2001xs}, and a flux of $10 \; \mathrm{nW} / \mathrm{m}^2 / \mathrm{sr}$ corresponds to $\omega \Fcal_\gamma \sim 6 \times 10^4 \erg / \mathrm{cm}^2 / \mathrm{sec}$ upon integrating over $2 \pi \;\mathrm{sr}$.  

An angular frequency of $\omega = 10 \meV$ corresponds to a linear frequency of $\nu = \omega / 2 \pi \approx 2.4 \THz$ and a wavelength of $\lambda = 1/\nu \approx 0.1 \; \mathrm{mm}$.  
For these fiducial parameters, the signal falls in the `terahertz gap' of the electromagnetic spectrum, between the microwave band and the far infrared band.  
Detection of terahertz radiation is particularly challenging for a receiver on Earth, since atmospheric attenuation strongly suppresses the signal.  

The authors of \rref{2016PASJ...68R...1H} discuss possible science cases for ground-based terrahertz telescopes.  
In particular they focus on the 12-m Greenland Telescope that would be deployed to the Greenland Summit Station.  
The dry atmospheric conditions in Greenland help to mitigate the attenuation due to atmospheric water vapor.  
In order to detect their various scientific objectives, the noise level of a terrahertz telescope must be lowered to a sensitivity of $0.1-10 \Jy$ at frequencies of $1-1.5 \THz$.  
This value motivates our fiducial parameter choices in \eref{eq:omega_F_gamma_estimate}: $(6 \times 10^{-11} \erg / \mathrm{cm}^2 / \mathrm{sec}) / (10 \meV) \approx 0.4 \Jy$.  

\section{Sources of axion dark radiation}
\label{sec:sources}

The phrase `axion dark radiation' refers to a population of relativistic axions, which we assume to be incident upon a neutron star's magnetosphere with flux $\Phi_a(\omega)$.  
Depending on their mass and interactions, relativistic axions may be produced in various ways. 
Here we briefly mention several possibilities and estimate $\Phi_a$.  

\subsection{Thermal relic axions}
\label{sub:thermal_relic}

Axions may have reached thermal equilibrium with the primordial plasma in the early universe if their interactions with the Standard Model particles are sufficiently strong. 
Later as the universe expanded and the plasma cooled, the axions' interactions would have become inadequate to maintain thermal equilibrium, and they would have frozen out of equilibrium~\cite{Berezhiani:1992rk,Salvio:2013iaa,Chacko:2015noa,Baumann:2016wac,Arias-Aragon:2020shv,Arias:2023wyg}.
In this scenario, the Universe today is expected to contain a thermal cosmic axion background (CaB)~\cite{Conlon:2013isa,Dror:2021nyr}, similar to the thermal photon background (CMB) and the thermal neutrino background (C$\nu$B).  

The predicted CaB energy per unit volume per unit angular frequency is 
\ba{
    \frac{\dd \rho_a}{\dd \omega} = \frac{1}{2\pi^2} \frac{\omega^3}{e^{\omega/T_a} - 1} 
    \com
}
assuming that the axion mass $m_a$ is small compared to the CaB temperature today $T_a$.  
If the CaB froze out when the photon temperature was $T_d$, then the conservation of comoving entropy implies that the CaB temperature today is $T_a \simeq [ g_{\ast,S}(T_0) / g_{\ast,S}(T_d) ]^{1/3} \, T_\gamma$ where $T_\gamma \approx 0.234 \meV$ is the CMB temperature today.  
We estimate the corresponding axion flux at the conversion radius $\radres$ as $\Phi_a(\omega) = \dd E_a / \dd^2 \nhat \dd t \dd \omega \sim (\dd E_a / \dd V \dd \omega) (c \radres^2 / \mathrm{sr}) = (\dd \rho_a / \dd \omega) \, (c \radres^2 / \mathrm{sr})$.  
For $T_a = 0.1 \meV$, $\radres = 100 \km \approx 10 \RNS$, and $\omega = 1 \meV$ this estimate gives $\Phi_a \sim 1 \times 10^9 \erg / \mathrm{sr} / \mathrm{sec} / \mathrm{meV}$, which is much smaller than the fiducial value used in the estimate of $\omega \Fcal_\gamma(\omega)$ from \eref{eq:omega_F_gamma_estimate}.  
Consequently, the thermal CaB is not expected to induce a detectable electromagnetic signal from resonant axion-photon conversion in a neutron star magnetosphere.  

\subsection{Stellar axion emission}
\label{sub:stellar}

Axion dark radiation may be created in the universe today by scattering processes taking place within stars~\cite{Raffelt:1996wa}.  
In particular, if axions couple to the constituents of a neutron star, and if the axion is lighter than the neutron star core temperature $T \sim 10 \keV$, then neutron stars create their own flux of relativistic axions.  
The dominant channels, typically nucleon bremstrahlung $NN \to NNa$ (where $N = p,n$) or Cooper pair breaking and formation, are controlled by the axion-nucleon coupling $\gaNN$.  
Since the star must not lose too much energy via axion emission, one can derive an upper limit on $\gaNN$ from luminosity and age measurements of isolated neutron stars. 
For example, \rref{Buschmann:2021juv} study axion emission from a set of five isolated neutron stars and derive limits on $\gaNN$, which are weakly model dependent, and generally at level of $|\gaNN| < 1 \times 10^{-9}$ at $95\%$ confidence.  
When this inequality is saturated, the corresponding axion luminosity is on the order of $L_a \sim 10^{32}-10^{33} \erg / \mathrm{sec}$ where, for reference, the photon luminosity of the Sun is $L_\odot \approx 3.8 \times 10^{33} \erg / \mathrm{sec}$.  
Assuming a thermal spectrum at $T \sim 10 \keV$, the axion flux at angular frequencies around $\omega_0 = 10 \meV$ is $\Phi_a(\omega) \approx \Phi_{a,0} \, (\omega / \omega_0)^{2}$ with $\Phi_{a,0} \sim 10^{12} \erg / \mathrm{sr} / \mathrm{sec} / \mathrm{meV}$.  
This flux is much smaller than the fiducial value of $\Phi_{a,0}$ used to estimate $\omega \Fcal_\gamma$ at \eref{eq:omega_F_gamma_estimate}, indicating a negligible terrahertz signal. 
There is a larger flux at higher energies ($\omega \sim T$), which may lead to X-ray emission via nonresonant conversion \cite{Fortin:2018ehg,Buschmann:2019pfp,Fortin:2021sst}, but these energies are outside the window \eqref{eq:EHres_omega_range} that experience resonant conversion via the EHR. 
Additional channels for axion emission include gap production in the polar cap regions~\cite{Prabhu:2021zve,Noordhuis:2022ljw,Prabhu:2023cgb}, electrobaryonic axion hair~\cite{Bai:2023bbg}, and axion stellar basin~\cite{VanTilburg:2020jvl}.  
However, for each of these channels the axion spectrum is peaked far away from $\omega \approx \mathrm{meV}$ and the flux in the \EHR{} energy window is suppressed. 
Consequently, the axion radiation produced by the neutron star itself is not expected to induce a detectable electromagnetic signal after resonant axion-photon conversion in the magnetosphere via the EHR.  

\subsection{From dark matter decay or annihilation}
\label{sub:decay}

Relativistic axions may also be created through the decay or annihilation of a heavier cosmological relic~\cite{Cicoli:2012aq,Higaki:2012ar,Conlon:2013isa,Conlon:2013txa,Higaki:2013lra,Hebecker:2014gka,Cicoli:2014bfa,Cui:2017ytb,Moroi:2020has,Jaeckel:2021ert}.  
For instance, the dark matter may be unstable but sufficiently long lived to survive in the Universe today.  
If dark matter particles have a decay channel into much lighter axions, then dark matter decays would produce axion dark radiation~\cite{Dror:2021nyr}.  
The spectrum of the axion dark radiation depends on the nature of the decay.  
If the dark matter decays to a two-body final state, such as $\chi \to aa$, then the axion spectrum would have two components:  a line centered at approximately half of the dark matter mass, $\omega \approx m_\chi / 2$, associated with dark matter decaying today, and a low-frequency power-law tail associated with earlier decays into axions that experienced cosmological redshift. 
Generating axions with energies around $1-100 \meV$ would require a dark matter particle with its mass around this scale.  
Alternatively if the dark matter decays to a three-body (or higher) final state, then a broader spectrum of axion emission would result.  
Similarly, if dark matter particles become trapped inside a neutron star and annihilate with one another~\cite{Bell:2023ysh}, then axion dark radiation could also be produced in this way.  

To estimate the associated energy flux spectrum $\Phi_a(\omega)$ we make the (wildly overly-generous) assumption that all of the dark matter energy is transformed into axion dark radiation at $\omega \approx 10 \meV$.  
In other words, we take $\dd\rho_a/\dd\omega \sim \rho_\mathrm{dm} / \omega$ and assume that the dark matter energy density in the vicinity of the neutron star is comparable to the value in the Milky Way nearby to our solar system, $\rho_\mathrm{dm} \sim 0.3 \GeV / \mathrm{cm}^3$.  
By adapting the estimates used in \sref{sub:thermal_relic}, we obtain $\Phi_a \sim 1 \times 10^{18} \erg / \mathrm{sr} / \mathrm{sec} / \mathrm{meV}$.  
Comparing with the fiducial axion flux used in the estimate of \eref{eq:omega_F_gamma_estimate}, we do not anticipate a detectable signal due to axion dark radiation produced from dark matter decay.  
It is worth mentioning that the dark matter density is much larger than the local value near the galactic center, but taking $d \approx 8 \kpc$ in \eref{eq:omega_F_gamma_estimate} would incur a suppression of approximately $10^{-6}$, and moreover attenuation during propagation would become increasingly important.  

\subsection{Topological defect production}
\label{sub:topo_defect}

Axion-like particles often arise in field theories as the pseudo-Goldstone boson of a spontaneously broken global symmetry.  
If the symmetry breaking proceeded through a cosmological phase transition in the early universe, causality arguments~\cite{Kibble:1976sj} imply that a cosmological network of topological defects, typically strings and domain walls, should have formed.  
A combination of analytical arguments and numerical simulations indicate that these defect networks evolve by exhausting energy into axions~\cite{Yamaguchi:1998gx,Hagmann:1998me,Yamaguchi:2002sh,Hiramatsu:2010yu,Hiramatsu:2012gg,Kawasaki:2014sqa,Lopez-Eiguren:2017dmc,Gorghetto:2018myk,Kawasaki:2018bzv,Vaquero:2018tib,Martins:2018dqg,Buschmann:2019icd,Hindmarsh:2019csc,Klaer:2019fxc,Gorghetto:2020qws,Hindmarsh:2021vih}.  
If the axion is sufficiently light, those particles constitute a component of the axion dark radiation today~\cite{Harari:1987ht,Long:2018nsl}.  
Assuming that strings with tension $\mu$ lose an order one fraction of their energy in each Hubble time into relativistic axions, then the accumulated axion dark radiation in the universe today can be estimated roughly as $\rho_a \sim \mu H_0^2 \log(t_\mathrm{eq} / t_\text{\sc pq})$ where logarithmic factor accounts for the build up of axion dark radiation from the time of string formation $t_\text{\sc pq}$ until the time of radiation-matter equality $t_\mathrm{eq}$.  
If the string tension were $\mu \sim (10^{14} \GeV)^2$ and the logarithimic factor were $\log \sim 10^2$, then the corresponding cosmological energy fraction in these axions is expected to be quite small $\rho_a / 3 M_\mathrm{pl}^2 H_0^2 \sim 10^{-6}$.  
The corresponding axion flux $\Phi_a(\omega)$ onto a neutron star magnetosphere would also be quite small, even in comparison with the estimates for thermal relic axions from \sref{sub:thermal_relic}.  

\section{Summary and conclusion}
\label{sec:conclusion}

Let us first summarize the key elements of this work.  
\begin{enumerate}
    \item  We are interested in the resonant conversion of axions into photons in a neutron star magnetosphere.  We identify two parametrically-distinct classes of resonant solutions; see \fref{fig:res_condit}.  The first, which we call the mass-matched resonance (\MMR{}), can occur if there is a region outside the star where the plasma frequency crosses the axion mass, $\wpl \approx m_a$.  The second, which we call the Euler-Heisenberg assisted resonance (\EHR{}), can occur at a location where the axion energy satisfies $\omega = [2 \wpl^2 / 7 \ggggg \barB^2]^{1/2}$.  This second resonance relies upon the Euler-Heisenberg four-photon self-interaction ($\ggggg = 8 \alpha^2 / 45 m_e^4$).  Both resonances collectively have been called `axion-photon resonance' by \rref{Lai:2006af}.  
    \item  We focus our study on the \EHR{}.  We find that it is generally only accessible for relativistic axions with energy $\omega$ in the range of $1 \meV$ to $100 \meV$; see \fref{fig:omega_range}.  Additionally, for relativistic axions, we find that the \EHR{} tends to induce a larger axion-photon conversion than either the MMR or nonresonant conversion; see \fref{fig:Pag_v_omega}.  
    \item  We suppose that a population of relativistic axions is incident on a strongly-magnetized neutron star.  We calculate the resultant electromagnetic radiation that would arise via the \EHR{} as the axions propagate through the neutron star's magnetosphere.  We estimate the spectrum of radiation that would reach a detector on Earth; see \eref{eq:omega_F_gamma_estimate}.  Obtaining a signal that is strong enough to detect would require a very large flux of axions.  Moreover, the signal typically falls into the `terahertz gap' between the microwave and infrared bands, which is attenuated by the atmosphere, making detection additionally challenging. 
    \item  We briefly survey a few possible sources of axion dark radiation.  These include a primordial population of thermal relic axions, axions produced in stars, axions produced from the decay of dark matter (or other long-lived moduli), and axions produced from topological defect networks.  These sources of axion dark radiation generally do not provide the requisite flux of axions to induce a detectable electromagnetic signal due to axion-photon conversion via the EHR.  
\end{enumerate}

Our main conclusions are that the \EHR{} can provide the dominant channel for axion-photon conversion in a neutron star magnetosphere for axions with energy from $\omega \sim 1$ to $100 \meV$, provided that the magnetic field is sufficiently strong $B_0 \gtrsim 10^{11} \Gauss$ and the axion is sufficiently light $m_a \lesssim \mu\mathrm{eV}$.  
For typical values of the polar magnetic field strength $B \gtrsim 10^{11} \Gauss$ and the neutron star rotation period $P \sim 1-10 \sec$, the resultant electromagnetic radiation would fall into the terahertz band, allowing for synergy with laboratory efforts to detect axions at terahertz masses~\cite{Barry:2021,BREAD:2021tpx}.  
However, given reasonable restrictions on the incident axion flux, the associated electromagnetic signal is not expected to be large enough to detect.  

In our study we have adopted several simplifying assumptions in order to arrive at simple analytical relations for the properties of the \EHR{}.  
Of course, it is possible to relax these assumptions and revisit the analysis with a more accurate modeling of the axion-photon system.  
For example, we assume axions propagate on radial trajectories out from the center of the star, but one can perform three-dimensional ray tracing as in Refs.~\cite{Safdi:2018oeu,Dessert:2019sgw,McDonald:2023shx}.  
Moreover, we assume that any electromagnetic radiation produced by axion-photon conversion is able to escape from the neutron star environment and reach Earth, but one can also consider photon-plasma interactions and propagation effects as in Refs.~\cite{Witte:2021arp,Tjemsland:2023vvc}.  
However, we do not expect a more refined analysis to alter the broader conclusions about the detectability of axion-induced terahertz radiation from the \EHR{}.  

In this work we have focused on the conversion of axions into photons.  
Although the \EHR{} allows for a large conversion probability, the strength of the signal is limited by the tiny incident flux of axion dark radiation.  
It may also be interesting to explore the conversion of photons into axions, which could lead to potentially detectable features in the spectrum or polarization of the background radiation.  
For magnetic white dwarf stars, in which the weaker magnetic field makes the \EHR{} inaccessible and the axion-photon conversion is nonresonant, similar studies have led to constraints on the axion-photon coupling from observations of polarized optical emission~\cite{Gill:2011yp,Dessert:2022yqq}. 
As for neutron stars, a pair of recent studies have already investigated the role of the \EHR{} to modulate the star's spectrum and polarization. 
\rref{Bondarenko:2022ngb} uses measurements of the high-frequency radio spectrum of a radio-loud magnetar to search for a kink-like feature (corresponding in our notation to $\omega \approx \omega_\mathrm{min}$ where the \EHR{} becomes accessible).  
They conclude that if such a feature could be detected with $O(5 - 20\%)$ sensitivity, it would provide a strong probe of axion-like particles at $\gagg \gtrsim (0.5-2) \times 10^{-12} \GeV^{-1}$ and $m_a \lesssim \mu\mathrm{eV}$.  
\rref{Song:2024rru} uses optical linear polarization measurements for a set of three neutron stars to search for an enhancement in polarized emission due to the \EHR{}.  
Using upper limits on the polarization degree for these stars, they obtain powerful constraints on the axion-photon coupling that can be as strong as $\gagg < \mathrm{few} \times 10^{-12} \GeV^{-1}$ for small $m_a$.  
In our view these studies demonstrate the potentially important role of the \EHR{} to probe axions in neutron star environments.  

\acknowledgments

We thank Brandon Khek for collaboration and discussion in the early stages of this work.  We are grateful to Christopher Dessert for thoughtful feedback on a draft of this paper.  We are particularly grateful to Kuver Sinha for an illuminating discussion of nonresonant conversion.  
A.J.L thanks Mustafa Amin, Kimberly Boddy, and Benjamin Lehmann for encouragement and moral support. 
E.D.S thanks Andreas Reisenegger (Metropolitan University of Educational Sciences, Chile) for fruitful discussions about neutron star properties.  A.J.L. is supported in part by the National Science Foundation under Award No.~PHY-2114024.  

\phantom{.}\newline 
\noindent$^*$\href{mailto:andrewjlong@rice.edu}{andrewjlong@rice.edu} \\
$^\dagger$\href{mailto:Enrico.Schiappacasse@uss.cl}{Enrico.Schiappacasse@uss.cl}

\appendix
\widetext

\section{Selected derivations}
\label{app:derivation}

This appendix provides details for the derivations of the equations of motion, the mixing matrix, the axion-photon conversion probability, and the resonance condition.  
In the second appendix, we explore the angular dependence and validate the analytical approximations against direct numerical integration of the equations of motion.  

\subsection{Equations of motion}
\label{app:EOM}

The field equations for the axion-photon system \eqref{eq:Lagrangian} are given by 
\bsa{eq:EOM_1}{
	\ddot{a} - \nabla^2 a + m_a^2 a & = \gagg \Evec \cdot \Bvec \\ 
	\dvec \cdot \Dvec & = \rho_f - \gagg \, \dvec a \cdot \Bvec 
	\\ & \qquad 
	- 2 \ggggg \, \dvec \cdot \Bigl[ \tfrac{1}{2} \bigl( |\Evec|^2 - |\Bvec|^2 \bigr) \Evec + \tfrac{7}{4} \bigl( \Evec \cdot \Bvec \bigr) \Bvec \Bigr] \nn 
	\dvec \times \Hvec & = \jvec_f + \dot{\Dvec} + \gagg \, \dot{a} \, \Bvec + \gagg \, \dvec a \times \Evec 
	\\ & \qquad 
	+ 2 \ggggg \, \partial_t \Bigl[ \tfrac{1}{2} \bigl( |\Evec|^2 - |\Bvec|^2 \bigr) \Evec + \tfrac{7}{4} \bigl( \Evec \cdot \Bvec \bigr) \Bvec \Bigr] 
	\nn & \qquad 
	- 2 \ggggg \, \dvec \times \Bigl[ \tfrac{1}{2} \bigl( |\Evec|^2 - |\Bvec|^2 \bigr) \Bvec - \tfrac{7}{4} \bigl( \Evec \cdot \Bvec \bigr) \Evec \Bigr] \nn 
	\dvec \cdot \Bvec & = 0 \\ 
	\dvec \times \Evec & = - \dot{\Bvec} 
	\com
}
which combine the Klein-Gordon equation with the in-medium Maxwell equations, and which further include the appropriate modifications to account for the axion-photon interaction ($\gagg$ terms)~\cite{Wilczek:1987mv} and the Euler-Heisenberg photon self-interaction ($\ggggg$ terms)~\cite{Heisenberg:1936nmg}. 
These equations contain the axion field $a(\rvec,t)$, the electric field $\Evec(\rvec,t)$, the magnetic field $\Bvec(\rvec,t)$, the electric displacement field $\Dvec(\rvec,t)$, the field $\Hvec(\rvec,t)$, the free charge density field $\rho_f(\rvec,t)$, and the free current density field $\jvec_f(\rvec,t)$.  
Since we are interested in axion-photon conversion in a neutron star magnetosphere, we assume Minkowski spacetime, but for cosmological applications one may want to assume an Friedmann-Robertson-Walker spacetime instead.  

\subsection{Electric and magnetic potentials}
\label{app:potentials}

The magnetic divergence condition $\dvec \cdot \Bvec = 0$ and Faraday's Law $\dvec \times \Evec = - \dot{\Bvec}$ are constraint equations.  
These constraints are solved by electric and magnetic fields that take the form 
\ba{\label{eq:introduce_potentials}
    \Evec(\rvec,t) = - \dvec \phi(\rvec,t) - \dot{\Avec}(\rvec,t) 
    \qquad \text{and} \qquad 
    \Bvec(\rvec,t) = \dvec \times \Avec(\rvec,t) 
    \com
}
where $\phi(\rvec,t)$ is the electric potential field and $\Avec(\rvec,t)$ is the magnetic potential field.  
The electric and magnetic fields are left invariant under the gauge transformations $\phi(\rvec,t) \mapsto \phi(\rvec,t) - \dot{\chi}(\rvec,t)$ and $\Avec(\rvec,t) \mapsto \Avec(\rvec,t) + \dvec \chi(\rvec,t)$ for any differentiable $\chi(\rvec,t)$.  
In studies of axion-photon interconversion, it is customary to use the gauge freedom to impose $\phi(\rvec,t) = 0$ for all $t$ and $\rvec$, and this gauge-fixing condition is known as the Weyl or temporal gauge.  

\subsection{Constitutive relations}
\label{app:medium}

A set of constitutive relations determine $\Dvec$ and $\Hvec$ in terms of $\Evec$ and $\Bvec$ for the medium of interest.  
At this point it is useful to anticipate that we intend to search for plane wave solutions that vary in time as $\propto \ee^{-\ii \omega t}$ where $\omega$ is the angular frequency of oscillation.  
The constitutive relations may depend on $\omega$ as well.  

For a linear medium the constitutive relations take the form $\Dvec = \epsilon \, \Evec$ and $\Hvec = \mu^{-1} \, \Bvec$ where $\epsilon(\rvec,t)$ is called the electric permittivity tensor and $\mu^{-1}(\rvec,t)$ is the inverse magnetic permeability tensor.  
If the medium were also isotropic and homogeneous, then the tensors would be independent of $\rvec$ and proportional to the identity matrix.  

For example, a linear, isotropic, and homogeneous plasma composed of nonrelativistic charge carriers has $\epsilon_{ij} = (1-\wpl^2 / \omega^2) \, \delta_{ij}$ and $\mu_{ij}^{-1} = \delta_{ij}$ where $\wpl$ is called the plasma frequency~\cite{Sitenko:1967}.  
The squared plasma frequency is calculated by summing the contributions from each particle species in the system: 
\ba{
    \wpl^2 = \sum_\psi \omega_{\mathrm{pl},s}^2 
    \qquad \text{where} \qquad 
    \omega_{\mathrm{pl},s}^2 = \frac{q_s^2 e^2 n_s}{m_s} 
    \com
}
where $q_s$ is the electric charge, $n_s$ is the number density, and $m_s$ is the mass of particle species $s$.  
Note $e = \sqrt{4 \pi \alpha} \approx 0.303$.  
In a nonrelativistic electron-ion plasma the plasma frequency contributions are $\omega_{\mathrm{pl},e}^2 = e^2 n_e / m_e$ for the electrons and $\omega_{\mathrm{pl},\mathrm{ion}}^2 = Z^2 e^2 n_\mathrm{ion} / m_\mathrm{ion}$ for the ions with atomic number $Z$ and mass number $A$.  
Since $m_e \ll m_\mathrm{ion}/Z^2 \approx A m_p / Z^2$ even for reasonably large $Z$, it follows that $\omega_{\mathrm{pl},\mathrm{ion}}^2 \ll \omega_{\mathrm{pl},e}^2$ if the system is approximately charge neutral $n_e \approx n_\mathrm{ion}$.  

For a neutron star magnetosphere, the strong dipolar magnetic field $\Bvec$ breaks the spatial isotropy of the system.  
Charged particles tend to move up and down the field lines while orbiting in the transverse direction at their cyclotron frequency.  
If we write $B = |\Bvec|$ and $\hat{\Bvec} = \Bvec / B$ then the cyclotron frequency for particles of species $s$ is $\omega_{B,s} = |q_s| e B / m_s$.  
Consequently the dielectric permittivity in the direction of $\Bvec$ is different from the permittivity in the directions normal to $\Bvec$.  
Since the particles can only move down the field lines, the conductivity in the transverse directions is approximately zero.  
In the high magnetization limit, \textit{i.e.} $\wB \gg \omega, \wpl$, the appropriate modification of the constitutive relations is~\cite{Beskin:1993xx} 
\bes{\label{eq:constitutive_relations}
    \Dvec(\rvec,t) 
    & = \epsilon(\rvec,t) \, \Evec(\rvec,t) 
    \qquad \text{where} \qquad 
    \epsilon_{ij} 
    = \delta_{ij} - \tfrac{\wpl^2}{\omega^2} \, \hat{B}_i \hat{B}_j \\ 
    \Hvec(\rvec,t) 
    & = \mu^{-1}(\rvec,t) \, \Bvec(\rvec,t) 
    \qquad \text{where} \qquad 
    \mu_{ij}^{-1} 
    = \delta_{ij} 
    \per 
}
In other words, the electric permittivity is $1 - \wpl^2 / \omega^2$ in the direction of the background magnetic field $\Bvec$, and it is $1$ in the orthogonal directions.  
Note that $\wpl(\rvec,t)$ varies throughout the neutron star magnetosphere, and we treat it as a part of the background solution. 
In the equatorial regions, the neutron star magnetosphere primarily consists of a nonrelativistic electron-proton plasma~\cite{Hook:2018iia}, and the plasma frequency is well approximated by just the electron contribution $\wpl^2 = e^2 n_e / m_e$.  
On the other hand, in the polar regions, a relativistic electron-positron plasma is present, and the calculation of axion-photon conversion near the poles requires a different analysis, which we do not consider here; see for example Refs.~\cite{Safdi:2018oeu,Noordhuis:2022ljw,Caputo:2023cpv,Noordhuis:2023wid}.   

\subsection{Background and perturbations}
\label{app:background_and_perturb}

A neutron star magnetosphere possesses a strong electromagnetic field associated with the star's large magnetic dipole moment.  
Additionally the axion dark radiation corresponds to an ambient axion field.  
However, we are interested in the much weaker electromagnetic and axion fields associated with radiation propagating through this environment. 
This motivates us to employ perturbation theory to derive an approximate solution.  
We decompose the various fields onto background and perturbations by writing 
\bsa{eq:bkg_expand}{
	\rho_f(\rvec,t) & = \bar{\rho}_f(\rvec,t) \\
    \jvec_f(\rvec,t) & = \bar{\jvec}_f(\rvec,t) \\
	a(\rvec,t) & = \bara(\rvec,t) + \da(\rvec,t) \\ 
	\Evec(\rvec,t) & = \barEvec(\rvec,t) + \dEvec(\rvec,t) \\ 
    \Dvec(\rvec,t) & = \barDvec(\rvec,t) + \dDvec(\rvec,t) \\ 
	\Bvec(\rvec,t) & = \barBvec(\rvec,t) + \dBvec(\rvec,t) \\ 
    \Hvec(\rvec,t) & = \barHvec(\rvec,t) + \dHvec(\rvec,t) \\ 
	\phi(\rvec,t) & = \bar{\phi}(\rvec,t) + \dphi(\rvec,t) \\ 
 	\Avec(\rvec,t) & = \barAvec(\rvec,t) + \dAvec(\rvec,t) 
    \com
}
where background quantities are denoted by a bar.  
Note that the free charge and current density are treated as entirely background quantities, since these are responsible for the neutron star's dipolar electromagnetic field.  
The barred background quantities are assumed to solve \eref{eq:EOM_1} exactly, although in practice the $\gagg$ and $\ggggg$ terms can be neglected.  
Since the perturbations are assumed to be small compared with the background quantities, we can neglect terms that are quadratic or cubic in the perturbations.  
Keeping only the linear terms, the equations become 
\bsa{eq:EOM_2}{
	\dotdotda - \nabla^2 \da + m_a^2 \da & = \gagg \barEvec \cdot \dBvec + \gagg \dEvec \cdot \barBvec \\ 
	\dvec \cdot \dDvec & = - \gagg \, \dvec \da \cdot \barBvec - \gagg \, \dvec \bara \cdot \dBvec 
	\\ & \qquad 
	- 2 \ggggg \, \dvec \cdot \Bigl[ \bigl( \barEvec \cdot \dEvec - \barBvec \cdot \dBvec \bigr) \barEvec + \tfrac{1}{2} \bigl( |\barEvec|^2 - |\barBvec|^2 \bigr) \dEvec 
    \nn & \hspace{3cm} 
    + \tfrac{7}{4} \bigl( \barEvec \cdot \barBvec \bigr) \dBvec + \tfrac{7}{4} \bigl( \barEvec \cdot \dBvec \bigr) \barBvec + \tfrac{7}{4} \bigl( \dEvec \cdot \barBvec \bigr) \barBvec \Bigr] \nn 
	\dvec \times \dHvec & = \dotdDvec + \gagg \, \dotbara \, \dBvec + \gagg \, \dotda \, \barBvec + \gagg \, \dvec \bara \times \dEvec + \gagg \, \dvec \da \times \barEvec 
	\\ & \qquad 
	+ 2 \ggggg \, \partial_t \Bigl[ \bigl( \barEvec \cdot \dEvec - \barBvec \cdot \dBvec \bigr) \barEvec + \tfrac{1}{2} \bigl( |\barEvec|^2 - |\barBvec|^2 \bigr) \dEvec 
    \nn & \hspace{3cm} 
    + \tfrac{7}{4} \bigl( \barEvec \cdot \barBvec \bigr) \dBvec + \tfrac{7}{4} \bigl( \barEvec \cdot \dBvec \bigr) \barBvec + \tfrac{7}{4} \bigl( \dEvec \cdot \barBvec \bigr) \barBvec \Bigr] 
	\nn & \qquad 
	- 2 \ggggg \, \dvec \times \Bigl[ \bigl( \barEvec \cdot \dEvec - \barBvec \cdot \dBvec \bigr) \barBvec + \tfrac{1}{2} \bigl( |\barEvec|^2 - |\barBvec|^2 \bigr) \dBvec 
    \nn & \hspace{3cm} 
    - \tfrac{7}{4} \bigl( \barEvec \cdot \barBvec \bigr) \dEvec - \tfrac{7}{4} \bigl( \barEvec \cdot \dBvec \bigr) \barEvec - \tfrac{7}{4} \bigl( \dEvec \cdot \barBvec \bigr) \barEvec \Bigr] \nn 
	\dvec \cdot \dBvec & = 0 \\ 
	\dvec \times \dEvec & = - \dotdBvec \\ 
    \dEvec & = - \dvec \dphi - \dotdAvec \\ 
    \dBvec & = \dvec \times \dAvec \\ 
    \dDvec & = \dEvec - \tfrac{\wpl^2}{\omega^2} \bigl( \barBhat \cdot \dEvec \bigr) \, \barBhat \\ 
    \dHvec & = \dBvec 
	\com
}
which govern the evolution of the field perturbations. 

\subsection{Goldreich-Julian background solution}
\label{app:background}

The background fields are required to solve 
\bes{
	& \ddot{\bara} - \nabla^2 \bara + m_a^2 \bara = 0 
    \ , \qquad 
	\dvec \cdot \barDvec = \bar{\rho}_f 
    \ , \qquad 
	\dvec \times \barHvec = \bar{\jvec}_f + \dot{\barDvec} 
    \\ & \quad 
	\dvec \cdot \barBvec = 0 
    \ , \qquad 
	\dvec \times \barEvec = - \dotbarBvec 
    \ , \qquad 
    \barDvec = \barEvec 
    \ , \qquad 
    \barHvec = \barBvec 
	\com
}
where we have neglected the $O(\gagg)$ and $O(\ggggg)$ terms.  
We take $\bara(\rvec,t) = 0$, since the background axion field associated with axion dark radiation is not expected to be very large.  
In the polar regions where $\barEvec \cdot \barBvec$ can be significant, a nonzero $\bara$ can be induced by the axion-photon interaction ($\gagg$ terms) \cite{Prabhu:2021zve,Noordhuis:2022ljw,Prabhu:2023cgb}, but we neglect this effect since it is not expected to significantly impact resonant axion-photon conversion. 

We model the background magnetic field in the region outside of the neutron star as a magnetic dipole that rotates at the same speed as the star. 
Here we use a Cartesian coordinate system and take the origin to be the center of the star such that the position vector is $\rvec = x \, \xhat + y \, \yhat + z \, \zhat$ with $r = |\rvec| = (x^2 + y^2 + z^2)^{1/2}$ and $\rhat = \rvec/r$.  
We write the star's angular velocity as $\Ovec = \Omega \, \Ohat$ where $\Omega = 2\pi/P$ and $P$ is the rotation period of the star and $\Ohat = \zhat$ is the orientation of the angular velocity.  
We further suppose that the magnetic dipole moment is 
\ba{
    \mvec(t) = m \, \sin\theta_m \cos(\phi_m + \Omega t) \, \xhat + m \, \sin\theta_m \sin(\phi_m + \Omega t) \, \yhat + m \, \cos\theta_m \zhat 
    \com
}
which remains constant in magnitude $m = |\mvec(t)|$, and which varies in orientation $\mhat(t) = \mvec(t) / m$ while maintaining a constant the angle with respect to the star's rotation axis: $\Ohat \cdot \mhat(t) = \cos\theta_m$. 
This idealized magnetic dipole moment is a contribution to the free current density $\bar{\jvec}_f(\rvec,t)$ that is only nonzero at the origin. 
This approximation is reliable in the magnetosphere outside the star $(\rad \geq \RNS)$, but it would break down inside the star.  
The background $\barHvec(\rvec,t)$ field must obey Ampere's Law $\dvec \times \barHvec = \bar{\jvec}_f + \dot{\barDvec}$, which implies 
\bes{\label{eq:barHvec}
    & \bar\Hvec(\rvec,t) = \frac{3 \mvec(t) \cdot \rvec}{\rad^5} \, \rvec - \frac{\mvec(t)}{\rad^3} 
    \qquad \text{for $\rad \geq \RNS$}
    \per
}
Note that $\dvec \times \barHvec = 0$ everywhere except at the origin, which is the location of the magnetic dipole.  
In order for Ampere's Law to be satisfied, we will need to adjust the free current such that $\bar{\jvec}_f + \dot{\barDvec} = 0$.  
The background magnetic field $\barBvec(\rvec,t)$ must obey the constitutive relation $\barHvec = \barBvec$ since $\mu^{-1} = 1$, which implies 
\bes{\label{eq:barBvec}
    & \bar\Bvec(\rvec,t) 
    = B_0 \, 
    \psivecB(\rhat,t) \, 
    \biggl( \frac{r}{\RNS} \biggr)^{\! \! -3} 
    \qquad \text{where} \qquad 
    \psivecB(\rhat,t) = \tfrac{3}{2} \bigl( \mhat(t) \cdot \rhat \bigr) \, \rhat - \tfrac{1}{2} \mhat(t) 
    \qquad \text{for $\rad \geq \RNS$}
    \per
}
Note that the magnetic field strength at the star's surface in the direction of $\mvec$ is $B_0 = \barBvec(\RNS \, \mhat,t) = 2 m / \RNS^3$, and it is useful to treat $B_0$ as the free parameter rather than $m$.  
Note that this expression for the background magnetic field obeys the magnetic divergence condition $\dvec \cdot \barBvec = 0$.  
The background electric field $\barEvec(\rvec,t)$ must obey Faraday's Law $\dvec \times \barEvec = - \dotbarBvec$.  
One solution is 
\ba{\label{eq:barEvec}
    \barEvec(\rvec,t) 
    = - \bigl( \Ovec \times \rvec \bigr) \times \barBvec(\rvec,t)
    \com
}
and other solutions may be obtained by adding the gradient of a scalar field, since the additional term would have zero curl. 
Note that $\barEvec \cdot \barBvec = 0$.  
The associated electric displacement field $\barDvec(\rvec,t)$ is given by the constitutive relation $\barDvec = \barEvec$, and thereby implies 
\ba{\label{eq:barDvec}
    \barDvec(\rvec,t)  
    = - \bigl( \Ovec \times \rvec \bigr) \times \barBvec(\rvec,t)
    \per
}
The free charge density $\bar{\rho}_f(\rvec,t)$ is given by the in-medium Gauss's Law, $\dvec \cdot \barDvec = \bar{\rho}_f$, which implies 
\ba{\label{eq:barrhof}
    \bar{\rho}_f(\rvec,t) 
    = -2 \Ovec \cdot \barBvec(\rvec,t) 
    = - 2 \Omega B_0 \ \Ohat \cdot \psivecB(\rhat,t) \ \biggl( \frac{r}{\RNS} \biggr)^{\! \! -3}
    \qquad \text{for $\rad \geq \RNS$} 
    \per
}
If the star rotates so rapidly that relativistic corrections near its surface become significant, then an additional factor of $[1 - |\Ovec \times \rvec|^2]^{-1}$ is required. 
Similarly the free current density $\bar{\jvec}_f(\rvec,t)$ is given by the condition $\bar{\jvec}_f + \dot{\barDvec} = 0$, implying 
\ba{\label{eq:barjf}
    \bar{\jvec}_f(\rvec,t) = \bigl( \Ovec \times \rvec \bigr) \times \tfrac{\dd}{\dd t} \barBvec(\rvec,t) 
    \per
}
Note that if the dipole were aligned with the rotation axis of the star, corresponding to $\mhat = \Ohat$, then the current would vanish $\bar{\jvec}_f = 0$.  
The free charge and current densities arise from the motion of the plasma's charged constituents: nonrelativistic electrons and protons.  
In particular we can write $\bar{\rho}_f(\rvec,t) = (-e) \bar{n}_e(\rvec,t) + (+e) \bar{n}_p(\rvec,t)$ where $\bar{n}_e$ and $\bar{n}_p$ are the number densities of electrons and protons, respectively.  
The squared plasma frequency arises primarily from the lighter electrons, and we have $\wpl^2(\rvec,t) = e^2 \bar{n}_e(\rvec,t) / m_e$.  
If we estimate $\bar{n}_e \approx |n_c|$ where $n_c \equiv \bar{\rho}_f / (-e)$ is the effective number density of electric charge then the plasma frequency is given by 
\ba{\label{eq:barwpl}
    \wpl(\rvec,t) 
    & = \omega_{\mathrm{pl},0} \ 
    \psiomega(\rhat,t) \ 
    \biggl( \frac{r}{\RNS} \biggr)^{\! \! -3/2}
    \quad \text{where} \quad 
    \omega_{\mathrm{pl},0} = \sqrt{\frac{e \Omega B_0}{m_e} \frac{n_e}{|n_c|}} 
    \\ & \hspace{4.5cm} 
    \quad \text{and} \quad 
    \psiomega(\rhat,t) = \Bigl| 2 \Ohat \cdot \psivecB(\rhat,t) \Bigr|^{1/2} 
    \qquad \text{for $\rad \geq \RNS$}
    \per
    \nonumber
}
We estimate
\ba{\label{eq:wpl0}
    \omega_{\mathrm{pl},0} \approx 
    \bigl( 70 \ueV \bigr) 
    \biggl( \frac{B_0}{10^{14} \Gauss} \biggr)^{\! \! 1/2} 
    \biggl( \frac{P}{1 \sec} \biggr)^{\! \! -1/2} 
    \biggl( \frac{n_e}{|n_c|} \biggr)^{\! \! 1/2} \, 
    \com
}
using a fiducial set of parameters that are appropriate for a magnetar. 

Note that the effective fluid speed $v \sim \bar{j}_f / \bar{\rho}_f \sim r \Omega$ grows with increasing distance from the star.  
The approximation of a dipolar magnetic field with a frozen-in plasma breaks down at the light cylinder where $r \approx \Omega^{-1} \equiv \RLC$.  
Our analysis focuses on $r \ll \RLC$ where speeds are much smaller and the Goldreich-Julian dipole model is reliable.  

For the purposes of our calculation, it is a good approximation to neglect the background electric field $\barEvec$.  
One is led to this conclusion from the following argument. 
Inspecting \eref{eq:barEvec}, on sees the parametric relationship $\barE \sim \Omega r \barB$.  
At the star's surface $r \sim \RNS$ and this estimate gives $\barE \sim \Omega \RNS \barB$ where $\Omega \RNS = 2 \pi \RNS / P$ is the speed of a point on the surface of the star that rotates with period $P$.  
Different types of neutron stars have different rotation periods, ranging from millisecond pulsars with $P \approx 0.001 \ \mathrm{sec}$ to magnetars with $P \approx 10 \ \mathrm{sec}$~\cite{Lorimer:2006qs,Kaspi:2017fwg,Nattila:2022evn}. 
For a magnetar with $P = 1 \ \mathrm{sec}$ and $\RNS = 10 \ \mathrm{km}$, the speed is $\Omega \RNS \approx 2 \times 10^{-4}$, much smaller than the speed of light, and consequently $\barE \ll \barB$.  
Since one also expects solutions with $|\dEvec| \sim |\dBvec|$, then it is a good approximation to neglect terms that contain factors of $\barEvec$.  

\subsection{Propagation on a fixed background}
\label{app:propagation_fixed_bkg}

\begin{figure}[t]
\centering
\includegraphics[scale=0.85]{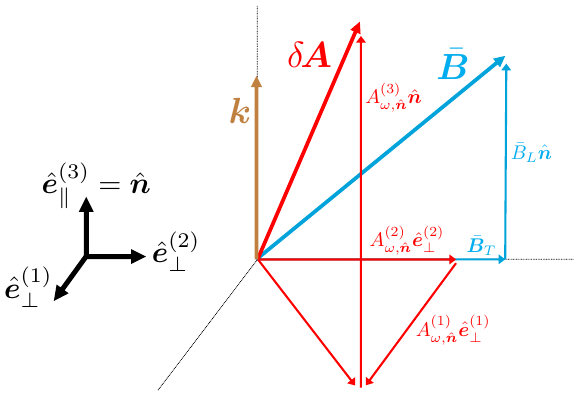}
\caption{\label{fig:coordinates}
An illustration of the coordinate system used in the text.  
}
\end{figure}

To build intuition, let us first study this system of equations under the assumption that the background is homogeneous and static.  
Upon dropping derivatives of $\bara$, $\barBvec$, and $\wpl$, the equations become 
\bsa{eq:EOM_4}{
	& \dotdotda - \nabla^2 \da + m_a^2 \da = \gagg \dEvec \cdot \barBvec \\ 
	& \dvec \cdot \dDvec = - \gagg \, \dvec \da \cdot \barBvec 
	+ 2 \ggggg \, \Bigl[ \tfrac{1}{2} |\barBvec|^2 \, \bigl( \dvec \cdot \dEvec \bigr) 
    - \tfrac{7}{4} \bigl( \barBvec \cdot \dvec \bigr) \bigl( \dEvec \cdot \barBvec \bigr) \Bigr] \\  
	& \dvec \times \dHvec = \dotdDvec + \gagg \, \dotda \, \barBvec 
	- 2 \ggggg \, \Bigl[ 
    \tfrac{1}{2} |\barBvec|^2 \dotdEvec 
    - \tfrac{7}{4} \bigl( \dotdEvec \cdot \barBvec \bigr) \barBvec 
	- \dvec \times \bigl( \barBvec \cdot \dBvec \bigr) \barBvec 
    -\tfrac{1}{2} |\barBvec|^2 \bigl( \dvec \times \dBvec \bigr) 
    \Bigr] \\ 
	& \dvec \cdot \dBvec = 0 \\ 
	& \dvec \times \dEvec = - \dotdBvec \\ 
    & \dDvec(\rvec,t) 
    = \dEvec(\rvec,t) - \tfrac{\wpl^2}{\omega^2} \, \bigl( \barBhat \cdot \dEvec \bigr) \, \barBhat \\ 
    & \dHvec(\rvec,t) 
    = \dBvec(\rvec,t) 
	\com
}
These equations admit plane wave solutions of the form 
\ba{\label{eq:Ansatz}
    \begin{pmatrix} \da(\rvec,t) \\ \dEvec(\rvec,t) \\ \dDvec(\rvec,t) \\ \dBvec(\rvec,t) \\ \dHvec(\rvec,t) \end{pmatrix} & = \vec{C}_{\omega,\nhat} \, \ee^{-\ii \omega t + \ii \kvec \cdot \rvec} + \mathrm{c.c.}  
    \qquad \text{where} \qquad 
    \vec{C}_{\omega,\nhat} = \begin{pmatrix} a_{\omega,\nhat} \\ E_{\omega,\nhat}^{(1)} \, \ehat_\perp^{(1)} + E_{\omega,\nhat}^{(2)} \, \ehat_\perp^{(2)} + E_{\omega,\nhat}^{(3)} \, \ehat_\parallel^{(3)} \\ D_{\omega,\nhat}^{(1)} \, \ehat_\perp^{(1)} + D_{\omega,\nhat}^{(2)} \, \ehat_\perp^{(2)} + D_{\omega,\nhat}^{(3)} \, \ehat_\parallel^{(3)} \\ B_{\omega,\nhat}^{(1)} \, \ehat_\perp^{(1)} + B_{\omega,\nhat}^{(2)} \, \ehat_\perp^{(2)} + B_{\omega,\nhat}^{(3)} \, \ehat_\parallel^{(3)} \\ H_{\omega,\nhat}^{(1)} \, \ehat_\perp^{(1)} + H_{\omega,\nhat}^{(2)} \, \ehat_\perp^{(2)} + H_{\omega,\nhat}^{(3)} \, \ehat_\parallel^{(3)} \end{pmatrix}
    \per 
}
Solutions are labeled by an angular frequency $\omega \in [0,\infty)$ and a unit vector $\nhat$, which gives the orientation of the wavevector $\kvec = k(\omega) \, \nhat$.  
The coefficients --- $a_{\omega,\nhat}$, $E_{\omega,\nhat}^{(1)}$, $E_{\omega,\nhat}^{(2)}$, and so on --- are complex-valued integration constants.  
The vector fields are further decomposed onto an orthonormal set of real basis 3-vectors $(\ehat_\perp^{(1)}, \, \ehat_\perp^{(2)}, \, \ehat_\parallel^{(3)})$.  
We take $\ehat_\parallel^{(3)} = \nhat$ to be the direction of the wavevector, and we arrange the other two unit vectors such that $\barBvec$ lies in the plane spanned by $\ehat_\parallel^{(3)}$ and $\ehat_\perp^{(2)}$; see \fref{fig:coordinates}. 
It is also convenient to define $\barB_L = \barBvec \cdot \ehat_\parallel^{(3)}$ and $\barBvec_T = \barBvec - \barB_L \, \ehat_\parallel^{(3)}$ and $\barB_T = \barBvec \cdot \ehat_\perp^{(2)}$ and $\barB = (\barBvec \cdot \barBvec)^{1/2} = (\barB_L^2 + \barB_T^2)^{1/2}$.  
Whereas $\barB_L$ may be either positive or negative (or zero), $\barB_T \geq 0$ must be nonnegative since $\ehat_\perp^{(2)}$ is defined with respect to the (fixed) orientation of $\barBvec$.  

For each choice of $\omega$ and $\nhat$, the solutions of \eref{eq:EOM_4} form a 6-dimensional vector space, which correspond to three families of solutions, each with its own c-number integration constant.
The first family of solutions has 
\ba{
    k(\omega) = k_\perp(\omega) \equiv \sqrt{ \frac{1 - \ggggg \barB^2}{1 - \ggggg (\barB^2 + 2 \barB_T^2)} } \, \omega \approx \omega + O(\ggggg \barB^2) 
    \qquad \text{and} \qquad 
    \vec{C}_{\omega,\nhat} = \begin{pmatrix} 0 \\ \ehat_\perp^{(1)} \\ \ehat_\perp^{(1)} \\ \frac{\ka}{\omega} \, \ehat_\perp^{(2)} \\ \frac{\ka}{\omega} \, \ehat_\perp^{(2)}
    \end{pmatrix} \, C_{\omega,\nhat,\perp} 
    \com
}
where $C_{\omega,\nhat,\perp}$ is the c-number integration constant. 
These solutions correspond to a plane wave in the electromagnetic field that propagates in the $\nhat = \ehat_\parallel^{(3)}$ direction, that is polarized in the $\ehat_\perp^{(1)}$ direction, and that has amplitude $C_{\omega,\nhat,\perp}$.  
The dispersion relation is modified away from free photon propagation ($k = \omega$) due to the Euler-Heisenberg interaction. 
The second family of solutions has 
\bes{
    k(\omega) & = k_a(\omega) \equiv \sqrt{ \frac{b - \sqrt{b^2 - 4 a c}}{2a} } \approx \sqrt{\omega^2 - m_a^2} + O(\gagg^2) + O(\gagg^2 \ggggg, \gagg^4) \\ 
    \vec{C}_{\omega,\nhat} & = \begin{pmatrix} C_{\omega,\nhat,a} \\ E_{\omega,\nhat}^{(2)} \, \ehat_\perp^{(2)} + E_{\omega,\nhat}^{(3)} \, \ehat_\parallel^{(3)} \\ D_{\omega,\nhat}^{(2)} \, \ehat_\perp^{(2)} + D_{\omega,\nhat}^{(3)} \, \ehat_\parallel^{(3)} \\ B_{\omega,\nhat}^{(1)} \, \ehat_\perp^{(1)} \\ H_{\omega,\nhat}^{(1)} \, \ehat_\perp^{(1)} \end{pmatrix}  
    \com
}
and the third family of solutions has 
 \bes{
    k(\omega) & = k_\parallel(\omega) \equiv \sqrt{ \frac{b + \sqrt{b^2 - 4 a c}}{2a} } \approx \sqrt{ \frac{1 - \tfrac{\wpl^2}{\omega^2}}{1 - \tfrac{\wpl^2}{\omega^2} \tfrac{\barB_L^2}{\barB^2}}} \, \omega + O(\gagg^2, \ggggg) \\ 
    \vec{C}_{\omega,\nhat} & = \text{same as second with $C_{\omega,\nhat,a} \to C_{\omega,\nhat,\parallel}$}
    \per
}
The squared wavevector solves a quadratic equation with positive coefficients 
\bes{
    a & = 1 - \frac{\wpl^2}{\omega^2} \, \frac{\barB_L^2}{\barB^2} - \ggggg \, \barB^2 \Bigl( 1 - \frac{7}{2} \frac{\barB_L^2}{\barB^2} \Bigr) \\ 
    b & = \bigl( \omega^2 - m_a^2 \bigr) \Bigl( 1 - \frac{\wpl^2}{\omega^2} \frac{\barB_L^2}{\barB^2} \Bigr) + \bigl( \omega^2 - \wpl^2 \bigr) - \gagg^2 \, \barB_L^2 + \ggggg \, \barB^2 \biggl[ \omega^2 \Bigl( \frac{3}{2} + \frac{7}{2} \, \frac{\barB_L^2}{\barB^2} \bigr) + m_a^2 \Bigl( 1 - \frac{7}{2} \, \frac{\barB_L^2}{\barB^2} \Bigr) \biggr] \\ 
    c & = \bigl( \omega^2 - m_a^2 \bigr) \bigl( \omega^2 - \wpl^2 \bigr) - \gagg^2 \, \barB^2 \, \omega^2 + \tfrac{5}{2} \ggggg \, \barB^2 \, \omega^2 \bigl( \omega^2 - m_a^2 \bigr) 
    \per
}
The nonzero entries in $\vec{C}_{\omega,\nhat}$ are 
\bes{ 
    E_{\omega,\nhat}^{(2)} & = - \frac{(\omega^2 - \ka^2 - m_a^2) \, \wpl^2 + \gagg^2 \barB^2 \omega^2 - \tfrac{7}{2} \ggggg \barB^2 \omega^2 (\omega^2 - \ka^2 - m_a^2)}{\gagg \, \barB \, (1 - \ggggg \, \barB^2) \, (\omega^2 - \ka^2)} \, \frac{\barB_T}{\barB} \, a_{\omega,\nhat} \\ 
    E_{\omega,\nhat}^{(3)} & = - \frac{\omega^2 - \ka^2 - m_a^2}{\gagg \, \barB_L} \, a_{\omega,\nhat} - \frac{\barB_T}{\barB_L} \, E_{\omega,\nhat}^{(2)} \\ 
    D_{\omega,\nhat}^{(2)} & = \Bigl( 1 - \tfrac{\wpl^2}{\omega^2} \tfrac{\barB_T^2}{\barB^2} \Bigr) \, E_{\omega,\nhat}^{(2)} - \tfrac{\wpl^2}{\omega^2} \, \tfrac{\barB_T \barB_L}{\barB^2} \, E_{\omega,\nhat}^{(3)} \\ 
    D_{\omega,\nhat}^{(3)} & = \Bigl( 1 - \tfrac{\wpl^2}{\omega^2} \tfrac{\barB_L^2}{\barB^2} \Bigr) \, E_{\omega,\nhat}^{(3)} - \tfrac{\wpl^2}{\omega^2} \, \tfrac{\barB_T \barB_L}{\barB^2} \, E_{\omega,\nhat}^{(2)} \\ 
    B_{\omega,\nhat}^{(1)} = H_{\omega,\nhat}^{(1)} & = - \frac{\ka}{\omega} \, E_{\omega,\nhat}^{(2)} 
    \per
}
If the couplings were zero, the second family of solutions would correspond to a plane wave propagating in the axion field (for $\omega > m_a$), and the third family would correspond to a plane wave propagating in the electromagnetic field (for $\omega > \wpl$) that is polarized in the plane spanned by $\ehat_\perp^{(2)}$ and $\ehat_\parallel^{(3)}$. 

It is interesting to investigate the conditions under which the wavenumbers for the different modes are equal, namely $k_a(\omega) = k_\parallel(\omega)$.  
For the fixed background, the plane wave solutions do not mix, but if the background varies slowly, then a plane wave in the axion field induces a plane wave in the electromagnetic field, or vice versa.  
The axion-photon interconversion may be resonantly enhanced when the momenta of the axion and photon are nearly equal, which corresponds to $k_a(\omega) = k_\parallel(\omega)$.  
Since these variables are the two roots of a quadratic equation, they are equal when the discriminant of the equation vanishes: $b^2 - 4 a c = 0$ implies $k_a(\omega) = k_\parallel(\omega) = \sqrt{b/2a} = \sqrt{2c/b}$.  
For general parameters, the discriminant equation $b^2 - 4 a c = 0$ is quartic in $\omega^2$.  
For the sake of illustration, consider the special cases: 
\bes{
    \text{if $\gagg = 0$ and $\ggggg = 0$ then: } 
    & \quad 
    \omega_\mathrm{res} = \biggl[ 1 - \frac{\barB^2}{\barB_L^2} \Bigl( 1 - \frac{m_a^2}{\wpl^2} \Bigr) \biggr]^{-1/2} \, m_a 
    \\ 
    \text{if $\gagg = 0$ and $m_a = 0$ then: } 
    & \quad 
    \omega_\mathrm{res} = k_\mathrm{res} = \sqrt{ \frac{2}{7 \ggggg \barB^2} } \, \wpl 
    \per
}
These two solution branches have the same resonance conditions as the \MMR{} and the \EHR{} from the main text.  

\subsection{Propagation on a slowly-varying background}
\label{app:propagation_varying_bkg}

Whereas the previous discussion was restricted to an idealized system with axion and electromagnetic waves propagating on a homogeneous and static background, we now turn our attention a more accurate modeling of the relevant physical system.  
As these waves propagate in a neutron star magnetosphere, the background fields can be modeled as a rotating magnetic dipole.  
Under this assumption, one can solve the linearized equations of motion using numerical methods and calculate the probability of axion-photon interconversion.  
However, if one seeks to derive analytical expressions, this background, which varies in both space and time, leads to equations that are intractable.  
Moreover, for the regime of interest, the calculation naturally simplifies in several respects, which we now enumerate.  
\begin{itemize}
    \item  To study the conversion of relativistic axions into photons, it is a good approximation to treat the background as static.  Since these waves travel at nearly the speed of light, the time it takes for them to traverse the neutron star's magnetosphere is short compared to the rotation period of the star, and on these times scales the background is effectively static.  Moreover, if the conversion is resonantly enhanced, then it typically occurs in a small region of space, and the light-crossing time for this region is generally much smaller than the rotation period of the star, which again justifies treating the background as static.  
    \item  Additionally, it is a good approximation to treat the background fields as homogeneous in the direction transverse to the axis of propagation.  In the framework of quantum mechanics, the incident axion can be viewed as a wavepacket.  The spatial extent of the wavepacket is on the order of the de Broglie wavelength for the particle, which set by its momentum $\pvec$ via $\lambda_p = 2 \pi / |\pvec|$.  If the momentum is sufficiently large, then the wavepacket is tiny, and the background fields are approximately homogeneous on the scale of the wavepacket, \textit{i.e.} $\lambda_p \ll |\barBvec| / |\nabla \barBvec| \sim \RNS$.  This justifies neglecting variations in the background fields transverse to the direction of motion.  See also \rref{Carenza:2023nck} for a discussion of when this approximation breaks down. 
    \item  It is a good approximation to neglect the background axion field.  The presence of axion dark radiation does not lead to a very large amplitude for the axion field.  
    \item  Finally we make one assumption for convenience.  The direction of the magnetic field is expected to vary along the path of the axion.  This is the case for any ray passing through a magnetic dipole, except ones passing through the origin.  In our study of propagation on a fixed background, we saw that the mode polarized normal to the plane spanned by $\barBvec$ and $\kvec$ decouples from the other two modes.  If the background magnetic field changes orientation, all three modes will couple during the propagation, which leads to a more challenging analysis.  As a matter of convenience, we assume that the orientation of the background magnetic field remains fixed along the axion's path of propagation.  This is accomplished by restricting $\rvec(t)$ to be a radial path that passes through the center of the star.  One can study the more general field configuration by solving the equations of motion numerically, and when this was done in earlier work the numerical and analytical results were found to agree with order one factors~\cite{Dessert:2019sgw}, and we expect a similar error here.  
\end{itemize}
With the preceding remarks in mind, we model the background as 
\ba{
    \barBvec(\rvec,t) = \barB(\rad) \, \barBhat = \barB_T(\rad) \, \ehat_\perp^{(2)} + \barB_L(\rad) \, \ehat_\parallel^{(3)}
    \qquad \text{and} \qquad 
    \wpl(\rvec,t) = \wpl(\rad) 
    \qquad \text{and} \qquad 
    \bara(\rvec,t) = 0 
    \com
}
where $\rad = \rvec \cdot \nhat$ is the spatial coordinate in the direction of propagation $\nhat$.  
Notice that $\dvec \cdot \barBvec = \barB^\prime(\rad) \, \barBhat \cdot \nhat$ need not be zero, and one may worry that this background does not solve Maxwell's equations.  
However, this is because we have neglected the variation of $\barB$ in the transverse direction, and if we had taken this into account, then we would have $\dvec \cdot \barBvec = 0$.  

We search for solutions of the form 
\ba{\label{eq:Ansatz2}
    \begin{pmatrix} \da(\rvec,t) \\ \dphi(\rvec,t) \\ \dAvec(\rvec,t) \end{pmatrix} & = \vec{C}_{\omega,\nhat}(\rad) \, \ee^{-\ii \omega t + \ii \kavec \cdot \rvec} + \mathrm{c.c.}  
    \qquad \text{where} \qquad 
    \vec{C}_{\omega,\nhat}(\rad) = \begin{pmatrix} 
    a_{\omega,\nhat}(\rad) \\ 
    \phi_{\omega,\nhat}(\rad) \\ 
    A_{\omega,\nhat}^{(1)}(\rad) \, \ehat_\perp^{(1)} + A_{\omega,\nhat}^{(2)}(\rad) \, \ehat_\perp^{(2)} + A_{\omega,\nhat}^{(3)}(\rad) \, \ehat_\parallel^{(3)} 
    \end{pmatrix}
    \per 
}
Note that $\kavec \cdot \rvec = \ka \rad$ where $\ka > 0$ is the wavenumber.  
Away from the magnetosphere where $\barB(\rad), \wpl(\rad) \to 0$, we want plane wave solutions to have constant amplitude $a_{\omega,\nhat}$.  
This is achieved by taking $\ka = (\omega^2 - m_a^2)^{1/2}$. 
Putting this Ansatz into the equations of motion and constitutive relations yields 
\begin{subequations}
\ba{
    \biggl[ 
    M_2(\rad) \frac{\dd^2}{\dd \rad^2} 
    + M_1(\rad) \frac{\dd}{\dd \rad} 
    + M_0(\rad) 
    \biggr] \begin{pmatrix} a_{\omega,\nhat}(\rad) \\ A_{\omega,\nhat}^{(2)}(\rad) \\ A_{\omega,\nhat}^{(1)}(\rad) \\ A_{\omega,\nhat}^{(3)}(\rad) \\ \phi_{\omega,\nhat}(\rad) \end{pmatrix} 
}
where the nonzero entries in the three matrices are given by 
\ba{
    (M_2)_{11} & = 1 \\ 
    (M_2)_{22} & = 1 - \ggggg \barB^2 \\ 
    (M_2)_{33} & = 1 - \ggggg \beta_L^2 \barB^2 - 3 \ggggg \beta_T^2 \barB^2 \\ 
    (M_2)_{45} & = 1 - \beta_L^2 \tfrac{\wpl^2}{\omega^2} + \tfrac{5}{2} \ggggg \beta_L^2 \barB^2 - \ggggg \beta_T^2 \barB^2 
    \com
}
\ba{
    (M_1)_{11} & = 2 \ii \ka \\ 
    (M_1)_{15} & = - \gagg \beta_L \barB \\ 
    (M_1)_{22} & = 2 \ii \ka - 2 \ii \ka \ggggg \barB^2 - \ggggg (\barB^2)^\prime \\
    (M_1)_{25} & = \tfrac{7}{2} \ii \omega \ggggg \beta_L \beta_T \barB^2 - \ii \omega \beta_L \beta_T \tfrac{\wpl^2}{\omega^2} \\
    (M_1)_{33} & = 2 \ii \ka - 2 \ii \ka \ggggg \beta_L^2 \barB^2 - 6 \ii \ka \ggggg \beta_T^2 \barB^2 - \ggggg \beta_L^2 (\barB^2)^\prime - 3 \ggggg \beta_T^2 (\barB^2)^\prime \\
    (M_1)_{41} & = - \gagg \beta_L \barB \\ 
    (M_1)_{42} & = - \tfrac{7}{2} \ii \omega \ggggg \beta_L \beta_T \barB^2 + \ii \omega \beta_L \beta_T \tfrac{\wpl^2}{\omega^2} \\ 
    (M_1)_{44} & = - \ii \omega - \tfrac{5}{2} \ii \omega \ggggg \beta_L^2 \barB^2 + \ii \omega \ggggg \beta_T^2 \barB^2 + \ii \omega \beta_L^2 \tfrac{\wpl^2}{\omega^2} \\
    (M_1)_{45} & = 2 \ii \ka + 5 \ii \ka \ggggg \beta_L^2 \barB^2 - 2 \ii \ka \ggggg \beta_T^2 \barB^2 - 2 \ii \ka \beta_L^2 \tfrac{\wpl^2}{\omega^2} 
    \\ & \qquad 
    + \tfrac{5}{2} \ggggg \beta_L^2 (\barB^2)^\prime - \ggggg \beta_T^2 (\barB^2)^\prime - \beta_L^2 \tfrac{(\wpl^2)^\prime}{\omega^2} \nn
    (M_1)_{55} & = \ii \omega + \tfrac{5}{2} \ii \omega \ggggg \beta_L^2 \barB^2 - \ii \omega \ggggg \beta_T^2 \barB^2 - \ii \omega \beta_L^2 \tfrac{\wpl^2}{\omega^2} 
    \com
}
\ba{
    (M_0)_{12} & = \ii \omega \gagg \beta_T \barB \\ 
    (M_0)_{14} & = \ii \omega  \gagg \beta_L \barB \\ 
    (M_0)_{15} & = - \ii \ka \gagg \beta_L \barB \\ 
    (M_0)_{21} & = - \ii \omega \gagg \beta_T \barB \\ 
    (M_0)_{22} & = m_a^2 - m_a^2 \ggggg \barB^2 + \tfrac{7}{2} \omega^2 \ggggg \beta_T^2 \barB^2 - \omega^2 \beta_T^2 \tfrac{\wpl^2}{\omega^2} - \ii \ka \ggggg (\barB^2)^\prime \\
    (M_0)_{24} & = \tfrac{7}{2} \omega^2 \ggggg \beta_L \beta_T \barB^2 - \omega^2 \beta_L \beta_T \tfrac{\wpl^2}{\omega^2} \\
    (M_0)_{25} & = - \tfrac{7}{2} \ka \omega \ggggg \beta_L \beta_T \barB^2 + \ka \omega \beta_L \beta_T \tfrac{\wpl^2}{\omega^2} \\
    (M_0)_{33} & = m_a^2 - m_a^2 \ggggg \beta_L^2 \barB^2 - 3 m_a^2 \ggggg \beta_T^2 \barB^2 + 2 \omega^2 \ggggg \beta_T^2 \barB^2 
    \\ & \qquad 
    - \ii \ka \ggggg \beta_L^2 (\barB^2)^\prime - 3 \ii \ka \ggggg \beta_T^2 (\barB^2)^\prime \nn 
    (M_0)_{41} & = - \ii \ka \gagg \beta_L \barB \\ 
    (M_0)_{42} & = \tfrac{7}{2} \ka \omega \ggggg \beta_L \beta_T \barB^2 - \ka \omega \beta_L \beta_T \tfrac{\wpl^2}{\omega^2} - \tfrac{7}{2} \ii \omega \ggggg \beta_L \beta_T (\barB^2)^\prime + \ii \omega \beta_L \beta_T \tfrac{(\wpl^2)^\prime}{\omega^2} \\ 
    (M_0)_{44} & = \ka \omega + \tfrac{5}{2} \ka \omega \ggggg \beta_L^2 \barB^2 - \ka \omega \ggggg \beta_T^2 \barB^2 - \ka \omega \beta_L^2 \tfrac{\wpl^2}{\omega^2} 
    \\ & \qquad 
    - \tfrac{5}{2} \ii \omega \ggggg \beta_L^2 (\barB^2)^\prime + \ii \omega \ggggg \beta_T^2 (\barB^2)^\prime + \ii \omega \beta_L^2 \tfrac{(\wpl^2)^\prime}{\omega^2} \nn 
    (M_0)_{45} & = - \ka^2 - \tfrac{5}{2} \ka^2 \ggggg \beta_L^2 \barB^2 + \ka^2 \ggggg \beta_T^2 \barB^2 + \ka^2 \beta_L^2 \tfrac{\wpl^2}{\omega^2} 
    \\ & \qquad 
    + \tfrac{5}{2} \ii \ka \ggggg \beta_L^2 (\barB^2)^\prime - \ii \ka \ggggg \beta_T^2 (\barB^2)^\prime - \ii \ka \beta_L^2 \tfrac{(\wpl^2)^\prime}{\omega^2} \nn 
    (M_0)_{51} & = - \ii \omega  \gagg \beta_L \barB \\ 
    (M_0)_{52} & = \tfrac{7}{2} \omega^2 \ggggg \beta_L \beta_T \barB^2 - \omega^2 \beta_L \beta_T \tfrac{\wpl^2}{\omega^2} \\
    (M_0)_{54} & = \omega^2 + \tfrac{5}{2} \omega^2 \ggggg \beta_L^2 \barB^2 - \omega^2 \ggggg \beta_T^2 \barB^2 - \omega^2 \beta_L^2 \tfrac{\wpl^2}{\omega^2} \\
    (M_0)_{55} & = - \ka \omega - \tfrac{5}{2} \ka \omega \ggggg \beta_L^2 \barB^2 + \ka \omega \ggggg \beta_T^2 \barB^2 + \ka \omega \beta_L^2 \tfrac{\wpl^2}{\omega^2}
    \per
}
\end{subequations}
Here we've defined $\beta_L = \barB_L(\rad) / \barB(\rad)$, $\beta_T = \barB_T(\rad) / \barB(\rad)$, $\beta_L^2 + \beta_T^2 = 1$, and used $\ka^2 = \omega^2 - m_a^2$.  
Note that we have ordered the variables in the 5-plet in order to block diagonalize the matrices as much as possible.  
The fifth equation can be solved to eliminate $A_{\omega,\nhat}^{(3)}$, and upon doing so $\phi_{\omega,\nhat}$ drops out of the remaining equations as a consequence of the gauge freedom.
Moreover, we now drop terms that go as derivatives of the background magnetic field and plasma frequency, since these are expected to be small.  
With these last simplifications, we finally have 
\ba{\label{eq:EOM_matrix}
    - \frac{1}{2\ka} \, 
    \frac{\dd^2}{\dd \rad^2} \begin{pmatrix} a_{\omega,\nhat}(\rad) \\ \ii A_{\omega,\nhat}^{(2)}(\rad) \\ \ii A_{\omega,\nhat}^{(1)}(\rad) \end{pmatrix} 
    - \ii \frac{\dd}{\dd \rad} \begin{pmatrix} a_{\omega,\nhat}(\rad) \\ \ii A_{\omega,\nhat}^{(2)}(\rad) \\ \ii A_{\omega,\nhat}^{(1)}(\rad) \end{pmatrix}
    = \begin{pmatrix} 
    \Delta_a(\rad) & (1 - \ggggg \barB^2) \, \Delta_B(\rad) & 0 \\ 
    \Delta_B(\rad) & \Delta_\parallel(\rad) & 0 \\ 
    0 & 0 & \Delta_\perp(\rad) 
    \end{pmatrix} 
    \begin{pmatrix} a_{\omega,\nhat}(\rad) \\ \ii A_{\omega,\nhat}^{(2)}(\rad) \\ \ii A_{\omega,\nhat}^{(1)}(\rad) \end{pmatrix} 
    \com
}
where 
\bsa{eq:Delta_defs}{
    \Delta_a(\rad) & = - \frac{1}{2\ka} \, \frac{\gagg^2 \beta_L^2 \barB^2}{1 - \beta_L^2 \tfrac{\wpl^2}{\omega^2} + \tfrac{5}{2} \ggggg \beta_L^2 \barB^2 - \ggggg \beta_T^2 \barB^2} \\ 
    \Delta_B(\rad) & = \frac{\omega}{2\ka} \frac{\gagg \beta_T \barB}{1 - \beta_L^2 \tfrac{\wpl^2}{\omega^2} + \tfrac{5}{2} \ggggg \beta_L^2 \barB^2 - \ggggg \beta_T^2 \barB^2} \\ 
    \Delta_\parallel(\rad) & = \frac{m_a^2}{2\ka} - \frac{\omega^2}{2\ka} \frac{\beta_T^2 \tfrac{\wpl^2}{\omega^2} - \tfrac{7}{2} \ggggg \beta_T^2 \barB^2}{1 - \beta_L^2 \tfrac{\wpl^2}{\omega^2} + \tfrac{5}{2} \ggggg \beta_L^2 \barB^2 - \ggggg \beta_T^2 \barB^2} \\ 
    \Delta_\perp(\rad) & = \frac{m_a^2}{2\ka} + \frac{\omega^2}{\ka} \frac{\ggggg \beta_T^2 \barB^2}{1 - \ggggg \beta_L^2 \barB^2 - 3 \ggggg \beta_T^2 \barB^2} 
    \per 
}
In practice terms of order $O(\gagg^2)$ and $O(\gagg \ggggg)$ can be neglected.  
Since the evolution of $A_{\omega,\nhat}^{(1)}$ is decoupled from $a_{\omega,\nhat}$ and $A_{\omega,\nhat}^{(2)}$, we are free to set $A_{\omega,\nhat}^{(1)}(\rad) = 0$ everywhere for the purposes of studying axion-photon interconversion.  
Additionally, provided that the profile functions do not vary rapidly, we can neglect the second derivative terms.  
With these additional simplifications, the equations of motion are reduced to 
\bes{\label{eq:EOM_iso}
	\frac{\dd}{\dd\rad} \Psi(\rad) = \ii \Mcal(\rad) \, \Psi(\rad)
}
where
\bes{
	\Psi(\rad) = \begin{pmatrix} a_{\omega,\nhat}(\rad) \\ \ii A_{\omega,\nhat}^{(2)}(\rad) \end{pmatrix} 
	\qquad \text{and} \qquad 
	\Mcal(\rad) \equiv \begin{pmatrix} \Delta_a(\rad) & (1 - \ggggg \barB^2) \, \Delta_B(\rad) \\ \Delta_B(\rad) & \Delta_\parallel(\rad) \end{pmatrix} 
    \per
}

\subsection{Axion-photon interconversion probability}
\label{app:mixing}

The equations of motion in \eref{eq:EOM_iso} govern the interconversion of axion waves and electromagnetic waves.  
In particular, to calculate the probability that an axion is converted into a photon as it passes through the neutron star's magnetosphere, we set a boundary condition at the star's surface where $A_{\omega,\nhat}^{(2)} = A_{\omega,\nhat}^{(1)} = 0$ and $a_{\omega,\nhat} = a_{\omega,\nhat,0} \neq 0$.  
Then the probability that a photon is detected after propagating a distance $\rad$ is calculated as \cite{Raffelt:1987im}
\bes{\label{eq:P_ag_def}
    \mathbb{P}_{a\to\gamma}(\rad) = \frac{|A_{\omega,\nhat}^{(2)}(\rad)|^2}{|a_{\omega,\nhat,0}|^2} 
    \per
}
We are especially interested in the behavior as $\rad \to \infty$.  

For pedagogical purposes, let us first suppose that the mixing matrix is homogeneous.  
That is to say, we treat the variables $\Delta_a$, $\Delta_B$, $\Delta_\parallel$, and $\barB$ as independent of $\rad$. 
With this assumption, \eref{eq:EOM_iso} is easily solved to obtain 
\bes{
    \mathbb{P}_{a\to\gamma}(\rad) = \sin^2(2\theta) \, \sin^2\Bigl( \frac{\rad - \RNS}{l_\mathrm{osc}} \Bigr) 
    \com
}
where the mixing angle $\theta$ and oscillation length $l_\mathrm{osc}$ are defined by 
\ba{\label{eq:tan2theta}
    \tan2\theta = \frac{2 \Delta_B}{\Delta_a - \Delta_\parallel} 
    \qquad \text{and} \qquad 
    l_\mathrm{osc} = \Bigl[ \Delta_{B}^2 + \tfrac{1}{4} \bigl( \Delta_a - \Delta_\parallel \bigr)^2 \Bigr]^{-1/2} 
    \per
}
Typically $|\Delta_B| \ll |\Delta_a - \Delta_\parallel| / 2$ such that $\sin^2(2\theta) \approx 4 \Delta_B^2 / (\Delta_a - \Delta_\parallel)^2 \ll 1$ and $l_\mathrm{osc} \approx 2 / |\Delta_a - \Delta_\parallel|$.  
However, if $\Delta_a(\rad)$, $\Delta_\parallel(\rad)$, and $\Delta_B(\rad)$ are varying such that temporarily $|\Delta_a - \Delta_\parallel| / 2 \ll \Delta_B$ then the mixing angle is near maximal $\theta \approx \pi/4$, and the conversion probability is resonantly enhanced. 
This resonant contribution to the probability can be calculated analytically, as we demonstrate in the following section.  

\subsection{Resonant axion-photon conversion }
\label{app:approximation}

Using perturbation theory, one can derive an approximate expression for the conversion probability, which is valid to leading order in the mixing $\Delta_B(\rad)$, and it takes the form~\cite{Raffelt:1987im} 
\bes{\label{eq:Pag_from_g_and_f}
    \Pag(\rad) \approx \biggl| \int_{\RNS}^{\rad} \! \dd\rad^\prime \, g(\rad') \, \ee^{\ii f(\rad^\prime)} \biggr|^2 
    \quad \text{where} \quad 
    g(\rad) = \Delta_B(\rad)
    \quad \text{and} \quad 
    f(\rad)= \int_{\RNS}^{\rad} \! \dd\rad^\pprime \, \Bigl( \Delta_a(\rad^\pprime) - \Delta_{\parallel}(\rad^\pprime) \Bigr) 
    \per
}
If there exists a radius $\radres$, which we call the resonance radius, where $f(\rad)$ is stationary, then the integral can be evaluated by employing the stationary phase approximation.  
The resonance condition is expressed as 
\bes{\label{eq:res_condit_v1}
    \frac{\dd}{\dd\rad} f(\rad) \Bigr|_{\rad = \radres} = 0 
    \qquad \Rightarrow \qquad 
    \Delta_a(\radres) = \Delta_\parallel(\radres) 
    \per
}
Using the expressions for $\Delta_a(\rad)$ and $\Delta_\parallel(\rad)$ from \eref{eq:Delta_defs} the resonance condition can also be written as\footnote{The term $\tfrac{7}{2} \ggggg \barB_T^2 \omega^2$ differs by a factor of $2$ from the analogous term in Eq.~(S10) of \cite{Hook:2018iia}.  To compare use $\ggggg \barB^2 = 2 \kappa$. } 
\ba{\label{eq:res_condit_v2}
    m_a^2 
    - m_a^2 \beta_L^2 \frac{\wpl^2}{\omega^2} 
    - \beta_T^2 \wpl^2 
    + \frac{7}{2} \ggggg \beta_T^2 \barB^2 \omega^2 
    \biggr|_{\rad=\radres}
    = 0 
    \per 
}
In this expression, we have dropped terms that are negligible for the parameters of interest: $\ggggg \barB^2$ is small compared to $1$, and $\gagg^2 \barB^2$ is small compared to $\wpl^2$.  
For a given angular frequency $\omega$ and for the specified varying backgrounds --- $\barB_T(\rad)$, $\barB_L(\rad)$, and $\wpl(\rad)$ --- one is interested in whether there exists a $\radres$ that satisfies this equation and obeys $\radres > \RNS$.  
Such a solution may not exist depending on the choice of parameters. 

The width of the resonance region is defined by 
\bes{\label{eq:dres_def}
    \dres 
    = \sqrt{2\pi} \, \Bigl| \frac{\dd^2}{\dd\rad^2} f(\rad) \bigr|_{\radres} \Bigr|^{-1/2} 
    = \sqrt{2\pi} \, \Bigl| \frac{\dd}{\dd\rad} \bigl( \Delta_a(\rad) - \Delta_\parallel(\rad) \bigr) \bigr|_{\radres} \Bigr|^{-1/2} 
    \per
}
Provided that $\RNS < \radres < \rad$ such that the resonance radius is within the integration domain, the integral may be evaluated in the stationary phase approximation, leading to an approximate analytic expression for the asymptotic probability 
\bes{\label{eq:Pag_res_approx}
    \Pag(\omega,\nhat)
    \equiv \lim_{\rad \gg \radres} \Pag(\rad) 
    \approx |g(\radres)|^2 \, \dres^2
    \approx \frac{2 \pi \, |\Delta_B(\rad)|^2}{\bigl| \tfrac{\dd}{\dd\rad} \Delta_a(\rad) - \tfrac{\dd}{\dd\rad} \Delta_\parallel(\rad) \bigr| } \biggr|_{\rad=\radres}
    \per
}
In the next two sections we explore two different solution branches of the resonance condition, and we provide expressions for $\radres$, $\dres$, and $\Pag$ in both cases.  
Then we discuss nonresonant axion-photon interconversion. 

\subsection{Mass-matched resonance}
\label{app:mass_match_resonance}

If there exists a solution of the resonance condition in \eref{eq:res_condit_v2} with $\radres > \RNS$, then resonant axion-photon interconversion occurs at radius $\rad = \radres$.  
Here we first identify the well-known resonance~\cite{Raffelt:1987im}, which we call the ``mass-matched resonance'' (\MMR{}) and in the next section we identify an additional resonance solution.  

To identify the \MMR{}, it is sufficient to set $\ggggg = 0$ and $\gagg = 0$.  
If the terms suppressed by these couplings are neglected, the resonance condition \eqref{eq:res_condit_v2} reduces to 
\begin{subequations}\label{eq:res_condit_MM}
\ba{
    \bigl( \omega^2 - m_a^2 \bigr) 
    \biggl( \beta_L^2 \frac{\wpl^2}{\omega^2} \biggr) 
    - \bigl( \wpl^2 - m_a^2 \bigr) 
    \biggr|_{\rad=\radres}
    = 0 
    \per
}
Resonance occurs if there exists a radius $\radres$ at which this equation is solved for a given angular frequency $\omega$ and for the specified varying backgrounds $\barB_T(\rad)$, $\barB_L(\rad)$, and $\wpl(\rad)$.  
If the axion is nonrelativistic such that $\omega \approx m_a$, then the equation reduces further to $\wpl(\radres) \approx m_a$ and the resonance occurs as the axion passes through a region of space where the plasma frequency equals the axion mass.  
Alternatively, if the axion is relativistic such that $\omega \gg m_a \beta_L / \beta_T$, then the equation reduces instead to $\wpl(\radres) \approx m_a / \beta_T$.  
More generally, the resonance condition can be written as 
\bes{
    \wpl(\radres) = m_a \biggl( \beta_T^2 + \beta_L^2 \frac{m_a^2}{\omega^2} \biggr)^{\! \! -1/2} 
    \per
}
\end{subequations}
Since $\wpl(\rad)$ acts like an effective mass for the dispersion relation of an electromagnetic wave in medium, we call this the ``mass-matched resonance.''  
Using the expressions for $\barB(\rad)$ and $\wpl(\rad)$ from \erefs{eq:barBvec}{eq:barwpl}, and then solving for $r = \radres$ gives the resonance radius 
\bsa{eq:rres_MM}{
    \radres(\omega,\nhat) 
    & = 
    \RNS 
    \biggl( \beta_T^2 \frac{\omega_{\mathrm{pl},0}^2}{m_a^2} + \beta_L^2 \frac{\omega_{\mathrm{pl},0}^2}{\omega^2} \biggr)^{\! \! 1/3} 
    |\psiomega|^{2/3} 
    \\ 
    & \approx 
    \bigl( 17 \, \RNS \bigr) 
    \biggl( \frac{B_0}{10^{14} \Gauss} \biggr)^{\! \! 1/3} 
    \biggl( \frac{P}{1 \sec} \biggr)^{\! \! -1/3} 
    \biggl( \frac{n_e}{|n_c|} \biggr)^{\! \! 1/3} \, 
    \\ & \qquad \times 
    \biggl( \frac{m_a}{1 \ueV} \biggr)^{\! \! -2/3} 
    |\psiomega(\nhat)|^{2/3} 
    \times \begin{cases}
    1 
    & \quad \text{if $\omega \approx m_a$} \\ 
    |\beta_T(\nhat)|^{2/3} 
    & \quad \text{if $\omega \gg m_a \, \beta_L / \beta_T$} 
    \end{cases}
    \per
    \nonumber
}
Here we've assumed that $\beta_T(\nhat)$ and $\beta_L(\nhat)$ are independent of $\rad$, which assumes that the orientation of the magnetic field is not changing, and only its strength varies.  
Since $\wpl(\rad) \propto \rad^{-3/2}$ is a decreasing function of $\rad$, a more massive axion will convert closer to the star's surface, but since $\radres$ must be larger than $\RNS$, there is a maximal mass (for a given $B_0$, $P$, $\nhat$, and $t$) above which no resonant conversion may occur; for the fiducial parameters used above, this is approximately $70 \ueV$.  
The width of the resonance region $\dres$ may be calculated by evaluating the derivatives in \eref{eq:dres_def}.  
Using the expressions for $\barB(\rad)$ and $\wpl(\rad)$ from \erefs{eq:barBvec}{eq:barwpl}, and using \eref{eq:res_condit_EH} to eliminate $\omega$, and setting $\ggggg = 0$ leads to 
\bsa{eq:dres_MM}{
    \dres(\omega,\nhat) 
    & = \frac{\sqrt{2\pi}}{\wpl} \, 
    |\beta_T|^{-1} \, 
    \biggl| 1 - \beta_L^2 \frac{\wpl^2}{\omega^2} \biggr| \, 
    \biggl| \frac{1}{\ka \, \wpl} \frac{\dd}{\dd \rad} \wpl \biggr|^{-1/2} 
    \biggr|_{\rad=\radres} \\ 
    & = \sqrt{\frac{4 \pi}{3}} \frac{\sqrt{\ka \radres}}{\wpl} \, 
    |\beta_T|^{-1} \, 
    \biggl| 1 - \beta_L^2 \frac{\wpl^2}{\omega^2} \biggr| \\ 
    & \approx 
    \biggl( \frac{B_0}{10^{14} \Gauss} \biggr)^{\! \! -1/6} 
    \biggl( \frac{P}{1 \sec} \biggr)^{\! \! 1/6} 
    \biggl( \frac{\RNS}{10 \km} \biggr)^{\! \! -1/2} 
    \biggl( \frac{n_e}{|n_c|} \biggr)^{\! \! -1/6}
    \ \bigl| \psiomega(\nhat) \bigr|^{-1/3} 
    \\ & \qquad \times 
    \begin{cases}
    \bigl( 7.0 \times 10^{-5} \, \radres \bigr) \, 
    \bigl( \frac{m_a}{1 \ueV} \bigr)^{-2/3} \, 
    \bigl( \frac{\ka}{1 \neV} \bigr)^{1/2} \, 
    \ |\beta_T(\nhat)|^{}
    & \quad \text{if $\omega \approx m_a$} \\ 
    \bigl( 2.2 \times 10^{-1} \, \radres \bigr) \, 
    \bigl( \frac{m_a}{1 \ueV} \bigr)^{-2/3} \, 
    \bigl( \frac{\ka}{10 \meV} \bigr)^{1/2} \, 
    \ |\beta_T(\nhat)|^{-1/3} 
    & \quad \text{if $\omega \gg m_a \, \beta_L / \beta_T$} 
    \end{cases}
    \per
}
For these fiducial parameters, the resonance region is very narrow, corresponding to $\dres \ll \radres$.  
However, decreasing $m_a$ tends to broaden the resonance region, and if $\dres \gtrsim \radres$ the stationary phase approximation is no longer applicable.  
The probability for resonant axion-photon conversion $\Pag$ may be calculated by evaluating \eref{eq:Pag_res_approx}. 
Using the expressions for $\barB(\rad)$ and $\wpl(\rad)$ from \erefs{eq:barBvec}{eq:barwpl}, and using \eref{eq:res_condit_EH} to eliminate $\omega$, and setting $\ggggg = 0$ leads to 
\bsa{eq:Pag_MM}{
    \Pag(\omega,\nhat) 
    & = \frac{\pi}{2} \frac{\barB^2 \omega^2}{\ka^2 \wpl^2} 
    \biggl| \frac{1}{\ka \, \wpl} \frac{\dd}{\dd \rad} \wpl \biggr|^{-1} 
    \biggr|_{\rad=\radres} 
    \gagg^2 
    + O(\gagg^4) \\ 
    & = \frac{\pi}{3} \frac{\gagg^2 \barB^2 \omega^2 \radres}{\ka \wpl^2} 
    + O(\gagg^4) \\ 
    & = \frac{1}{4} \gagg^2 \beta_T^2 \barB^2 \dres^2 \times \frac{\omega^2}{\ka^2} \biggl( 1 - \beta_L^2 \frac{\wpl^2}{\omega^2} \biggr)^{\! \! -2}  
    + O(\gagg^4) \\ 
    & \approx 
    \biggl( \frac{\gagg}{10^{-12} \GeV^{-1}} \biggr)^{\! \! 2} 
    \biggl( \frac{B_0}{10^{14} \Gauss} \biggr)^{\! \! 1/3} 
    \biggl( \frac{P}{1 \sec} \biggr)^{\! \! 5/3} 
    \biggl( \frac{\RNS}{10 \km} \biggr)^{} 
    \biggl( \frac{n_e}{|n_c|} \biggr)^{\! \! -5/3} \, 
    \bigl| \psivecB(\nhat) \bigr|^{2}  
    \bigl| \psiomega(\nhat) \bigr|^{-10/3} 
    \\ & \qquad \times 
    \begin{cases}
    \bigl( 1.5 \times 10^{-4} \bigr) \, 
    \bigl( \frac{m_a}{1 \ueV} \bigr)^{10/3} \, 
    \bigl( \frac{\ka}{1 \neV} \bigr)^{-1} 
    & \quad \text{if $\omega \approx m_a$} \\ 
    \bigl( 1.5 \times 10^{-3} \bigr) \, 
    \bigl( \frac{m_a}{1 \ueV} \bigr)^{4/3} \, 
    \bigl( \frac{\omega}{10 \meV} \bigr)^{2} \, 
    \bigl( \frac{\ka}{10 \meV} \bigr)^{-1} \, 
    |\beta_T(\nhat)|^{-4/3} 
    & \quad \text{if $\omega \gg m_a \, \beta_L/\beta_T$} 
    \end{cases}
    \per 
    \nonumber 
}
For the case with $\omega \approx m_a$, we have fiducialized to $\ka \approx 10^{-3} m_a$, since this calculation is partly motivated by searches for axion dark matter, and the speed of dark matter in a galactic halo is typically $v_\mathrm{dm} \sim 10^{-3}$.  
However, for accurate calculations one should bear in mind that dark matter particles accelerate as they fall into a neutron star's gravitational potential.  
For the case with $\omega \gg m_a$, we have fiducialized to $\ka \approx \omega \approx 10^{4} m_a$. 
For relativistic axions, the probability is an increasing function of axion energy $\Pag \propto \omega^2 \ka^{-1} \approx \omega$.  
Regardless of whether the axions are nonrelativsitic or relativistic, the probability scales with a positive power of the axion mass $m_a$.  
Since the plasma frequency decreases with distance from the star $\wpl \propto \rad^{-3/2}$, reducing the axion mass causes the resonance condition $\wpl(\radres) \approx m_a$ to be satisfied as a larger radius $\radres \propto m_a^{-2/3}$ where the magnetic field is weaker $\barB \propto \rad^{-3}$, leading to a suppression of the probability.  

\subsection{Euler-Heisenberg assisted resonance}
\label{app:EH_assisted_resonance}

Now we explore a second solution of the resonance condition, which is made possible through the Euler-Heisenberg four-photon self-interaction.  
To identify this solution, it suffices to set $m_a = 0$ in the resonance condition \eqref{eq:res_condit_v2}, which then reduces to 
\begin{subequations}\label{eq:res_condit_EH}
\ba{
    - \beta_T^2 \wpl^2 
    + \frac{7}{2} \ggggg \beta_T^2 \barB^2 \omega^2 
    \biggr|_{\rad=\radres}
    = 0 
    \per
}
Rearranging allows the resonance condition to be written as 
\ba{
    \omega 
    = \sqrt{ \frac{2}{7 \ggggg \barB^2} } \, \wpl \biggr|_{\rad=\radres} 
    \per
}
\end{subequations}  
Since this resonance requires $\ggggg \neq 0$, we call this the ``Euler-Heisenberg assisted resonance'' (\EHR{}).    
Using the expressions for $\barB(\rad)$ and $\wpl(\rad)$ from \erefs{eq:barBvec}{eq:barwpl}, and then solving for $r = \radres$ gives the resonance radius 
\bsa{eq:rres_EH}{
    \radres(\omega,\nhat) & = 
    \frac{7^{1/3}}{2^{1/3}} \ggggg^{1/3} B_0^{2/3} \frac{\omega^{2/3}}{\omega_{\mathrm{pl},0}^{2/3}} \frac{|\psivecB|^{2/3}}{|\psiomega|^{2/3}} \, \RNS
    \\ 
    & \approx 
    \bigl( 3.4 \RNS \bigr) 
    \biggl( \frac{B_0}{10^{14} \Gauss} \biggr)^{\! \! 1/3} 
    \biggl( \frac{P}{1 \sec} \biggr)^{\! \! 1/3}
    \biggl( \frac{n_e}{|n_c|} \biggr)^{\! \! -1/3} \, 
    \biggl( \frac{\omega}{10 \meV} \biggr)^{\! \! 2/3} \, 
    |\psivecB(\nhat)|^{2/3} \, 
    |\psiomega(\nhat)|^{-2/3}
    \per
}
Since $\radres$ grows with increasing $\omega$, it implies that more energetic axions, have their resonance condition satisfied further from the star.  
The lowest energy axions that participate in this resonance are the ones that meet the resonance condition closest to the star, which corresponds to $\omega \approx 1.6 \meV$ for the fiducial parameters used above (assuming $m_a \ll 1.6 \meV$).  
The width of the resonance region $\dres$ may be calculated by evaluating the derivatives in \eref{eq:dres_def}.  
Using the expressions for $\barB(\rad)$ and $\wpl(\rad)$ from \erefs{eq:barBvec}{eq:barwpl}, and using \eref{eq:res_condit_EH} to eliminate $\omega$, and setting $m_a = 0$ and $\gagg = 0$ leads to 
\bsa{eq:dres_EH}{
    \dres(\omega,\nhat) 
    & = \frac{\sqrt{2\pi}}{\wpl} \, 
    |\beta_T|^{-1} \, 
    \biggl| \frac{1}{\ka \, \wpl} \frac{\dd}{\dd \rad} \wpl - \frac{1}{\ka \, \barB} \frac{\dd}{\dd \rad} \barB \biggr|^{-1/2} 
    \biggr|_{\rad=\radres} \\ 
    & = \sqrt{\frac{4\pi}{3}} \frac{\sqrt{\ka \radres}}{\wpl} \, 
    |\beta_T|^{-1} \\ 
    & \approx 
    \bigl( 4.5 \times 10^{-2} \, \radres \bigr) 
    \biggl( \frac{B_0}{10^{14} \Gauss} \biggr)^{\! \! -1/6} 
    \biggl( \frac{P}{1 \sec} \biggr)^{\! \! 5/6} 
    \biggl( \frac{\RNS}{10 \km} \biggr)^{\! \! -1/2}
    \biggl( \frac{n_e}{|n_c|} \biggr)^{\! \! -5/6} \, 
    \\ & \qquad \times 
    \biggl( \frac{\omega}{10 \meV} \biggr)^{\! \! 2/3} 
    \biggl( \frac{\ka}{10 \meV} \biggr)^{\! \! 1/2} 
    |\beta_T(\nhat)|^{-1} \, 
    |\psivecB(\nhat)|^{2/3} \, 
    |\psiomega(\nhat)|^{-5/3}
    \per 
    \nonumber
}
For these fiducial parameters, the resonance region is very narrow, corresponding to $\dres \ll \radres$.  
However, increasing $\omega \approx \ka$ tends to broaden the resonance region, and if $\dres \gtrsim \radres$ the stationary phase approximation is no longer applicable.  
The probability for resonant axion-photon conversion $\Pag$ may be calculated by evaluating \eref{eq:Pag_res_approx}. 
Using the expressions for $\barB(\rad)$ and $\wpl(\rad)$ from \erefs{eq:barBvec}{eq:barwpl}, and using \eref{eq:res_condit_EH} to eliminate $\omega$, and setting $m_a = 0$ leads to 
\bsa{eq:Pag_EH}{
    \Pag(\omega,\nhat) 
    & = \frac{\pi}{7} \frac{1}{\ggggg \ka^2} 
    \biggl| \frac{1}{\ka \, \wpl} \frac{\dd}{\dd \rad} \wpl - \frac{1}{\ka \, \barB} \frac{\dd}{\dd \rad} \barB \biggr|^{-1} 
    \biggr|_{\rad=\radres} 
    \gagg^2 
    + O(\gagg^4) \\ 
    & = \frac{2\pi}{21} \frac{\gagg^2 \radres}{\ggggg \ka} + O(\gagg^4) \\ 
    & = \frac{1}{4} \gagg^2 \beta_T^2 \barB^2 \dres^2 \times \frac{\omega^2}{\ka^2} + O(\gagg^4) \\ 
    & \approx 
    \bigl( 3.7 \times 10^{-2} \bigr) 
    \biggl( \frac{\gagg}{10^{-12} \GeV^{-1}} \biggr)^{\! \! 2} 
    \biggl( \frac{B_0}{10^{14} \Gauss} \biggr)^{\! \! 1/3} 
    \biggl( \frac{P}{1 \sec} \biggr)^{\! \! 1/3} 
    \biggl( \frac{\RNS}{10 \km} \biggr)^{} 
    \biggl( \frac{n_e}{|n_c|} \biggr)^{\! \! -1/3} \, 
    \\ & \qquad \times 
    \biggl( \frac{\omega}{10 \meV} \biggr)^{\! \! 2/3} 
    \biggl( \frac{\ka}{10 \meV} \biggr)^{\! \! -1} \, 
    |\psivecB(\nhat)|^{2/3} \, 
    |\psiomega(\nhat)|^{-2/3} 
    \per 
    \nonumber 
}
Notice that the coupling $\ggggg$ appears in the denominator, since the resonance condition implies $\omega^2 \propto 1 / \ggggg$.  
Here we've distinguished $\omega$ and $\ka$, but since the formula is derived by taking $m_a = 0$, one should set $\ka = \omega$.   
For relativistic axions, the probability is a decreasing function of axion energy $\Pag \propto \omega^{2/3} \ka^{-1} \approx \omega^{-1/3}$.  
Additional discussion is provided in \sref{sub:comparison}.  

\subsection{Nonresonant conversion}
\label{app:nonres}

If the resonance condition \eqref{eq:res_condit_v2} does not admit any real solutions with $\radres > \RNS$ then we say that the axion-photon interconversion is nonresonant.  
For instance, this happens when the axion energy is high, such that $\ggggg B_0^2 \omega^2$ is larger than $m_a^2$ and $\wpl^2$, since then the fourth term in \eref{eq:res_condit_v2} is larger than the first three terms for all $\rad$ (except for special orientations along which $\beta_T(\nhat) = 0$).  
In the regime where the axion mass and plasma frequency can be neglected, then the conversion probability admits a particularly simple expression.  
We set $m_a = 0$ and $\wpl = 0$ in the axion-photon equations of motion \eqref{eq:EOM_iso}, and we neglect the $O(\ggggg \barB^2)$ terms in the denominators.  
Additionally, if the conversion is ineffective, then it is a good approximation to hold the axion field fixed to its initial condition $a_{\omega,\nhat}(r) = a_{\omega,\nhat,0}$ and solve only the equation for the photon field.  
The resultant equation can be integrated to yield an exact analytic expression for $\ii A_{\omega,\nhat}^{(2)}(r)$.  
The corresponding conversion probability \eqref{eq:P_ag_def} is given by~\cite{Fortin:2018aom,Dessert:2019sgw,Fortin:2021sst} 
\bsa{eq:Pag_NR}{
    \Pag(\omega,\nhat) 
    & = \frac{c_1^2}{c_2^{4/5}} \biggl| \frac{\Gamma(\tfrac{2}{5}) - \Gamma(\tfrac{2}{5}, - \frac{1}{5} c_2)}{5^{3/5}} \biggr|^2 
    \quad \text{with} \quad 
    c_1 = \tfrac{1}{2\ka} \, \gagg \beta_T B_0 |\psivecB| \omega \RNS 
    \\ & \hspace{4.5cm}
    \quad \text{and} \quad 
    c_2 = \tfrac{7}{4\ka} \, \ggggg \beta_T^2 B_0^2 |\psivecB|^2 \omega^2 \RNS \nn 
    & \approx \frac{\Gamma(2/5)^2}{2^{2/5} 5^{6/5} 7^{4/5}} \frac{B_0^{2/5} \gagg^2 \RNS^{6/5} \omega^{2/5}}{\ggggg^{4/5} \ka^{6/5}} \, |\beta_T|^{2/5} |\psivecB|^{2/5} 
    \quad \text{for $c_2 \gg 1$} \\ 
    & \approx 
    \bigl( 5.1 \times 10^{-6} \bigr) 
    \biggl( \frac{\gagg}{10^{-12} \GeV^{-1}} \biggr)^{\! \! 2} 
    \biggl( \frac{B_0}{10^{14} \Gauss} \biggr)^{\! \! 2/5} 
    \biggl( \frac{\RNS}{10 \km} \biggr)^{\! \! 6/5} 
    \\ & \qquad \times 
    \biggl( \frac{\omega}{1 \keV} \biggr)^{\! \! 2/5} 
    \biggl( \frac{\ka}{1 \keV} \biggr)^{\! \! -6/5} 
    |\beta_T(\nhat)|^{2/5} \, 
    |\psivecB(\nhat)|^{2/5} 
    \per 
    \nonumber 
}
Moving from the first line to the second line, we assumed a sufficiently strong surface magnetic field $B_0$ and sufficiently large axion energy $\omega$ such that $c_2 \gg 1$; see \rref{Dessert:2019sgw} for additional details on the $c_2 \ll 1$ regime.  
For relativistic axions $\ka = (\omega^2 - m_a^2)^{1/2} \approx \omega$ and $\Pag \propto \omega^{2/5} \ka^{-6/5} \approx \omega^{-4/5}$.  
This derivation took $m_a = 0$, and for larger values of the axion mass the probability is eventually exponentially suppressed; see \rref{Fortin:2021sst} for additional details. 

\section{Numerical validation}
\label{app:numerical}

This short appendix contains numerical checks of the analytical approximations presented in the main text.  
It also includes a discussion of the angular dependence of the resonance radius and conversion probability.  

\subsection{Angular dependence of resonance radius}
\label{app:angular}

In this appendix we explore the angular dependence of the resonance radius for both resonances.  
For simplicity we focus on the aligned neutron star, which has $\mvec \, \parallel \, \Ohat$, such that the angular dependence only enters through the polar angle $\theta = \mathrm{arccos}(\Ohat \cdot \rhat)$.  
\begin{figure}[t!]
\centering
\includegraphics[scale=0.7]{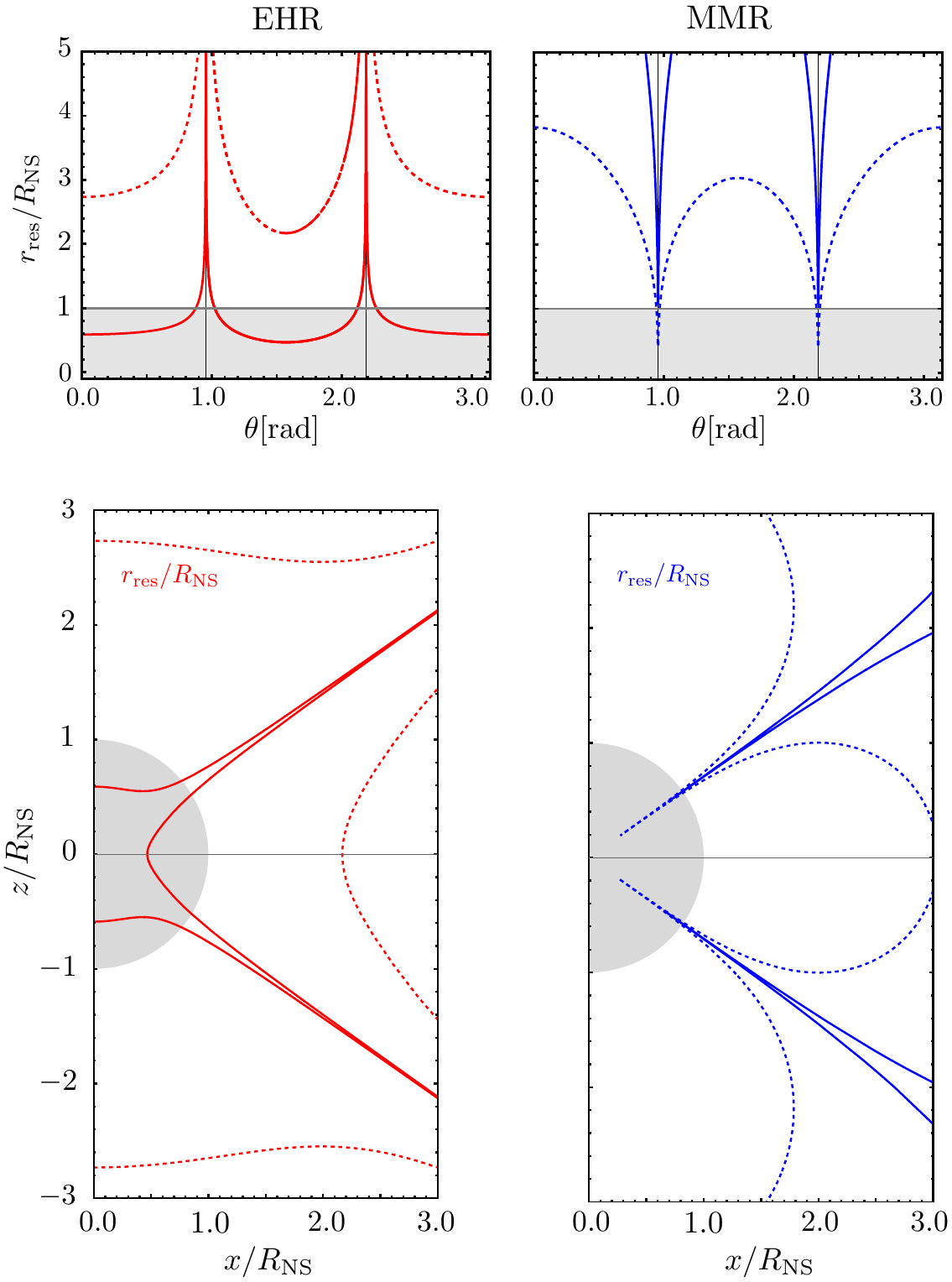}
\caption{\label{fig:rc_over_RNS}
Resonance radius as a function of the polar angle $\theta$ between the axis of the neutron star (assumed rotation and magnetic axes align $\Ohat = \mhat$) and the propagation direction of the axion (assumed radial $\nhat = \rhat$).  For the Euler-Heisenberg assisted resonance, the red curves correspond to $\omega = \ka = 1 \meV$ (solid) and $10 \meV$ (dashed).  For the mass-matched resonance, the blue curves correspond to $m_a = 3.3 \ueV$ (solid) and $13 \ueV$ (dashed).  We take $B_0 = 10^{14} \Gauss$, $P = 1 \sec$, and $n_e = |n_c|$. The upper panels show plots of $\radres/\RNS$ as a function of $\theta$ and the lower panels show the corresponding regions of space in cylindrical coordinates. The gray shaded region shows the region of space within the neutron star.  
}
\end{figure}

First we consider the \EHR{}, which has a resonance radius $\radres$ given by \eref{eq:rres_EH}.  
Depending on the propagation direction of the axion (assumed radial $\nhat = \rhat$), the resonance radius varies through the factors of $\psivecB(\nhat)$ and $\psiomega(\nhat)$; this behavior is shown in \fref{fig:rc_over_RNS}. 
Requiring that the resonant conversion occurs outside of the neutron star imposes $\radres > \RNS$.  
This inequality is solved for two ranges of polar angles: $\theta \in (\theta_1,\theta_2)\,\cup \,(\theta_3,\theta_4)$ where 
\bes{
    & \theta_{1,2} = \mathrm{arccos}\left[\frac{-1\mp 0.1\left( \frac{ 10^{-2} \eV}{\omega} \right)^{2} \left( \frac{10^{14} \Gauss}{B_0} \right)
    \left( \frac{1 \sec}{P} \right) }{3\left(1\mp 0.1\left( \frac{10^{-2} \eV}{\omega} \right)^{2} \left( \frac{10^{14} \Gauss}{B_0} \right)
    \left( \frac{1 \sec}{P} \right) \right)}  \right] 
    \\ & \qquad 
    \theta_3 = \pi-\theta_2
    \ , \quad 
    \theta_4 = \pi-\theta_1 
    \ , \quad \text{and} \quad 
    0 \leq \theta_1 < \theta_2 < \theta_3 < \theta_4 \leq \pi 
    \per
}
However, the resonance radius diverges at $\theta = \theta_{+/-} \equiv \mathrm{arccos}(\pm\sqrt{1/3})$ where the plasma frequency in the Goldreich-Julian model crosses through zero.  
Indeed, the model breaks down for sufficiently large resonance radius when the plasma frequency is the order of interstellar medium (ISM) plasma frequency, \textit{i.e.} $\omega_{p,\text{ISM}} = (4\pi \alpha n_e/m_e)^{1/2} \approx (6.4 \times 10^{-12} \eV) (n_e/0.03 \cm^{-3})^{1/2}$, where $n_e$ and $m_e$ are the electron number density and mass, respectively. 
Since the  ISM plasma frequency is many order of magnitudes smaller than the typical frequency that we are interested in and the effect of the combination $(B_0 P)$ is suppressed to the 1/3-power, we can reduce the set of allowed angles for resonance conversion approximately to $\theta \in (\theta_1,\theta_{+}) \cup (\theta_{+},\theta_2) \cup (\theta_3,\theta_{-}) \cup (\theta_{-},\theta_4)$.  

Second we consider the \MMR{}, which has the resonance radius given by \eref{eq:rres_MM}.  
For aligned NSs, the set of angles which satisfy the condition $\radres > \RNS$ reads now as $\theta \in (0,\theta_1) \cup (\theta_2,\theta_3) \cup (\theta_4,\pi)$, where \cite{Nurmi:2021xds}
\bes{
    & \theta_{1,2} = \mathrm{arccos}\left[ \sqrt{\frac{1}{3} \bigg(
    1 \pm \frac{1}{22.4^3}\left(\frac{m_a}{0.66 \ueV}\right)^2\left(\frac{P}{1 \sec} \right)\left(\frac{10^{14} \Gauss}{B_0}\right)\bigg )} \right] 
    \\ & \qquad 
    \theta_3 = \pi-\theta_2
    \ , \quad 
    \theta_4 = \pi-\theta_1 
    \ , \quad \text{and} \quad 
    0 \leq \theta_1 < \theta_2 < \theta_3 < \theta_4 \leq \pi 
    \per
}
For the \MMR{}, the resonance radius vanishes near $\theta_{+/-}$, rather than diverging. 

\subsection{Validation by direct numerical integration}
\label{app:validation}

In this section we seek to validate the analytical expression \eqref{eq:Pag_EH} for the axion-photon conversion probability in the regime of the \EHR{}.  
We do so by comparing \eref{eq:Pag_EH} against the result of a direct numerical integration of the mode equations \eqref{eq:EOM_iso}.  
We provide three figures, which show how $\Pag$ depends on the polar angle $\theta$, the radial distance $\rad$, and the axion energy $\omega$. 
For this numerical check, in all three figures we assume: that the rotation axis and the magnetic polar axis of the neutron star align ($\Ohat = \mhat$), that the axion follows a radial trajectory out from the center of the star ($\nhat = \rhat$), that the electron density equals the Goldreich-Julian density $n_e = |n_c|$, that $B_0 = 10^{14} \Gauss$, that $P = 10 \sec$, that $\gagg = 10^{12} \GeV^{-1}$, and that $m_a = 0$ so $\omega = \ka$.  
For these parameters, the probability is dominated by the \EHR{}, and we expect to find a good agreement between the analytical and numerical methods. 

\begin{figure}[t!]
\centering
\includegraphics[scale=0.5]{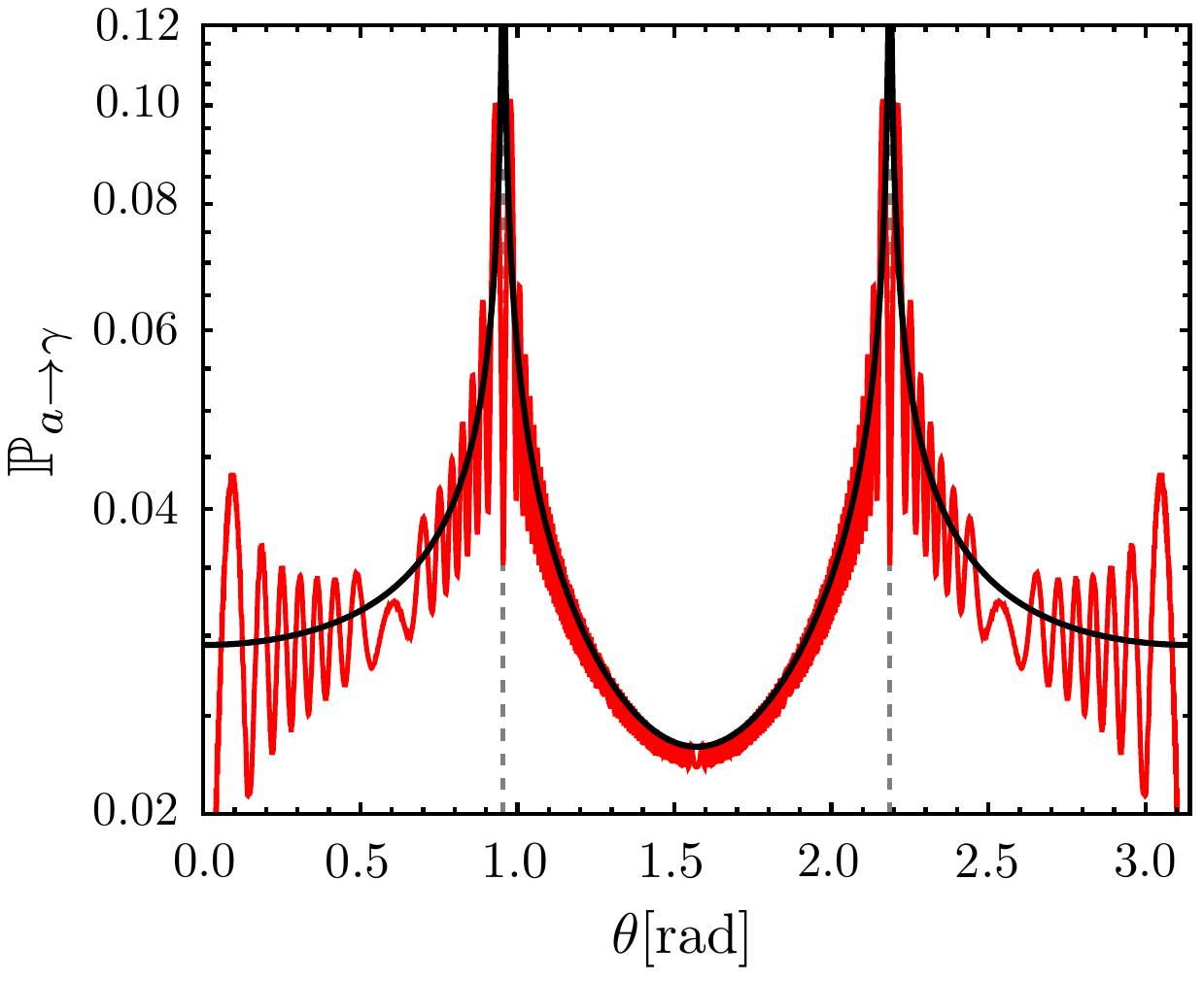}
\caption{\label{fig:probangles}
Axion-photon conversion probability $\Pbb_{a\to\gamma}$ as function of polar angle $\theta$ in the regime of the Euler-Heisenberg assisted resonance.  We assume that the rotation axis and the magnetic polar axis of the star align ($\Ohat = \mhat$), and $\theta$ is the angle that this axis makes with the trajectory of the axion (assumed radial $\nhat = \rhat$).  We show the result of direct numerical integration (red) and analytical resonance approximation (black).  Vertical dashed lines indicate the angles $\theta_{+/-}$ at which the resonance radius diverges.  The model parameters are chosen to be $B_0 = 10^{14} \Gauss$, $P = 1 \sec$, $\gagg = 10^{-12} \GeV^{-1}$, $n_e = |n_c|$, $m_a = 0$, and $\omega = \ka = 10^{-2} \eV$.  
}
\end{figure}

In \fref{fig:probangles} we show how the axion-photon conversion probability $\Pag$ varies as a function of the polar angle $\theta$. 
The probability calculated by direct integration (red) displays a rapidly oscillating component, and its mean agrees well with the analytical approximation (black), which we interpret as a sign of reliability for the approximation. 
Presumably, the oscillations are an artefact of assuming perfectly radial trajectories, and we expect that the oscillatory behavior would be removed if we had averaged over non-radial trajectories as well.  
For both the analytical and numerical calculations, the probability is seen to spike upward at $\theta_{+/-}$, corresponding to the region in \fref{fig:rc_over_RNS} where the resonance radius diverges.  

\begin{figure}[t]
\centering
\includegraphics[scale=0.45]{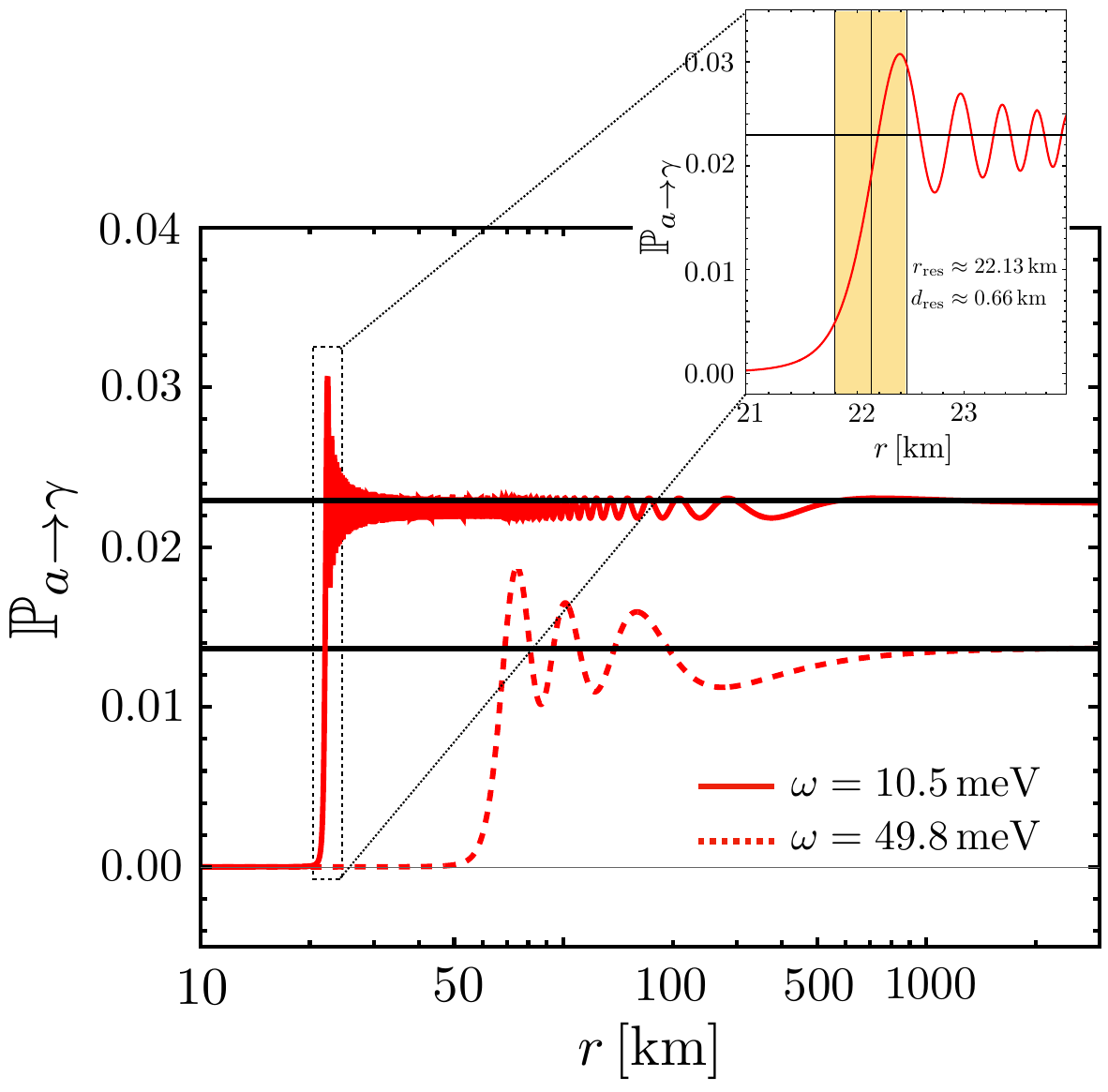}
\caption{\label{fig:prob0}
Axion-photon conversion probability $\Pbb_{a\to\gamma}$ as function of radial coordinate $\rad$ in the regime of the Euler-Heisenberg assisted resonance.  We show the result of direct numerical integration (red curves) as well as the analytical resonance approximation (horizontal black lines), which should agree toward asymptotically large $\rad$.  We consider an axion trajectory that lies in the neutron star's equatorial plane where $|\psivecB(\rhat)| = 1/2$, $|\psiomega(\rhat)| = 1$, $|\beta_L(\rhat)| = 0$, and $|\beta_T(\rhat)| = 1$.  The model parameters are chosen to be $B_0=10^{14} \Gauss$, $P = 1 \sec$, $\gagg = 10^{-12} \GeV^{-1}$, $n_e = |n_c|$, $m_a = 0$, and either $\omega = \ka = 10.5 \meV$ (solid red) or $49.8\meV$ (dashed red).  The inset in the upper-right corner zooms into the resonance region and shows the resonance radius (black vertical line) and the resonance width (yellow band).  
}
\end{figure}

In \fref{fig:prob0} we show how the axion-photon conversion probability $\Pag$ varies as a function of the radial distance $\rad$ from the center of the star.  
We focus on the equatorial plane where $\theta = \pi/2$, $|\psivecB(\nhat)| = 1/2$, $|\psiomega(\nhat)| = 1$, $|\beta_L(\nhat)| = 0$, and $|\beta_T(\nhat)| = 1$.  
The analytical approximation (black horizontal lines) agrees well with the asymptotic value of the probability calculated by direct integration (red) at large $\rad$. 
As the axion energy is increased from $\omega = 10.5 \meV$ (solid red line) to $49.8 \meV$ (dashed red line), the conversion radius increases in agreement with \eref{eq:rres_EH} and the conversion probability decreases in agreement with \eref{eq:Pag_EH}.  
The inset zooms into the resonance region, and shows how the resonance radius $\radres$ from \eref{eq:rres_EH} and the resonance width $\dres$ from \eref{eq:dres_EH} agree well with the numerical integration.  

\begin{figure}[t]
\centering
\includegraphics[scale=0.5]{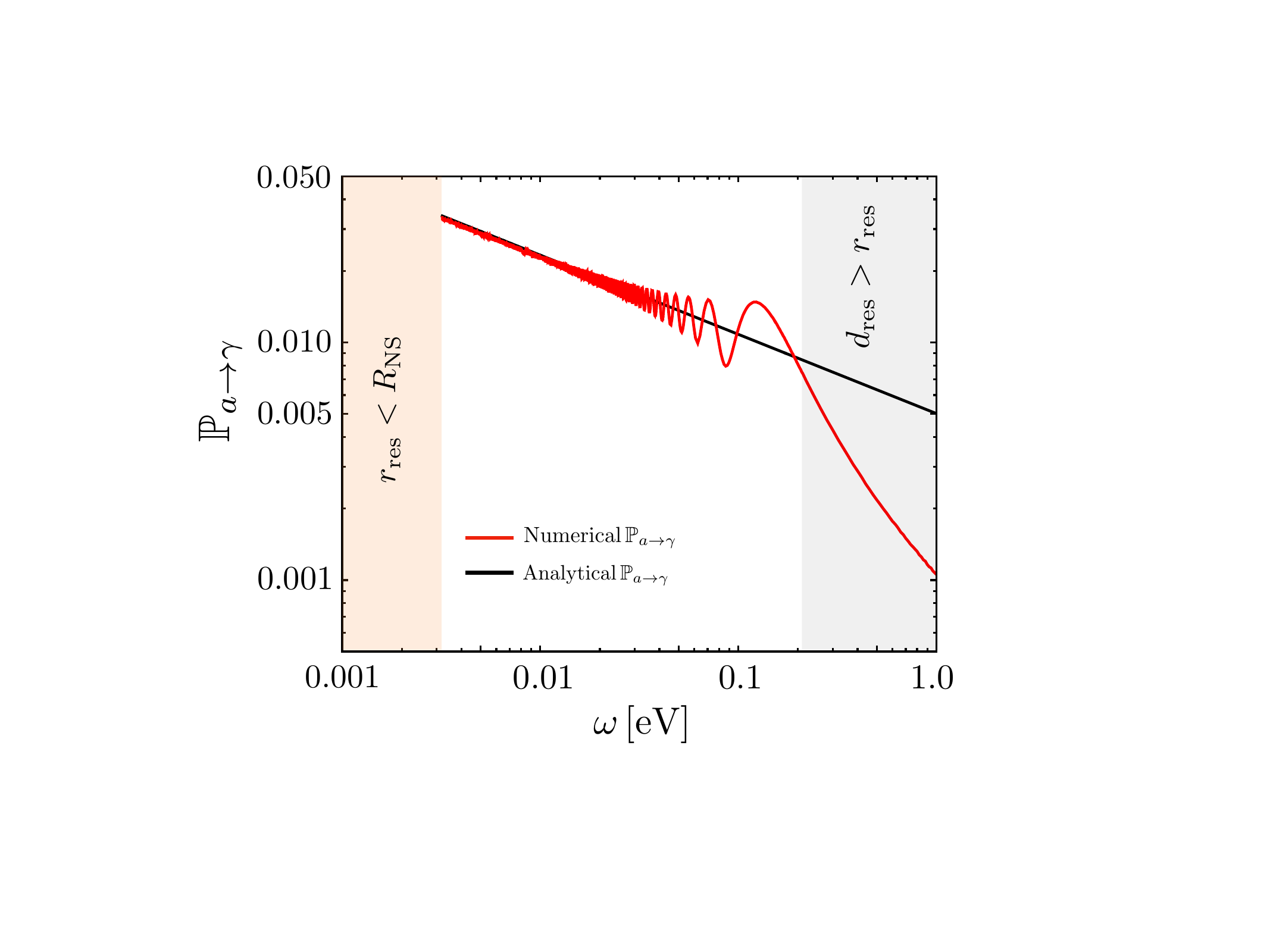}
\caption{\label{fig:prob}
Axion-photon conversion probability $\Pbb_{a\to\gamma}$ as function of the axion/photon energy $\omega$ in the regime of the Euler-Heisenberg assisted resonance.  We show the result of direct numerical integration (red) as well as the analytical approximation (black) from \eref{eq:Pag_EH}, which agrees with a power law having $\Pag \propto \omega^{-1/3}$.  The model parameters are chosen to be $B_0 = 10^{14} \Gauss$, $P = 1 \sec$, $\gagg = 10^{-12} \GeV^{-1}$, $n_e = |n_c|$, and $m_a = 0$.  We consider an axion trajectory that lies in the neutron star's equatorial plane where $|\psivecB(\rhat)| =1/2$, $|\psiomega(\rhat)| = 1$, $|\beta_L(\rhat)| = 0$, and $|\beta_T(\rhat)| = 1$. The shaded tan region indicates where the resonance radius is less than the neutron star radius ($\radres < \RNS$).  The shaded gray region indicates where the resonance width is larger than the resonance radius ($\dres > \radres$), and the analytical approximation is not expected to be reliable.  
}
\end{figure}

Finally in \fref{fig:prob} we show how the axion-photon  conversion probability $\Pbb_{a\to\gamma}$ varies as a function of the axion energy $\omega$.  
As in the previous figure, for this numerical check we focus on the equatorial plane and take the same model parameters as before.  
The probability calculated by direct numerical integration (black) agrees well with the analytical approximation (red) for axion energies between approximately $\omega = 0.003 \eV$ and $0.1 \eV$.  
For smaller values of $\omega$, the resonance radius $\radres$ drops below the neutron star radius $\RNS$, where the our model of the neutron star magnetosphere is inapplicable.  
For larger values of $\omega$, the resonance radius $\radres$ exceeds the resonance width $\dres$, and the narrow resonance approximation breaks down, which can be seen on the figure as the growing deviation between the red and black curves.  
As expected from \eref{eq:Pag_EH}, the resonant conversion probability follows a power-law behavior in terms of the axion energy as $\mathbb{P}_{a\gamma\gamma}(\omega) \propto \omega^{-1/3}$, where $m_a = 0$. 

\newpage
\bibliography{main}

\begin{thebibliography}{143}%
\makeatletter
\providecommand \@ifxundefined [1]{%
 \@ifx{#1\undefined}
}%
\providecommand \@ifnum [1]{%
 \ifnum #1\expandafter \@firstoftwo
 \else \expandafter \@secondoftwo
 \fi
}%
\providecommand \@ifx [1]{%
 \ifx #1\expandafter \@firstoftwo
 \else \expandafter \@secondoftwo
 \fi
}%
\providecommand \natexlab [1]{#1}%
\providecommand \enquote  [1]{``#1''}%
\providecommand \bibnamefont  [1]{#1}%
\providecommand \bibfnamefont [1]{#1}%
\providecommand \citenamefont [1]{#1}%
\providecommand \href@noop [0]{\@secondoftwo}%
\providecommand \href [0]{\begingroup \@sanitize@url \@href}%
\providecommand \@href[1]{\@@startlink{#1}\@@href}%
\providecommand \@@href[1]{\endgroup#1\@@endlink}%
\providecommand \@sanitize@url [0]{\catcode `\\12\catcode `\$12\catcode
  `\&12\catcode `\#12\catcode `\^12\catcode `\_12\catcode `\%12\relax}%
\providecommand \@@startlink[1]{}%
\providecommand \@@endlink[0]{}%
\providecommand \url  [0]{\begingroup\@sanitize@url \@url }%
\providecommand \@url [1]{\endgroup\@href {#1}{\urlprefix }}%
\providecommand \urlprefix  [0]{URL }%
\providecommand \Eprint [0]{\href }%
\providecommand \doibase [0]{http://dx.doi.org/}%
\providecommand \selectlanguage [0]{\@gobble}%
\providecommand \bibinfo  [0]{\@secondoftwo}%
\providecommand \bibfield  [0]{\@secondoftwo}%
\providecommand \translation [1]{[#1]}%
\providecommand \BibitemOpen [0]{}%
\providecommand \bibitemStop [0]{}%
\providecommand \bibitemNoStop [0]{.\EOS\space}%
\providecommand \EOS [0]{\spacefactor3000\relax}%
\providecommand \BibitemShut  [1]{\csname bibitem#1\endcsname}%
\let\auto@bib@innerbib\@empty
\bibitem [{\citenamefont {Peccei}\ and\ \citenamefont
  {Quinn}(1977{\natexlab{a}})}]{Peccei:1977hh}%
  \BibitemOpen
  \bibfield  {author} {\bibinfo {author} {\bibfnamefont {R.~D.}\ \bibnamefont
  {Peccei}}\ and\ \bibinfo {author} {\bibfnamefont {Helen~R.}\ \bibnamefont
  {Quinn}},\ }\bibfield  {title} {\enquote {\bibinfo {title} {{CP Conservation
  in the Presence of Instantons}},}\ }\href {\doibase
  10.1103/PhysRevLett.38.1440} {\bibfield  {journal} {\bibinfo  {journal}
  {Phys. Rev. Lett.}\ }\textbf {\bibinfo {volume} {38}},\ \bibinfo {pages}
  {1440--1443} (\bibinfo {year} {1977}{\natexlab{a}})}\BibitemShut {NoStop}%
\bibitem [{\citenamefont {Peccei}\ and\ \citenamefont
  {Quinn}(1977{\natexlab{b}})}]{Peccei:1977ur}%
  \BibitemOpen
  \bibfield  {author} {\bibinfo {author} {\bibfnamefont {R.~D.}\ \bibnamefont
  {Peccei}}\ and\ \bibinfo {author} {\bibfnamefont {Helen~R.}\ \bibnamefont
  {Quinn}},\ }\bibfield  {title} {\enquote {\bibinfo {title} {{Constraints
  Imposed by CP Conservation in the Presence of Instantons}},}\ }\href
  {\doibase 10.1103/PhysRevD.16.1791} {\bibfield  {journal} {\bibinfo
  {journal} {Phys. Rev. D}\ }\textbf {\bibinfo {volume} {16}},\ \bibinfo
  {pages} {1791--1797} (\bibinfo {year} {1977}{\natexlab{b}})}\BibitemShut
  {NoStop}%
\bibitem [{\citenamefont {Weinberg}(1978)}]{Weinberg:1977ma}%
  \BibitemOpen
  \bibfield  {author} {\bibinfo {author} {\bibfnamefont {Steven}\ \bibnamefont
  {Weinberg}},\ }\bibfield  {title} {\enquote {\bibinfo {title} {{A New Light
  Boson?}}}\ }\href {\doibase 10.1103/PhysRevLett.40.223} {\bibfield  {journal}
  {\bibinfo  {journal} {Phys. Rev. Lett.}\ }\textbf {\bibinfo {volume} {40}},\
  \bibinfo {pages} {223--226} (\bibinfo {year} {1978})}\BibitemShut {NoStop}%
\bibitem [{\citenamefont {Marsh}(2016)}]{Marsh:2015xka}%
  \BibitemOpen
  \bibfield  {author} {\bibinfo {author} {\bibfnamefont {David J.~E.}\
  \bibnamefont {Marsh}},\ }\bibfield  {title} {\enquote {\bibinfo {title}
  {{Axion Cosmology}},}\ }\href {\doibase 10.1016/j.physrep.2016.06.005}
  {\bibfield  {journal} {\bibinfo  {journal} {Phys. Rept.}\ }\textbf {\bibinfo
  {volume} {643}},\ \bibinfo {pages} {1--79} (\bibinfo {year} {2016})},\
  \Eprint {http://arxiv.org/abs/1510.07633} {arXiv:1510.07633 [astro-ph.CO]}
  \BibitemShut {NoStop}%
\bibitem [{\citenamefont {Irastorza}\ and\ \citenamefont
  {Redondo}(2018)}]{Irastorza:2018dyq}%
  \BibitemOpen
  \bibfield  {author} {\bibinfo {author} {\bibfnamefont {Igor~G.}\ \bibnamefont
  {Irastorza}}\ and\ \bibinfo {author} {\bibfnamefont {Javier}\ \bibnamefont
  {Redondo}},\ }\bibfield  {title} {\enquote {\bibinfo {title} {{New
  experimental approaches in the search for axion-like particles}},}\ }\href
  {\doibase 10.1016/j.ppnp.2018.05.003} {\bibfield  {journal} {\bibinfo
  {journal} {Prog. Part. Nucl. Phys.}\ }\textbf {\bibinfo {volume} {102}},\
  \bibinfo {pages} {89--159} (\bibinfo {year} {2018})},\ \Eprint
  {http://arxiv.org/abs/1801.08127} {arXiv:1801.08127 [hep-ph]} \BibitemShut
  {NoStop}%
\bibitem [{\citenamefont {Choi}\ \emph {et~al.}(2021)\citenamefont {Choi},
  \citenamefont {Im},\ and\ \citenamefont {Sub~Shin}}]{Choi:2020rgn}%
  \BibitemOpen
  \bibfield  {author} {\bibinfo {author} {\bibfnamefont {Kiwoon}\ \bibnamefont
  {Choi}}, \bibinfo {author} {\bibfnamefont {Sang~Hui}\ \bibnamefont {Im}}, \
  and\ \bibinfo {author} {\bibfnamefont {Chang}\ \bibnamefont {Sub~Shin}},\
  }\bibfield  {title} {\enquote {\bibinfo {title} {{Recent Progress in the
  Physics of Axions and Axion-Like Particles}},}\ }\href {\doibase
  10.1146/annurev-nucl-120720-031147} {\bibfield  {journal} {\bibinfo
  {journal} {Ann. Rev. Nucl. Part. Sci.}\ }\textbf {\bibinfo {volume} {71}},\
  \bibinfo {pages} {225--252} (\bibinfo {year} {2021})},\ \Eprint
  {http://arxiv.org/abs/2012.05029} {arXiv:2012.05029 [hep-ph]} \BibitemShut
  {NoStop}%
\bibitem [{\citenamefont {Raffelt}(1996)}]{Raffelt:1996wa}%
  \BibitemOpen
  \bibfield  {author} {\bibinfo {author} {\bibfnamefont {G.G.}\ \bibnamefont
  {Raffelt}},\ }\href@noop {} {\emph {\bibinfo {title} {{Stars as laboratories
  for fundamental physics}: {The astrophysics of neutrinos, axions, and other
  weakly interacting particles}}}}\ (\bibinfo {year} {1996})\BibitemShut
  {NoStop}%
\bibitem [{\citenamefont {Maiani}\ \emph {et~al.}(1986)\citenamefont {Maiani},
  \citenamefont {Petronzio},\ and\ \citenamefont {Zavattini}}]{Maiani:1986md}%
  \BibitemOpen
  \bibfield  {author} {\bibinfo {author} {\bibfnamefont {L.}~\bibnamefont
  {Maiani}}, \bibinfo {author} {\bibfnamefont {R.}~\bibnamefont {Petronzio}}, \
  and\ \bibinfo {author} {\bibfnamefont {E.}~\bibnamefont {Zavattini}},\
  }\bibfield  {title} {\enquote {\bibinfo {title} {{Effects of Nearly Massless,
  Spin Zero Particles on Light Propagation in a Magnetic Field}},}\ }\href
  {\doibase 10.1016/0370-2693(86)90869-5} {\bibfield  {journal} {\bibinfo
  {journal} {Phys. Lett. B}\ }\textbf {\bibinfo {volume} {175}},\ \bibinfo
  {pages} {359--363} (\bibinfo {year} {1986})}\BibitemShut {NoStop}%
\bibitem [{\citenamefont {Raffelt}\ and\ \citenamefont
  {Stodolsky}(1988)}]{Raffelt:1987im}%
  \BibitemOpen
  \bibfield  {author} {\bibinfo {author} {\bibfnamefont {Georg}\ \bibnamefont
  {Raffelt}}\ and\ \bibinfo {author} {\bibfnamefont {Leo}\ \bibnamefont
  {Stodolsky}},\ }\bibfield  {title} {\enquote {\bibinfo {title} {{Mixing of
  the Photon with Low Mass Particles}},}\ }\href {\doibase
  10.1103/PhysRevD.37.1237} {\bibfield  {journal} {\bibinfo  {journal} {Phys.
  Rev. D}\ }\textbf {\bibinfo {volume} {37}},\ \bibinfo {pages} {1237}
  (\bibinfo {year} {1988})}\BibitemShut {NoStop}%
\bibitem [{\citenamefont {Carosi}\ \emph {et~al.}(2013)\citenamefont {Carosi},
  \citenamefont {Friedland}, \citenamefont {Giannotti}, \citenamefont
  {Pivovaroff}, \citenamefont {Ruz},\ and\ \citenamefont
  {Vogel}}]{Carosi:2013rla}%
  \BibitemOpen
  \bibfield  {author} {\bibinfo {author} {\bibfnamefont {G.}~\bibnamefont
  {Carosi}}, \bibinfo {author} {\bibfnamefont {A.}~\bibnamefont {Friedland}},
  \bibinfo {author} {\bibfnamefont {M.}~\bibnamefont {Giannotti}}, \bibinfo
  {author} {\bibfnamefont {M.~J.}\ \bibnamefont {Pivovaroff}}, \bibinfo
  {author} {\bibfnamefont {J.}~\bibnamefont {Ruz}}, \ and\ \bibinfo {author}
  {\bibfnamefont {J.~K.}\ \bibnamefont {Vogel}},\ }\bibfield  {title} {\enquote
  {\bibinfo {title} {{Probing the axion-photon coupling: phenomenological and
  experimental perspectives. A snowmass white paper}},}\ }in\ \href@noop {}
  {\emph {\bibinfo {booktitle} {{Snowmass 2013}: {Snowmass on the
  Mississippi}}}}\ (\bibinfo {year} {2013})\ \Eprint
  {http://arxiv.org/abs/1309.7035} {arXiv:1309.7035 [hep-ph]} \BibitemShut
  {NoStop}%
\bibitem [{\citenamefont {Graham}\ \emph {et~al.}(2015)\citenamefont {Graham},
  \citenamefont {Irastorza}, \citenamefont {Lamoreaux}, \citenamefont
  {Lindner},\ and\ \citenamefont {van Bibber}}]{Graham:2015ouw}%
  \BibitemOpen
  \bibfield  {author} {\bibinfo {author} {\bibfnamefont {Peter~W.}\
  \bibnamefont {Graham}}, \bibinfo {author} {\bibfnamefont {Igor~G.}\
  \bibnamefont {Irastorza}}, \bibinfo {author} {\bibfnamefont {Steven~K.}\
  \bibnamefont {Lamoreaux}}, \bibinfo {author} {\bibfnamefont {Axel}\
  \bibnamefont {Lindner}}, \ and\ \bibinfo {author} {\bibfnamefont {Karl~A.}\
  \bibnamefont {van Bibber}},\ }\bibfield  {title} {\enquote {\bibinfo {title}
  {{Experimental Searches for the Axion and Axion-Like Particles}},}\ }\href
  {\doibase 10.1146/annurev-nucl-102014-022120} {\bibfield  {journal} {\bibinfo
   {journal} {Ann. Rev. Nucl. Part. Sci.}\ }\textbf {\bibinfo {volume} {65}},\
  \bibinfo {pages} {485--514} (\bibinfo {year} {2015})},\ \Eprint
  {http://arxiv.org/abs/1602.00039} {arXiv:1602.00039 [hep-ex]} \BibitemShut
  {NoStop}%
\bibitem [{\citenamefont {Adams}\ \emph {et~al.}(2022)\citenamefont {Adams}
  \emph {et~al.}}]{Adams:2022pbo}%
  \BibitemOpen
  \bibfield  {author} {\bibinfo {author} {\bibfnamefont {C.~B.}\ \bibnamefont
  {Adams}} \emph {et~al.},\ }\bibfield  {title} {\enquote {\bibinfo {title}
  {{Axion Dark Matter}},}\ }in\ \href@noop {} {\emph {\bibinfo {booktitle}
  {{Snowmass 2021}}}}\ (\bibinfo {year} {2022})\ \Eprint
  {http://arxiv.org/abs/2203.14923} {arXiv:2203.14923 [hep-ex]} \BibitemShut
  {NoStop}%
\bibitem [{\citenamefont {Yoshimura}(1988)}]{Yoshimura:1987ma}%
  \BibitemOpen
  \bibfield  {author} {\bibinfo {author} {\bibfnamefont {M.}~\bibnamefont
  {Yoshimura}},\ }\bibfield  {title} {\enquote {\bibinfo {title} {{Resonant
  axion - photon conversion in magnetized plasma}},}\ }\href {\doibase
  10.1103/PhysRevD.37.2039} {\bibfield  {journal} {\bibinfo  {journal} {Phys.
  Rev. D}\ }\textbf {\bibinfo {volume} {37}},\ \bibinfo {pages} {2039}
  (\bibinfo {year} {1988})}\BibitemShut {NoStop}%
\bibitem [{\citenamefont {Lai}\ and\ \citenamefont {Heyl}(2006)}]{Lai:2006af}%
  \BibitemOpen
  \bibfield  {author} {\bibinfo {author} {\bibfnamefont {Dong}\ \bibnamefont
  {Lai}}\ and\ \bibinfo {author} {\bibfnamefont {Jeremy}\ \bibnamefont
  {Heyl}},\ }\bibfield  {title} {\enquote {\bibinfo {title} {{Probing Axions
  with Radiation from Magnetic Stars}},}\ }\href {\doibase
  10.1103/PhysRevD.74.123003} {\bibfield  {journal} {\bibinfo  {journal} {Phys.
  Rev. D}\ }\textbf {\bibinfo {volume} {74}},\ \bibinfo {pages} {123003}
  (\bibinfo {year} {2006})},\ \Eprint {http://arxiv.org/abs/astro-ph/0609775}
  {arXiv:astro-ph/0609775} \BibitemShut {NoStop}%
\bibitem [{\citenamefont {Bondarenko}\ \emph {et~al.}(2023)\citenamefont
  {Bondarenko}, \citenamefont {Boyarsky}, \citenamefont {Pradler},\ and\
  \citenamefont {Sokolenko}}]{Bondarenko:2022ngb}%
  \BibitemOpen
  \bibfield  {author} {\bibinfo {author} {\bibfnamefont {Kyrylo}\ \bibnamefont
  {Bondarenko}}, \bibinfo {author} {\bibfnamefont {Alexey}\ \bibnamefont
  {Boyarsky}}, \bibinfo {author} {\bibfnamefont {Josef}\ \bibnamefont
  {Pradler}}, \ and\ \bibinfo {author} {\bibfnamefont {Anastasia}\ \bibnamefont
  {Sokolenko}},\ }\bibfield  {title} {\enquote {\bibinfo {title} {{Neutron
  stars as photon double-lenses: Constraining resonant conversion into
  ALPs}},}\ }\href {\doibase 10.1016/j.physletb.2023.138238} {\bibfield
  {journal} {\bibinfo  {journal} {Phys. Lett. B}\ }\textbf {\bibinfo {volume}
  {846}},\ \bibinfo {pages} {138238} (\bibinfo {year} {2023})},\ \Eprint
  {http://arxiv.org/abs/2203.08663} {arXiv:2203.08663 [hep-ph]} \BibitemShut
  {NoStop}%
\bibitem [{\citenamefont {Song}\ \emph {et~al.}(2024)\citenamefont {Song},
  \citenamefont {Su},\ and\ \citenamefont {Wu}}]{Song:2024rru}%
  \BibitemOpen
  \bibfield  {author} {\bibinfo {author} {\bibfnamefont {Ningqiang}\
  \bibnamefont {Song}}, \bibinfo {author} {\bibfnamefont {Liangliang}\
  \bibnamefont {Su}}, \ and\ \bibinfo {author} {\bibfnamefont {Lei}\
  \bibnamefont {Wu}},\ }\bibfield  {title} {\enquote {\bibinfo {title}
  {{Polarization Signals from Axion-Photon Resonant Conversion in Neutron Star
  Magnetosphere}},}\ }\href@noop {} {\  (\bibinfo {year} {2024})},\ \Eprint
  {http://arxiv.org/abs/2402.15144} {arXiv:2402.15144 [hep-ph]} \BibitemShut
  {NoStop}%
\bibitem [{\citenamefont {Yanagida}\ and\ \citenamefont
  {Yoshimura}(1988)}]{Yanagida:1987nf}%
  \BibitemOpen
  \bibfield  {author} {\bibinfo {author} {\bibfnamefont {T.}~\bibnamefont
  {Yanagida}}\ and\ \bibinfo {author} {\bibfnamefont {M.}~\bibnamefont
  {Yoshimura}},\ }\bibfield  {title} {\enquote {\bibinfo {title} {{Resonant
  Axion - Photon Conversion in the Early Universe}},}\ }\href {\doibase
  10.1016/0370-2693(88)90475-3} {\bibfield  {journal} {\bibinfo  {journal}
  {Phys. Lett. B}\ }\textbf {\bibinfo {volume} {202}},\ \bibinfo {pages}
  {301--306} (\bibinfo {year} {1988})}\BibitemShut {NoStop}%
\bibitem [{\citenamefont {Higaki}\ \emph
  {et~al.}(2013{\natexlab{a}})\citenamefont {Higaki}, \citenamefont
  {Nakayama},\ and\ \citenamefont {Takahashi}}]{Higaki:2013qka}%
  \BibitemOpen
  \bibfield  {author} {\bibinfo {author} {\bibfnamefont {Tetsutaro}\
  \bibnamefont {Higaki}}, \bibinfo {author} {\bibfnamefont {Kazunori}\
  \bibnamefont {Nakayama}}, \ and\ \bibinfo {author} {\bibfnamefont {Fuminobu}\
  \bibnamefont {Takahashi}},\ }\bibfield  {title} {\enquote {\bibinfo {title}
  {{Cosmological constraints on axionic dark radiation from axion-photon
  conversion in the early Universe}},}\ }\href {\doibase
  10.1088/1475-7516/2013/09/030} {\bibfield  {journal} {\bibinfo  {journal}
  {JCAP}\ }\textbf {\bibinfo {volume} {09}},\ \bibinfo {pages} {030} (\bibinfo
  {year} {2013}{\natexlab{a}})},\ \Eprint {http://arxiv.org/abs/1306.6518}
  {arXiv:1306.6518 [hep-ph]} \BibitemShut {NoStop}%
\bibitem [{\citenamefont {Tashiro}\ \emph {et~al.}(2013)\citenamefont
  {Tashiro}, \citenamefont {Silk},\ and\ \citenamefont
  {Marsh}}]{Tashiro:2013yea}%
  \BibitemOpen
  \bibfield  {author} {\bibinfo {author} {\bibfnamefont {Hiroyuki}\
  \bibnamefont {Tashiro}}, \bibinfo {author} {\bibfnamefont {Joseph}\
  \bibnamefont {Silk}}, \ and\ \bibinfo {author} {\bibfnamefont {David J.~E.}\
  \bibnamefont {Marsh}},\ }\bibfield  {title} {\enquote {\bibinfo {title}
  {{Constraints on primordial magnetic fields from CMB distortions in the
  axiverse}},}\ }\href {\doibase 10.1103/PhysRevD.88.125024} {\bibfield
  {journal} {\bibinfo  {journal} {Phys. Rev. D}\ }\textbf {\bibinfo {volume}
  {88}},\ \bibinfo {pages} {125024} (\bibinfo {year} {2013})},\ \Eprint
  {http://arxiv.org/abs/1308.0314} {arXiv:1308.0314 [astro-ph.CO]} \BibitemShut
  {NoStop}%
\bibitem [{\citenamefont {Angus}\ \emph {et~al.}(2014)\citenamefont {Angus},
  \citenamefont {Conlon}, \citenamefont {Marsh}, \citenamefont {Powell},\ and\
  \citenamefont {Witkowski}}]{Angus:2013sua}%
  \BibitemOpen
  \bibfield  {author} {\bibinfo {author} {\bibfnamefont {Stephen}\ \bibnamefont
  {Angus}}, \bibinfo {author} {\bibfnamefont {Joseph~P.}\ \bibnamefont
  {Conlon}}, \bibinfo {author} {\bibfnamefont {M.~C.~David}\ \bibnamefont
  {Marsh}}, \bibinfo {author} {\bibfnamefont {Andrew~J.}\ \bibnamefont
  {Powell}}, \ and\ \bibinfo {author} {\bibfnamefont {Lukas~T.}\ \bibnamefont
  {Witkowski}},\ }\bibfield  {title} {\enquote {\bibinfo {title} {{Soft X-ray
  Excess in the Coma Cluster from a Cosmic Axion Background}},}\ }\href
  {\doibase 10.1088/1475-7516/2014/09/026} {\bibfield  {journal} {\bibinfo
  {journal} {JCAP}\ }\textbf {\bibinfo {volume} {09}},\ \bibinfo {pages} {026}
  (\bibinfo {year} {2014})},\ \Eprint {http://arxiv.org/abs/1312.3947}
  {arXiv:1312.3947 [astro-ph.HE]} \BibitemShut {NoStop}%
\bibitem [{\citenamefont {Kraljic}\ \emph {et~al.}(2015)\citenamefont
  {Kraljic}, \citenamefont {Rummel},\ and\ \citenamefont
  {Conlon}}]{Kraljic:2014yta}%
  \BibitemOpen
  \bibfield  {author} {\bibinfo {author} {\bibfnamefont {David}\ \bibnamefont
  {Kraljic}}, \bibinfo {author} {\bibfnamefont {Markus}\ \bibnamefont
  {Rummel}}, \ and\ \bibinfo {author} {\bibfnamefont {Joseph~P.}\ \bibnamefont
  {Conlon}},\ }\bibfield  {title} {\enquote {\bibinfo {title} {{ALP Conversion
  and the Soft X-ray Excess in the Outskirts of the Coma Cluster}},}\ }\href
  {\doibase 10.1088/1475-7516/2015/01/011} {\bibfield  {journal} {\bibinfo
  {journal} {JCAP}\ }\textbf {\bibinfo {volume} {01}},\ \bibinfo {pages} {011}
  (\bibinfo {year} {2015})},\ \Eprint {http://arxiv.org/abs/1406.5188}
  {arXiv:1406.5188 [hep-ph]} \BibitemShut {NoStop}%
\bibitem [{\citenamefont {Evoli}\ \emph {et~al.}(2016)\citenamefont {Evoli},
  \citenamefont {Leo}, \citenamefont {Mirizzi},\ and\ \citenamefont
  {Montanino}}]{Evoli:2016zhj}%
  \BibitemOpen
  \bibfield  {author} {\bibinfo {author} {\bibfnamefont {Carmelo}\ \bibnamefont
  {Evoli}}, \bibinfo {author} {\bibfnamefont {Matteo}\ \bibnamefont {Leo}},
  \bibinfo {author} {\bibfnamefont {Alessandro}\ \bibnamefont {Mirizzi}}, \
  and\ \bibinfo {author} {\bibfnamefont {Daniele}\ \bibnamefont {Montanino}},\
  }\bibfield  {title} {\enquote {\bibinfo {title} {{Reionization during the
  dark ages from a cosmic axion background}},}\ }\href {\doibase
  10.1088/1475-7516/2016/05/006} {\bibfield  {journal} {\bibinfo  {journal}
  {JCAP}\ }\textbf {\bibinfo {volume} {05}},\ \bibinfo {pages} {006} (\bibinfo
  {year} {2016})},\ \Eprint {http://arxiv.org/abs/1602.08433} {arXiv:1602.08433
  [astro-ph.CO]} \BibitemShut {NoStop}%
\bibitem [{\citenamefont {Fortin}\ and\ \citenamefont
  {Sinha}(2018)}]{Fortin:2018ehg}%
  \BibitemOpen
  \bibfield  {author} {\bibinfo {author} {\bibfnamefont {Jean-Fran\c{c}ois}\
  \bibnamefont {Fortin}}\ and\ \bibinfo {author} {\bibfnamefont {Kuver}\
  \bibnamefont {Sinha}},\ }\bibfield  {title} {\enquote {\bibinfo {title}
  {{Constraining Axion-Like-Particles with Hard X-ray Emission from
  Magnetars}},}\ }\href {\doibase 10.1007/JHEP06(2018)048} {\bibfield
  {journal} {\bibinfo  {journal} {JHEP}\ }\textbf {\bibinfo {volume} {06}},\
  \bibinfo {pages} {048} (\bibinfo {year} {2018})},\ \Eprint
  {http://arxiv.org/abs/1804.01992} {arXiv:1804.01992 [hep-ph]} \BibitemShut
  {NoStop}%
\bibitem [{\citenamefont {Harris}\ \emph {et~al.}(2020)\citenamefont {Harris},
  \citenamefont {Fortin}, \citenamefont {Sinha},\ and\ \citenamefont
  {Alford}}]{Harris:2020qim}%
  \BibitemOpen
  \bibfield  {author} {\bibinfo {author} {\bibfnamefont {Steven~P.}\
  \bibnamefont {Harris}}, \bibinfo {author} {\bibfnamefont {Jean-Francois}\
  \bibnamefont {Fortin}}, \bibinfo {author} {\bibfnamefont {Kuver}\
  \bibnamefont {Sinha}}, \ and\ \bibinfo {author} {\bibfnamefont {Mark~G.}\
  \bibnamefont {Alford}},\ }\bibfield  {title} {\enquote {\bibinfo {title}
  {{Axions in neutron star mergers}},}\ }\href {\doibase
  10.1088/1475-7516/2020/07/023} {\bibfield  {journal} {\bibinfo  {journal}
  {JCAP}\ }\textbf {\bibinfo {volume} {07}},\ \bibinfo {pages} {023} (\bibinfo
  {year} {2020})},\ \Eprint {http://arxiv.org/abs/2003.09768} {arXiv:2003.09768
  [hep-ph]} \BibitemShut {NoStop}%
\bibitem [{\citenamefont {Buen-Abad}\ \emph {et~al.}(2022)\citenamefont
  {Buen-Abad}, \citenamefont {Fan},\ and\ \citenamefont
  {Sun}}]{Buen-Abad:2020zbd}%
  \BibitemOpen
  \bibfield  {author} {\bibinfo {author} {\bibfnamefont {Manuel~A.}\
  \bibnamefont {Buen-Abad}}, \bibinfo {author} {\bibfnamefont {JiJi}\
  \bibnamefont {Fan}}, \ and\ \bibinfo {author} {\bibfnamefont {Chen}\
  \bibnamefont {Sun}},\ }\bibfield  {title} {\enquote {\bibinfo {title}
  {{Constraints on axions from cosmic distance measurements}},}\ }\href
  {\doibase 10.1007/JHEP02(2022)103} {\bibfield  {journal} {\bibinfo  {journal}
  {JHEP}\ }\textbf {\bibinfo {volume} {02}},\ \bibinfo {pages} {103} (\bibinfo
  {year} {2022})},\ \Eprint {http://arxiv.org/abs/2011.05993} {arXiv:2011.05993
  [hep-ph]} \BibitemShut {NoStop}%
\bibitem [{\citenamefont {Fortin}\ \emph
  {et~al.}(2021{\natexlab{a}})\citenamefont {Fortin}, \citenamefont {Guo},
  \citenamefont {Harris}, \citenamefont {Kim}, \citenamefont {Sinha},\ and\
  \citenamefont {Sun}}]{Fortin:2021cog}%
  \BibitemOpen
  \bibfield  {author} {\bibinfo {author} {\bibfnamefont {Jean-Fran\c{c}ois}\
  \bibnamefont {Fortin}}, \bibinfo {author} {\bibfnamefont {Huai-Ke}\
  \bibnamefont {Guo}}, \bibinfo {author} {\bibfnamefont {Steven~P.}\
  \bibnamefont {Harris}}, \bibinfo {author} {\bibfnamefont {Doojin}\
  \bibnamefont {Kim}}, \bibinfo {author} {\bibfnamefont {Kuver}\ \bibnamefont
  {Sinha}}, \ and\ \bibinfo {author} {\bibfnamefont {Chen}\ \bibnamefont
  {Sun}},\ }\bibfield  {title} {\enquote {\bibinfo {title} {{Axions: From
  magnetars and neutron star mergers to beam dumps and BECs}},}\ }\href
  {\doibase 10.1142/S0218271821300020} {\bibfield  {journal} {\bibinfo
  {journal} {Int. J. Mod. Phys. D}\ }\textbf {\bibinfo {volume} {30}},\
  \bibinfo {pages} {2130002} (\bibinfo {year} {2021}{\natexlab{a}})},\ \Eprint
  {http://arxiv.org/abs/2102.12503} {arXiv:2102.12503 [hep-ph]} \BibitemShut
  {NoStop}%
\bibitem [{\citenamefont {Schiavone}\ \emph {et~al.}(2021)\citenamefont
  {Schiavone}, \citenamefont {Montanino}, \citenamefont {Mirizzi},\ and\
  \citenamefont {Capozzi}}]{Schiavone:2021imu}%
  \BibitemOpen
  \bibfield  {author} {\bibinfo {author} {\bibfnamefont {Francesco}\
  \bibnamefont {Schiavone}}, \bibinfo {author} {\bibfnamefont {Daniele}\
  \bibnamefont {Montanino}}, \bibinfo {author} {\bibfnamefont {Alessandro}\
  \bibnamefont {Mirizzi}}, \ and\ \bibinfo {author} {\bibfnamefont {Francesco}\
  \bibnamefont {Capozzi}},\ }\bibfield  {title} {\enquote {\bibinfo {title}
  {{Axion-like particles from primordial black holes shining through the
  Universe}},}\ }\href {\doibase 10.1088/1475-7516/2021/08/063} {\bibfield
  {journal} {\bibinfo  {journal} {JCAP}\ }\textbf {\bibinfo {volume} {08}},\
  \bibinfo {pages} {063} (\bibinfo {year} {2021})},\ \Eprint
  {http://arxiv.org/abs/2107.03420} {arXiv:2107.03420 [hep-ph]} \BibitemShut
  {NoStop}%
\bibitem [{\citenamefont {Kar}\ \emph {et~al.}(2023)\citenamefont {Kar},
  \citenamefont {Kumar}, \citenamefont {Roy},\ and\ \citenamefont
  {Zupan}}]{Kar:2022ngx}%
  \BibitemOpen
  \bibfield  {author} {\bibinfo {author} {\bibfnamefont {Arpan}\ \bibnamefont
  {Kar}}, \bibinfo {author} {\bibfnamefont {Tanmoy}\ \bibnamefont {Kumar}},
  \bibinfo {author} {\bibfnamefont {Sourov}\ \bibnamefont {Roy}}, \ and\
  \bibinfo {author} {\bibfnamefont {Jure}\ \bibnamefont {Zupan}},\ }\bibfield
  {title} {\enquote {\bibinfo {title} {{Searching for relativistic axions in
  the~sky}},}\ }\href {\doibase 10.1088/1475-7516/2023/08/056} {\bibfield
  {journal} {\bibinfo  {journal} {JCAP}\ }\textbf {\bibinfo {volume} {08}},\
  \bibinfo {pages} {056} (\bibinfo {year} {2023})},\ \Eprint
  {http://arxiv.org/abs/2212.04647} {arXiv:2212.04647 [hep-ph]} \BibitemShut
  {NoStop}%
\bibitem [{\citenamefont {Carenza}\ and\ \citenamefont
  {Marsh}(2023)}]{Carenza:2023nck}%
  \BibitemOpen
  \bibfield  {author} {\bibinfo {author} {\bibfnamefont {Pierluca}\
  \bibnamefont {Carenza}}\ and\ \bibinfo {author} {\bibfnamefont {M.~C.~David}\
  \bibnamefont {Marsh}},\ }\bibfield  {title} {\enquote {\bibinfo {title} {{On
  the applicability of the Landau-Zener formula to axion-photon conversion}},}\
  }\href {\doibase 10.1088/1475-7516/2023/04/021} {\bibfield  {journal}
  {\bibinfo  {journal} {JCAP}\ }\textbf {\bibinfo {volume} {04}},\ \bibinfo
  {pages} {021} (\bibinfo {year} {2023})},\ \Eprint
  {http://arxiv.org/abs/2302.02700} {arXiv:2302.02700 [hep-ph]} \BibitemShut
  {NoStop}%
\bibitem [{\citenamefont {Fortin}\ and\ \citenamefont
  {Sinha}(2023)}]{Fortin:2023jlg}%
  \BibitemOpen
  \bibfield  {author} {\bibinfo {author} {\bibfnamefont {Jean-Fran\c{c}ois}\
  \bibnamefont {Fortin}}\ and\ \bibinfo {author} {\bibfnamefont {Kuver}\
  \bibnamefont {Sinha}},\ }\bibfield  {title} {\enquote {\bibinfo {title}
  {{Polarization formalism for ALP-induced X-ray emission from magnetars}},}\
  }\href {\doibase 10.1088/1475-7516/2023/08/042} {\bibfield  {journal}
  {\bibinfo  {journal} {JCAP}\ }\textbf {\bibinfo {volume} {08}},\ \bibinfo
  {pages} {042} (\bibinfo {year} {2023})},\ \Eprint
  {http://arxiv.org/abs/2303.17641} {arXiv:2303.17641 [astro-ph.HE]}
  \BibitemShut {NoStop}%
\bibitem [{\citenamefont {Sun}\ \emph {et~al.}(2023)\citenamefont {Sun},
  \citenamefont {Buen-Abad},\ and\ \citenamefont {Fan}}]{Sun:2023wqq}%
  \BibitemOpen
  \bibfield  {author} {\bibinfo {author} {\bibfnamefont {Chen}\ \bibnamefont
  {Sun}}, \bibinfo {author} {\bibfnamefont {Manuel~A.}\ \bibnamefont
  {Buen-Abad}}, \ and\ \bibinfo {author} {\bibfnamefont {JiJi}\ \bibnamefont
  {Fan}},\ }\bibfield  {title} {\enquote {\bibinfo {title} {{Probing New
  Physics with High-Redshift Quasars: Axions and Non-standard Cosmology}},}\
  }\href@noop {} {\  (\bibinfo {year} {2023})},\ \Eprint
  {http://arxiv.org/abs/2309.07212} {arXiv:2309.07212 [astro-ph.CO]}
  \BibitemShut {NoStop}%
\bibitem [{\citenamefont {Mondino}\ \emph {et~al.}(2024)\citenamefont
  {Mondino}, \citenamefont {P\^\i{}rvu}, \citenamefont {Huang},\ and\
  \citenamefont {Johnson}}]{Mondino:2024rif}%
  \BibitemOpen
  \bibfield  {author} {\bibinfo {author} {\bibfnamefont {Cristina}\
  \bibnamefont {Mondino}}, \bibinfo {author} {\bibfnamefont {Dalila}\
  \bibnamefont {P\^\i{}rvu}}, \bibinfo {author} {\bibfnamefont {Junwu}\
  \bibnamefont {Huang}}, \ and\ \bibinfo {author} {\bibfnamefont {Matthew~C.}\
  \bibnamefont {Johnson}},\ }\bibfield  {title} {\enquote {\bibinfo {title}
  {{Axion-Induced Patchy Screening of the Cosmic Microwave Background}},}\
  }\href@noop {} {\  (\bibinfo {year} {2024})},\ \Eprint
  {http://arxiv.org/abs/2405.08059} {arXiv:2405.08059 [hep-ph]} \BibitemShut
  {NoStop}%
\bibitem [{\citenamefont {Beadle}\ \emph {et~al.}(2024)\citenamefont {Beadle},
  \citenamefont {Caputo},\ and\ \citenamefont {Ellis}}]{Beadle:2024jlr}%
  \BibitemOpen
  \bibfield  {author} {\bibinfo {author} {\bibfnamefont {Carl}\ \bibnamefont
  {Beadle}}, \bibinfo {author} {\bibfnamefont {Andrea}\ \bibnamefont {Caputo}},
  \ and\ \bibinfo {author} {\bibfnamefont {Sebastian A.~R.}\ \bibnamefont
  {Ellis}},\ }\bibfield  {title} {\enquote {\bibinfo {title} {{Resonant
  Conversion of Wave Dark Matter in the Ionosphere}},}\ }\href@noop {} {\
  (\bibinfo {year} {2024})},\ \Eprint {http://arxiv.org/abs/2405.13882}
  {arXiv:2405.13882 [hep-ph]} \BibitemShut {NoStop}%
\bibitem [{\citenamefont {Ferreira}\ and\ \citenamefont
  {Gil~Muyor}(2024)}]{Ferreira:2024ktd}%
  \BibitemOpen
  \bibfield  {author} {\bibinfo {author} {\bibfnamefont {Ricardo~Z.}\
  \bibnamefont {Ferreira}}\ and\ \bibinfo {author} {\bibfnamefont {\'Angel}\
  \bibnamefont {Gil~Muyor}},\ }\bibfield  {title} {\enquote {\bibinfo {title}
  {{Lighten up Primordial Black Holes in the Galaxy with the QCD axion: Signals
  at the LOFAR Telescope}},}\ }\href@noop {} {\  (\bibinfo {year} {2024})},\
  \Eprint {http://arxiv.org/abs/2404.12437} {arXiv:2404.12437 [hep-ph]}
  \BibitemShut {NoStop}%
\bibitem [{\citenamefont {Pshirkov}\ and\ \citenamefont
  {Popov}(2009)}]{Pshirkov:2007st}%
  \BibitemOpen
  \bibfield  {author} {\bibinfo {author} {\bibfnamefont {M.~S.}\ \bibnamefont
  {Pshirkov}}\ and\ \bibinfo {author} {\bibfnamefont {S.~B.}\ \bibnamefont
  {Popov}},\ }\bibfield  {title} {\enquote {\bibinfo {title} {{Conversion of
  Dark matter axions to photons in magnetospheres of neutron stars}},}\ }\href
  {\doibase 10.1134/S1063776109030030} {\bibfield  {journal} {\bibinfo
  {journal} {J. Exp. Theor. Phys.}\ }\textbf {\bibinfo {volume} {108}},\
  \bibinfo {pages} {384--388} (\bibinfo {year} {2009})},\ \Eprint
  {http://arxiv.org/abs/0711.1264} {arXiv:0711.1264 [astro-ph]} \BibitemShut
  {NoStop}%
\bibitem [{\citenamefont {Huang}\ \emph {et~al.}(2018)\citenamefont {Huang},
  \citenamefont {Kadota}, \citenamefont {Sekiguchi},\ and\ \citenamefont
  {Tashiro}}]{Huang:2018lxq}%
  \BibitemOpen
  \bibfield  {author} {\bibinfo {author} {\bibfnamefont {Fa~Peng}\ \bibnamefont
  {Huang}}, \bibinfo {author} {\bibfnamefont {Kenji}\ \bibnamefont {Kadota}},
  \bibinfo {author} {\bibfnamefont {Toyokazu}\ \bibnamefont {Sekiguchi}}, \
  and\ \bibinfo {author} {\bibfnamefont {Hiroyuki}\ \bibnamefont {Tashiro}},\
  }\bibfield  {title} {\enquote {\bibinfo {title} {{Radio telescope search for
  the resonant conversion of cold dark matter axions from the magnetized
  astrophysical sources}},}\ }\href {\doibase 10.1103/PhysRevD.97.123001}
  {\bibfield  {journal} {\bibinfo  {journal} {Phys. Rev. D}\ }\textbf {\bibinfo
  {volume} {97}},\ \bibinfo {pages} {123001} (\bibinfo {year} {2018})},\
  \Eprint {http://arxiv.org/abs/1803.08230} {arXiv:1803.08230 [hep-ph]}
  \BibitemShut {NoStop}%
\bibitem [{\citenamefont {Hook}\ \emph {et~al.}(2018)\citenamefont {Hook},
  \citenamefont {Kahn}, \citenamefont {Safdi},\ and\ \citenamefont
  {Sun}}]{Hook:2018iia}%
  \BibitemOpen
  \bibfield  {author} {\bibinfo {author} {\bibfnamefont {Anson}\ \bibnamefont
  {Hook}}, \bibinfo {author} {\bibfnamefont {Yonatan}\ \bibnamefont {Kahn}},
  \bibinfo {author} {\bibfnamefont {Benjamin~R.}\ \bibnamefont {Safdi}}, \ and\
  \bibinfo {author} {\bibfnamefont {Zhiquan}\ \bibnamefont {Sun}},\ }\bibfield
  {title} {\enquote {\bibinfo {title} {{Radio Signals from Axion Dark Matter
  Conversion in Neutron Star Magnetospheres}},}\ }\href {\doibase
  10.1103/PhysRevLett.121.241102} {\bibfield  {journal} {\bibinfo  {journal}
  {Phys. Rev. Lett.}\ }\textbf {\bibinfo {volume} {121}},\ \bibinfo {pages}
  {241102} (\bibinfo {year} {2018})},\ \Eprint
  {http://arxiv.org/abs/1804.03145} {arXiv:1804.03145 [hep-ph]} \BibitemShut
  {NoStop}%
\bibitem [{\citenamefont {Safdi}\ \emph {et~al.}(2019)\citenamefont {Safdi},
  \citenamefont {Sun},\ and\ \citenamefont {Chen}}]{Safdi:2018oeu}%
  \BibitemOpen
  \bibfield  {author} {\bibinfo {author} {\bibfnamefont {Benjamin~R.}\
  \bibnamefont {Safdi}}, \bibinfo {author} {\bibfnamefont {Zhiquan}\
  \bibnamefont {Sun}}, \ and\ \bibinfo {author} {\bibfnamefont {Alexander~Y.}\
  \bibnamefont {Chen}},\ }\bibfield  {title} {\enquote {\bibinfo {title}
  {{Detecting Axion Dark Matter with Radio Lines from Neutron Star
  Populations}},}\ }\href {\doibase 10.1103/PhysRevD.99.123021} {\bibfield
  {journal} {\bibinfo  {journal} {Phys. Rev. D}\ }\textbf {\bibinfo {volume}
  {99}},\ \bibinfo {pages} {123021} (\bibinfo {year} {2019})},\ \Eprint
  {http://arxiv.org/abs/1811.01020} {arXiv:1811.01020 [astro-ph.CO]}
  \BibitemShut {NoStop}%
\bibitem [{\citenamefont {Foster}\ \emph {et~al.}(2020)\citenamefont {Foster},
  \citenamefont {Kahn}, \citenamefont {Macias}, \citenamefont {Sun},
  \citenamefont {Eatough}, \citenamefont {Kondratiev}, \citenamefont {Peters},
  \citenamefont {Weniger},\ and\ \citenamefont {Safdi}}]{Foster:2020pgt}%
  \BibitemOpen
  \bibfield  {author} {\bibinfo {author} {\bibfnamefont {Joshua~W.}\
  \bibnamefont {Foster}}, \bibinfo {author} {\bibfnamefont {Yonatan}\
  \bibnamefont {Kahn}}, \bibinfo {author} {\bibfnamefont {Oscar}\ \bibnamefont
  {Macias}}, \bibinfo {author} {\bibfnamefont {Zhiquan}\ \bibnamefont {Sun}},
  \bibinfo {author} {\bibfnamefont {Ralph~P.}\ \bibnamefont {Eatough}},
  \bibinfo {author} {\bibfnamefont {Vladislav~I.}\ \bibnamefont {Kondratiev}},
  \bibinfo {author} {\bibfnamefont {Wendy~M.}\ \bibnamefont {Peters}}, \bibinfo
  {author} {\bibfnamefont {Christoph}\ \bibnamefont {Weniger}}, \ and\ \bibinfo
  {author} {\bibfnamefont {Benjamin~R.}\ \bibnamefont {Safdi}},\ }\bibfield
  {title} {\enquote {\bibinfo {title} {{Green Bank and Effelsberg Radio
  Telescope Searches for Axion Dark Matter Conversion in Neutron Star
  Magnetospheres}},}\ }\href {\doibase 10.1103/PhysRevLett.125.171301}
  {\bibfield  {journal} {\bibinfo  {journal} {Phys. Rev. Lett.}\ }\textbf
  {\bibinfo {volume} {125}},\ \bibinfo {pages} {171301} (\bibinfo {year}
  {2020})},\ \Eprint {http://arxiv.org/abs/2004.00011} {arXiv:2004.00011
  [astro-ph.CO]} \BibitemShut {NoStop}%
\bibitem [{\citenamefont {Buckley}\ \emph {et~al.}(2021)\citenamefont
  {Buckley}, \citenamefont {Dev}, \citenamefont {Ferrer},\ and\ \citenamefont
  {Huang}}]{Buckley:2020fmh}%
  \BibitemOpen
  \bibfield  {author} {\bibinfo {author} {\bibfnamefont {James~H.}\
  \bibnamefont {Buckley}}, \bibinfo {author} {\bibfnamefont {P.~S.~Bhupal}\
  \bibnamefont {Dev}}, \bibinfo {author} {\bibfnamefont {Francesc}\
  \bibnamefont {Ferrer}}, \ and\ \bibinfo {author} {\bibfnamefont {Fa~Peng}\
  \bibnamefont {Huang}},\ }\bibfield  {title} {\enquote {\bibinfo {title}
  {{Fast radio bursts from axion stars moving through pulsar
  magnetospheres}},}\ }\href {\doibase 10.1103/PhysRevD.103.043015} {\bibfield
  {journal} {\bibinfo  {journal} {Phys. Rev. D}\ }\textbf {\bibinfo {volume}
  {103}},\ \bibinfo {pages} {043015} (\bibinfo {year} {2021})},\ \Eprint
  {http://arxiv.org/abs/2004.06486} {arXiv:2004.06486 [astro-ph.HE]}
  \BibitemShut {NoStop}%
\bibitem [{\citenamefont {Edwards}\ \emph {et~al.}(2021)\citenamefont
  {Edwards}, \citenamefont {Kavanagh}, \citenamefont {Visinelli},\ and\
  \citenamefont {Weniger}}]{Edwards:2020afl}%
  \BibitemOpen
  \bibfield  {author} {\bibinfo {author} {\bibfnamefont {Thomas D.~P.}\
  \bibnamefont {Edwards}}, \bibinfo {author} {\bibfnamefont {Bradley~J.}\
  \bibnamefont {Kavanagh}}, \bibinfo {author} {\bibfnamefont {Luca}\
  \bibnamefont {Visinelli}}, \ and\ \bibinfo {author} {\bibfnamefont
  {Christoph}\ \bibnamefont {Weniger}},\ }\bibfield  {title} {\enquote
  {\bibinfo {title} {{Transient Radio Signatures from Neutron Star Encounters
  with QCD Axion Miniclusters}},}\ }\href {\doibase
  10.1103/PhysRevLett.127.131103} {\bibfield  {journal} {\bibinfo  {journal}
  {Phys. Rev. Lett.}\ }\textbf {\bibinfo {volume} {127}},\ \bibinfo {pages}
  {131103} (\bibinfo {year} {2021})},\ \Eprint
  {http://arxiv.org/abs/2011.05378} {arXiv:2011.05378 [hep-ph]} \BibitemShut
  {NoStop}%
\bibitem [{\citenamefont {Nurmi}\ \emph {et~al.}(2021)\citenamefont {Nurmi},
  \citenamefont {Schiappacasse},\ and\ \citenamefont
  {Yanagida}}]{Nurmi:2021xds}%
  \BibitemOpen
  \bibfield  {author} {\bibinfo {author} {\bibfnamefont {Sami}\ \bibnamefont
  {Nurmi}}, \bibinfo {author} {\bibfnamefont {Enrico~D.}\ \bibnamefont
  {Schiappacasse}}, \ and\ \bibinfo {author} {\bibfnamefont {Tsutomu~T.}\
  \bibnamefont {Yanagida}},\ }\bibfield  {title} {\enquote {\bibinfo {title}
  {{Radio signatures from encounters between neutron stars and QCD-axion
  minihalos around primordial black~holes}},}\ }\href {\doibase
  10.1088/1475-7516/2021/09/004} {\bibfield  {journal} {\bibinfo  {journal}
  {JCAP}\ }\textbf {\bibinfo {volume} {09}},\ \bibinfo {pages} {004} (\bibinfo
  {year} {2021})},\ \Eprint {http://arxiv.org/abs/2102.05680} {arXiv:2102.05680
  [hep-ph]} \BibitemShut {NoStop}%
\bibitem [{\citenamefont {Witte}\ \emph {et~al.}(2021)\citenamefont {Witte},
  \citenamefont {Noordhuis}, \citenamefont {Edwards},\ and\ \citenamefont
  {Weniger}}]{Witte:2021arp}%
  \BibitemOpen
  \bibfield  {author} {\bibinfo {author} {\bibfnamefont {Samuel~J.}\
  \bibnamefont {Witte}}, \bibinfo {author} {\bibfnamefont {Dion}\ \bibnamefont
  {Noordhuis}}, \bibinfo {author} {\bibfnamefont {Thomas D.~P.}\ \bibnamefont
  {Edwards}}, \ and\ \bibinfo {author} {\bibfnamefont {Christoph}\ \bibnamefont
  {Weniger}},\ }\bibfield  {title} {\enquote {\bibinfo {title} {{Axion-photon
  conversion in neutron star magnetospheres: The role of the plasma in the
  Goldreich-Julian model}},}\ }\href {\doibase 10.1103/PhysRevD.104.103030}
  {\bibfield  {journal} {\bibinfo  {journal} {Phys. Rev. D}\ }\textbf {\bibinfo
  {volume} {104}},\ \bibinfo {pages} {103030} (\bibinfo {year} {2021})},\
  \Eprint {http://arxiv.org/abs/2104.07670} {arXiv:2104.07670 [hep-ph]}
  \BibitemShut {NoStop}%
\bibitem [{\citenamefont {Millar}\ \emph {et~al.}(2021)\citenamefont {Millar},
  \citenamefont {Baum}, \citenamefont {Lawson},\ and\ \citenamefont
  {Marsh}}]{Millar:2021gzs}%
  \BibitemOpen
  \bibfield  {author} {\bibinfo {author} {\bibfnamefont {Alexander~J.}\
  \bibnamefont {Millar}}, \bibinfo {author} {\bibfnamefont {Sebastian}\
  \bibnamefont {Baum}}, \bibinfo {author} {\bibfnamefont {Matthew}\
  \bibnamefont {Lawson}}, \ and\ \bibinfo {author} {\bibfnamefont
  {M.~C.~David}\ \bibnamefont {Marsh}},\ }\bibfield  {title} {\enquote
  {\bibinfo {title} {{Axion-photon conversion in strongly magnetised
  plasmas}},}\ }\href {\doibase 10.1088/1475-7516/2021/11/013} {\bibfield
  {journal} {\bibinfo  {journal} {JCAP}\ }\textbf {\bibinfo {volume} {11}},\
  \bibinfo {pages} {013} (\bibinfo {year} {2021})},\ \Eprint
  {http://arxiv.org/abs/2107.07399} {arXiv:2107.07399 [hep-ph]} \BibitemShut
  {NoStop}%
\bibitem [{\citenamefont {Choi}\ and\ \citenamefont
  {Schiappacasse}(2022)}]{Choi:2022btl}%
  \BibitemOpen
  \bibfield  {author} {\bibinfo {author} {\bibfnamefont {Gongjun}\ \bibnamefont
  {Choi}}\ and\ \bibinfo {author} {\bibfnamefont {Enrico~D.}\ \bibnamefont
  {Schiappacasse}},\ }\bibfield  {title} {\enquote {\bibinfo {title} {{PBH
  assisted search for QCD axion dark matter}},}\ }\href {\doibase
  10.1088/1475-7516/2022/09/072} {\bibfield  {journal} {\bibinfo  {journal}
  {JCAP}\ }\textbf {\bibinfo {volume} {09}},\ \bibinfo {pages} {072} (\bibinfo
  {year} {2022})},\ \Eprint {http://arxiv.org/abs/2205.02255} {arXiv:2205.02255
  [hep-ph]} \BibitemShut {NoStop}%
\bibitem [{\citenamefont {Witte}\ \emph {et~al.}(2023)\citenamefont {Witte},
  \citenamefont {Baum}, \citenamefont {Lawson}, \citenamefont {Marsh},
  \citenamefont {Millar},\ and\ \citenamefont {Salinas}}]{Witte:2022cjj}%
  \BibitemOpen
  \bibfield  {author} {\bibinfo {author} {\bibfnamefont {Samuel~J.}\
  \bibnamefont {Witte}}, \bibinfo {author} {\bibfnamefont {Sebastian}\
  \bibnamefont {Baum}}, \bibinfo {author} {\bibfnamefont {Matthew}\
  \bibnamefont {Lawson}}, \bibinfo {author} {\bibfnamefont {M.~C.~David}\
  \bibnamefont {Marsh}}, \bibinfo {author} {\bibfnamefont {Alexander~J.}\
  \bibnamefont {Millar}}, \ and\ \bibinfo {author} {\bibfnamefont {Gustavo}\
  \bibnamefont {Salinas}},\ }\bibfield  {title} {\enquote {\bibinfo {title}
  {{Transient radio lines from axion miniclusters and axion stars}},}\ }\href
  {\doibase 10.1103/PhysRevD.107.063013} {\bibfield  {journal} {\bibinfo
  {journal} {Phys. Rev. D}\ }\textbf {\bibinfo {volume} {107}},\ \bibinfo
  {pages} {063013} (\bibinfo {year} {2023})},\ \Eprint
  {http://arxiv.org/abs/2212.08079} {arXiv:2212.08079 [hep-ph]} \BibitemShut
  {NoStop}%
\bibitem [{\citenamefont {McDonald}\ \emph {et~al.}(2023)\citenamefont
  {McDonald}, \citenamefont {Garbrecht},\ and\ \citenamefont
  {Millington}}]{McDonald:2023ohd}%
  \BibitemOpen
  \bibfield  {author} {\bibinfo {author} {\bibfnamefont {J.~I.}\ \bibnamefont
  {McDonald}}, \bibinfo {author} {\bibfnamefont {B.}~\bibnamefont {Garbrecht}},
  \ and\ \bibinfo {author} {\bibfnamefont {P.}~\bibnamefont {Millington}},\
  }\bibfield  {title} {\enquote {\bibinfo {title} {{Axion-photon conversion in
  3D media and astrophysical plasmas}},}\ }\href {\doibase
  10.1088/1475-7516/2023/12/031} {\bibfield  {journal} {\bibinfo  {journal}
  {JCAP}\ }\textbf {\bibinfo {volume} {12}},\ \bibinfo {pages} {031} (\bibinfo
  {year} {2023})},\ \Eprint {http://arxiv.org/abs/2307.11812} {arXiv:2307.11812
  [hep-ph]} \BibitemShut {NoStop}%
\bibitem [{\citenamefont {McDonald}\ and\ \citenamefont
  {Witte}(2023)}]{McDonald:2023shx}%
  \BibitemOpen
  \bibfield  {author} {\bibinfo {author} {\bibfnamefont {J.~I.}\ \bibnamefont
  {McDonald}}\ and\ \bibinfo {author} {\bibfnamefont {S.~J.}\ \bibnamefont
  {Witte}},\ }\bibfield  {title} {\enquote {\bibinfo {title} {{Generalized ray
  tracing for axions in astrophysical plasmas}},}\ }\href {\doibase
  10.1103/PhysRevD.108.103021} {\bibfield  {journal} {\bibinfo  {journal}
  {Phys. Rev. D}\ }\textbf {\bibinfo {volume} {108}},\ \bibinfo {pages}
  {103021} (\bibinfo {year} {2023})},\ \Eprint
  {http://arxiv.org/abs/2309.08655} {arXiv:2309.08655 [hep-ph]} \BibitemShut
  {NoStop}%
\bibitem [{\citenamefont {Tjemsland}\ \emph {et~al.}(2024)\citenamefont
  {Tjemsland}, \citenamefont {McDonald},\ and\ \citenamefont
  {Witte}}]{Tjemsland:2023vvc}%
  \BibitemOpen
  \bibfield  {author} {\bibinfo {author} {\bibfnamefont {Jonas}\ \bibnamefont
  {Tjemsland}}, \bibinfo {author} {\bibfnamefont {Jamie}\ \bibnamefont
  {McDonald}}, \ and\ \bibinfo {author} {\bibfnamefont {Samuel~J.}\
  \bibnamefont {Witte}},\ }\bibfield  {title} {\enquote {\bibinfo {title}
  {{Adiabatic axion-photon mixing near neutron stars}},}\ }\href {\doibase
  10.1103/PhysRevD.109.023015} {\bibfield  {journal} {\bibinfo  {journal}
  {Phys. Rev. D}\ }\textbf {\bibinfo {volume} {109}},\ \bibinfo {pages}
  {023015} (\bibinfo {year} {2024})},\ \Eprint
  {http://arxiv.org/abs/2310.18403} {arXiv:2310.18403 [hep-ph]} \BibitemShut
  {NoStop}%
\bibitem [{\citenamefont {Gin\'es}\ \emph {et~al.}(2024)\citenamefont
  {Gin\'es}, \citenamefont {Noordhuis}, \citenamefont {Weniger},\ and\
  \citenamefont {Witte}}]{Gines:2024ekm}%
  \BibitemOpen
  \bibfield  {author} {\bibinfo {author} {\bibfnamefont {Estanis~Utrilla}\
  \bibnamefont {Gin\'es}}, \bibinfo {author} {\bibfnamefont {Dion}\
  \bibnamefont {Noordhuis}}, \bibinfo {author} {\bibfnamefont {Christoph}\
  \bibnamefont {Weniger}}, \ and\ \bibinfo {author} {\bibfnamefont {Samuel~J.}\
  \bibnamefont {Witte}},\ }\bibfield  {title} {\enquote {\bibinfo {title}
  {{Numerical Analysis of Resonant Axion-Photon Mixing: Part I}},}\ }\href@noop
  {} {\  (\bibinfo {year} {2024})},\ \Eprint {http://arxiv.org/abs/2405.08865}
  {arXiv:2405.08865 [hep-ph]} \BibitemShut {NoStop}%
\bibitem [{\citenamefont {McDonald}\ and\ \citenamefont
  {Millington}(2024)}]{McDonald:2024uuh}%
  \BibitemOpen
  \bibfield  {author} {\bibinfo {author} {\bibfnamefont {J.~I.}\ \bibnamefont
  {McDonald}}\ and\ \bibinfo {author} {\bibfnamefont {P.}~\bibnamefont
  {Millington}},\ }\bibfield  {title} {\enquote {\bibinfo {title}
  {{Axion-Photon Mixing in 3D: Classical Equations and Geometric Optics}},}\
  }\href@noop {} {\  (\bibinfo {year} {2024})},\ \Eprint
  {http://arxiv.org/abs/2407.11192} {arXiv:2407.11192 [hep-ph]} \BibitemShut
  {NoStop}%
\bibitem [{\citenamefont {Goldreich}\ and\ \citenamefont
  {Julian}(1969)}]{Goldreich:1969sb}%
  \BibitemOpen
  \bibfield  {author} {\bibinfo {author} {\bibfnamefont {Peter}\ \bibnamefont
  {Goldreich}}\ and\ \bibinfo {author} {\bibfnamefont {William~H.}\
  \bibnamefont {Julian}},\ }\bibfield  {title} {\enquote {\bibinfo {title}
  {{Pulsar electrodynamics}},}\ }\href {\doibase 10.1086/150119} {\bibfield
  {journal} {\bibinfo  {journal} {Astrophys. J.}\ }\textbf {\bibinfo {volume}
  {157}},\ \bibinfo {pages} {869} (\bibinfo {year} {1969})}\BibitemShut
  {NoStop}%
\bibitem [{\citenamefont {Haensel}\ \emph {et~al.}(2007)\citenamefont
  {Haensel}, \citenamefont {Potekhin},\ and\ \citenamefont
  {Yakovlev}}]{Haensel:2007yy}%
  \BibitemOpen
  \bibfield  {author} {\bibinfo {author} {\bibfnamefont {P.}~\bibnamefont
  {Haensel}}, \bibinfo {author} {\bibfnamefont {A.~Y.}\ \bibnamefont
  {Potekhin}}, \ and\ \bibinfo {author} {\bibfnamefont {D.~G.}\ \bibnamefont
  {Yakovlev}},\ }\href {\doibase 10.1007/978-0-387-47301-7} {\emph {\bibinfo
  {title} {{Neutron stars 1: Equation of state and structure}}}},\ Vol.\
  \bibinfo {volume} {326}\ (\bibinfo  {publisher} {Springer},\ \bibinfo
  {address} {New York, USA},\ \bibinfo {year} {2007})\BibitemShut {NoStop}%
\bibitem [{\citenamefont {{Krause-Polstorff}}\ and\ \citenamefont
  {{Michel}}(1985)}]{1985MNRAS.213P..43K}%
  \BibitemOpen
  \bibfield  {author} {\bibinfo {author} {\bibfnamefont {J.}~\bibnamefont
  {{Krause-Polstorff}}}\ and\ \bibinfo {author} {\bibfnamefont {F.~C.}\
  \bibnamefont {{Michel}}},\ }\bibfield  {title} {\enquote {\bibinfo {title}
  {{Electrosphere of an aligned magnetized neutron star.}}}\ }\href {\doibase
  10.1093/mnras/213.1.43P} {\bibfield  {journal} {\bibinfo  {journal} {Monthly
  Notices of the Royal Astronomical Society}\ }\textbf {\bibinfo {volume}
  {213}},\ \bibinfo {pages} {43} (\bibinfo {year} {1985})}\BibitemShut
  {NoStop}%
\bibitem [{\citenamefont {Spitkovsky}\ and\ \citenamefont
  {Arons}(2002)}]{Spitkovsky:2002wg}%
  \BibitemOpen
  \bibfield  {author} {\bibinfo {author} {\bibfnamefont {Anatoly}\ \bibnamefont
  {Spitkovsky}}\ and\ \bibinfo {author} {\bibfnamefont {Jonathan}\ \bibnamefont
  {Arons}},\ }\bibfield  {title} {\enquote {\bibinfo {title} {{Simulations of
  pulsar wind formation}},}\ }\href@noop {} {\bibfield  {journal} {\bibinfo
  {journal} {ASP Conf. Ser.}\ }\textbf {\bibinfo {volume} {271}},\ \bibinfo
  {pages} {81} (\bibinfo {year} {2002})},\ \Eprint
  {http://arxiv.org/abs/astro-ph/0201360} {arXiv:astro-ph/0201360} \BibitemShut
  {NoStop}%
\bibitem [{\citenamefont {Cerutti}\ and\ \citenamefont
  {Beloborodov}(2017)}]{Cerutti:2016ttn}%
  \BibitemOpen
  \bibfield  {author} {\bibinfo {author} {\bibfnamefont {Beno\^\i{}t}\
  \bibnamefont {Cerutti}}\ and\ \bibinfo {author} {\bibfnamefont {Andrei}\
  \bibnamefont {Beloborodov}},\ }\bibfield  {title} {\enquote {\bibinfo {title}
  {{Electrodynamics of pulsar magnetospheres}},}\ }\href {\doibase
  10.1007/s11214-016-0315-7} {\bibfield  {journal} {\bibinfo  {journal} {Space
  Sci. Rev.}\ }\textbf {\bibinfo {volume} {207}},\ \bibinfo {pages} {111--136}
  (\bibinfo {year} {2017})},\ \Eprint {http://arxiv.org/abs/1611.04331}
  {arXiv:1611.04331 [astro-ph.HE]} \BibitemShut {NoStop}%
\bibitem [{\citenamefont {Faucher-Giguere}\ and\ \citenamefont
  {Kaspi}(2006)}]{Faucher-Giguere:2005dxp}%
  \BibitemOpen
  \bibfield  {author} {\bibinfo {author} {\bibfnamefont {Claude-Andre}\
  \bibnamefont {Faucher-Giguere}}\ and\ \bibinfo {author} {\bibfnamefont
  {Victoria~M.}\ \bibnamefont {Kaspi}},\ }\bibfield  {title} {\enquote
  {\bibinfo {title} {{Birth and evolution of isolated radio pulsars}},}\ }\href
  {\doibase 10.1086/501516} {\bibfield  {journal} {\bibinfo  {journal}
  {Astrophys. J.}\ }\textbf {\bibinfo {volume} {643}},\ \bibinfo {pages}
  {332--355} (\bibinfo {year} {2006})},\ \Eprint
  {http://arxiv.org/abs/astro-ph/0512585} {arXiv:astro-ph/0512585} \BibitemShut
  {NoStop}%
\bibitem [{\citenamefont {Harding}\ and\ \citenamefont
  {Lai}(2006)}]{Harding:2006qn}%
  \BibitemOpen
  \bibfield  {author} {\bibinfo {author} {\bibfnamefont {Alice~K.}\
  \bibnamefont {Harding}}\ and\ \bibinfo {author} {\bibfnamefont {Dong}\
  \bibnamefont {Lai}},\ }\bibfield  {title} {\enquote {\bibinfo {title}
  {{Physics of Strongly Magnetized Neutron Stars}},}\ }\href {\doibase
  10.1088/0034-4885/69/9/R03} {\bibfield  {journal} {\bibinfo  {journal} {Rept.
  Prog. Phys.}\ }\textbf {\bibinfo {volume} {69}},\ \bibinfo {pages} {2631}
  (\bibinfo {year} {2006})},\ \Eprint {http://arxiv.org/abs/astro-ph/0606674}
  {arXiv:astro-ph/0606674} \BibitemShut {NoStop}%
\bibitem [{\citenamefont {Olausen}\ and\ \citenamefont
  {Kaspi}(2014)}]{Olausen:2013bpa}%
  \BibitemOpen
  \bibfield  {author} {\bibinfo {author} {\bibfnamefont {S.~A.}\ \bibnamefont
  {Olausen}}\ and\ \bibinfo {author} {\bibfnamefont {V.~M.}\ \bibnamefont
  {Kaspi}},\ }\bibfield  {title} {\enquote {\bibinfo {title} {{The McGill
  Magnetar Catalog}},}\ }\href {\doibase 10.1088/0067-0049/212/1/6} {\bibfield
  {journal} {\bibinfo  {journal} {Astrophys. J. Suppl.}\ }\textbf {\bibinfo
  {volume} {212}},\ \bibinfo {pages} {6} (\bibinfo {year} {2014})},\ \Eprint
  {http://arxiv.org/abs/1309.4167} {arXiv:1309.4167 [astro-ph.HE]} \BibitemShut
  {NoStop}%
\bibitem [{\citenamefont {Kaspi}\ and\ \citenamefont
  {Beloborodov}(2017)}]{Kaspi:2017fwg}%
  \BibitemOpen
  \bibfield  {author} {\bibinfo {author} {\bibfnamefont {Victoria~M.}\
  \bibnamefont {Kaspi}}\ and\ \bibinfo {author} {\bibfnamefont {Andrei}\
  \bibnamefont {Beloborodov}},\ }\bibfield  {title} {\enquote {\bibinfo {title}
  {{Magnetars}},}\ }\href {\doibase 10.1146/annurev-astro-081915-023329}
  {\bibfield  {journal} {\bibinfo  {journal} {Ann. Rev. Astron. Astrophys.}\
  }\textbf {\bibinfo {volume} {55}},\ \bibinfo {pages} {261--301} (\bibinfo
  {year} {2017})},\ \Eprint {http://arxiv.org/abs/1703.00068} {arXiv:1703.00068
  [astro-ph.HE]} \BibitemShut {NoStop}%
\bibitem [{\citenamefont {Eilek}\ and\ \citenamefont
  {Hankins}(2016)}]{Eilek:2016hms}%
  \BibitemOpen
  \bibfield  {author} {\bibinfo {author} {\bibfnamefont {J.~A.}\ \bibnamefont
  {Eilek}}\ and\ \bibinfo {author} {\bibfnamefont {T.~H.}\ \bibnamefont
  {Hankins}},\ }\bibfield  {title} {\enquote {\bibinfo {title} {{Radio Emission
  Physics in the Crab Pulsar}},}\ }\href {\doibase 10.1017/S002237781600043X}
  {\bibfield  {journal} {\bibinfo  {journal} {J. Plasma Phys.}\ }\textbf
  {\bibinfo {volume} {82}},\ \bibinfo {pages} {036302} (\bibinfo {year}
  {2016})},\ \Eprint {http://arxiv.org/abs/1604.02472} {arXiv:1604.02472
  [astro-ph.HE]} \BibitemShut {NoStop}%
\bibitem [{\citenamefont {Istomin}\ and\ \citenamefont
  {Sobyanin}(2007)}]{Istomin:2007ge}%
  \BibitemOpen
  \bibfield  {author} {\bibinfo {author} {\bibfnamefont {Ya.~N.}\ \bibnamefont
  {Istomin}}\ and\ \bibinfo {author} {\bibfnamefont {D.~N.}\ \bibnamefont
  {Sobyanin}},\ }\bibfield  {title} {\enquote {\bibinfo {title}
  {{Electron-Positron Plasma Generation in a Magnetar Magnetosphere}},}\ }\href
  {\doibase 10.1134/S1063773707100040} {\bibfield  {journal} {\bibinfo
  {journal} {Astron. Lett.}\ }\textbf {\bibinfo {volume} {33}},\ \bibinfo
  {pages} {660--672} (\bibinfo {year} {2007})},\ \Eprint
  {http://arxiv.org/abs/0710.1000} {arXiv:0710.1000 [astro-ph]} \BibitemShut
  {NoStop}%
\bibitem [{\citenamefont {Kim}(1979)}]{Kim:1979if}%
  \BibitemOpen
  \bibfield  {author} {\bibinfo {author} {\bibfnamefont {Jihn~E.}\ \bibnamefont
  {Kim}},\ }\bibfield  {title} {\enquote {\bibinfo {title} {{Weak Interaction
  Singlet and Strong CP Invariance}},}\ }\href {\doibase
  10.1103/PhysRevLett.43.103} {\bibfield  {journal} {\bibinfo  {journal} {Phys.
  Rev. Lett.}\ }\textbf {\bibinfo {volume} {43}},\ \bibinfo {pages} {103}
  (\bibinfo {year} {1979})}\BibitemShut {NoStop}%
\bibitem [{\citenamefont {Shifman}\ \emph {et~al.}(1980)\citenamefont
  {Shifman}, \citenamefont {Vainshtein},\ and\ \citenamefont
  {Zakharov}}]{Shifman:1979if}%
  \BibitemOpen
  \bibfield  {author} {\bibinfo {author} {\bibfnamefont {Mikhail~A.}\
  \bibnamefont {Shifman}}, \bibinfo {author} {\bibfnamefont {A.~I.}\
  \bibnamefont {Vainshtein}}, \ and\ \bibinfo {author} {\bibfnamefont
  {Valentin~I.}\ \bibnamefont {Zakharov}},\ }\bibfield  {title} {\enquote
  {\bibinfo {title} {{Can Confinement Ensure Natural CP Invariance of Strong
  Interactions?}}}\ }\href {\doibase 10.1016/0550-3213(80)90209-6} {\bibfield
  {journal} {\bibinfo  {journal} {Nucl. Phys. B}\ }\textbf {\bibinfo {volume}
  {166}},\ \bibinfo {pages} {493--506} (\bibinfo {year} {1980})}\BibitemShut
  {NoStop}%
\bibitem [{\citenamefont {Dine}\ \emph {et~al.}(1981)\citenamefont {Dine},
  \citenamefont {Fischler},\ and\ \citenamefont {Srednicki}}]{Dine:1981rt}%
  \BibitemOpen
  \bibfield  {author} {\bibinfo {author} {\bibfnamefont {Michael}\ \bibnamefont
  {Dine}}, \bibinfo {author} {\bibfnamefont {Willy}\ \bibnamefont {Fischler}},
  \ and\ \bibinfo {author} {\bibfnamefont {Mark}\ \bibnamefont {Srednicki}},\
  }\bibfield  {title} {\enquote {\bibinfo {title} {{A Simple Solution to the
  Strong CP Problem with a Harmless Axion}},}\ }\href {\doibase
  10.1016/0370-2693(81)90590-6} {\bibfield  {journal} {\bibinfo  {journal}
  {Phys. Lett. B}\ }\textbf {\bibinfo {volume} {104}},\ \bibinfo {pages}
  {199--202} (\bibinfo {year} {1981})}\BibitemShut {NoStop}%
\bibitem [{\citenamefont {Zhitnitsky}(1980)}]{Zhitnitsky:1980tq}%
  \BibitemOpen
  \bibfield  {author} {\bibinfo {author} {\bibfnamefont {A.~R.}\ \bibnamefont
  {Zhitnitsky}},\ }\bibfield  {title} {\enquote {\bibinfo {title} {{On Possible
  Suppression of the Axion Hadron Interactions. (In Russian)}},}\ }\href@noop
  {} {\bibfield  {journal} {\bibinfo  {journal} {Sov. J. Nucl. Phys.}\ }\textbf
  {\bibinfo {volume} {31}},\ \bibinfo {pages} {260} (\bibinfo {year}
  {1980})}\BibitemShut {NoStop}%
\bibitem [{\citenamefont {Gendler}\ \emph {et~al.}(2023)\citenamefont
  {Gendler}, \citenamefont {Marsh}, \citenamefont {McAllister},\ and\
  \citenamefont {Moritz}}]{Gendler:2023kjt}%
  \BibitemOpen
  \bibfield  {author} {\bibinfo {author} {\bibfnamefont {Naomi}\ \bibnamefont
  {Gendler}}, \bibinfo {author} {\bibfnamefont {David J.~E.}\ \bibnamefont
  {Marsh}}, \bibinfo {author} {\bibfnamefont {Liam}\ \bibnamefont
  {McAllister}}, \ and\ \bibinfo {author} {\bibfnamefont {Jakob}\ \bibnamefont
  {Moritz}},\ }\bibfield  {title} {\enquote {\bibinfo {title} {{Glimmers from
  the Axiverse}},}\ }\href@noop {} {\  (\bibinfo {year} {2023})},\ \Eprint
  {http://arxiv.org/abs/2309.13145} {arXiv:2309.13145 [hep-th]} \BibitemShut
  {NoStop}%
\bibitem [{\citenamefont {Anastassopoulos}\ \emph {et~al.}(2017)\citenamefont
  {Anastassopoulos} \emph {et~al.}}]{CAST:2017uph}%
  \BibitemOpen
  \bibfield  {author} {\bibinfo {author} {\bibfnamefont {V.}~\bibnamefont
  {Anastassopoulos}} \emph {et~al.} (\bibinfo {collaboration} {CAST}),\
  }\bibfield  {title} {\enquote {\bibinfo {title} {{New CAST Limit on the
  Axion-Photon Interaction}},}\ }\href {\doibase 10.1038/nphys4109} {\bibfield
  {journal} {\bibinfo  {journal} {Nature Phys.}\ }\textbf {\bibinfo {volume}
  {13}},\ \bibinfo {pages} {584--590} (\bibinfo {year} {2017})},\ \Eprint
  {http://arxiv.org/abs/1705.02290} {arXiv:1705.02290 [hep-ex]} \BibitemShut
  {NoStop}%
\bibitem [{\citenamefont {Reynolds}\ \emph {et~al.}(2020)\citenamefont
  {Reynolds}, \citenamefont {Marsh}, \citenamefont {Russell}, \citenamefont
  {Fabian}, \citenamefont {Smith}, \citenamefont {Tombesi},\ and\ \citenamefont
  {Veilleux}}]{Reynolds:2019uqt}%
  \BibitemOpen
  \bibfield  {author} {\bibinfo {author} {\bibfnamefont {Christopher~S.}\
  \bibnamefont {Reynolds}}, \bibinfo {author} {\bibfnamefont {M.~C.~David}\
  \bibnamefont {Marsh}}, \bibinfo {author} {\bibfnamefont {Helen~R.}\
  \bibnamefont {Russell}}, \bibinfo {author} {\bibfnamefont {Andrew~C.}\
  \bibnamefont {Fabian}}, \bibinfo {author} {\bibfnamefont {Robyn}\
  \bibnamefont {Smith}}, \bibinfo {author} {\bibfnamefont {Francesco}\
  \bibnamefont {Tombesi}}, \ and\ \bibinfo {author} {\bibfnamefont {Sylvain}\
  \bibnamefont {Veilleux}},\ }\bibfield  {title} {\enquote {\bibinfo {title}
  {{Astrophysical limits on very light axion-like particles from Chandra
  grating spectroscopy of NGC 1275}},}\ }\href {\doibase
  10.3847/1538-4357/ab6a0c} {\bibfield  {journal} {\bibinfo  {journal}
  {Astrophys. J.}\ }\textbf {\bibinfo {volume} {890}},\ \bibinfo {pages} {59}
  (\bibinfo {year} {2020})},\ \Eprint {http://arxiv.org/abs/1907.05475}
  {arXiv:1907.05475 [hep-ph]} \BibitemShut {NoStop}%
\bibitem [{\citenamefont {Dessert}\ \emph {et~al.}(2020)\citenamefont
  {Dessert}, \citenamefont {Foster},\ and\ \citenamefont
  {Safdi}}]{Dessert:2020lil}%
  \BibitemOpen
  \bibfield  {author} {\bibinfo {author} {\bibfnamefont {Christopher}\
  \bibnamefont {Dessert}}, \bibinfo {author} {\bibfnamefont {Joshua~W.}\
  \bibnamefont {Foster}}, \ and\ \bibinfo {author} {\bibfnamefont
  {Benjamin~R.}\ \bibnamefont {Safdi}},\ }\bibfield  {title} {\enquote
  {\bibinfo {title} {{X-ray Searches for Axions from Super Star Clusters}},}\
  }\href {\doibase 10.1103/PhysRevLett.125.261102} {\bibfield  {journal}
  {\bibinfo  {journal} {Phys. Rev. Lett.}\ }\textbf {\bibinfo {volume} {125}},\
  \bibinfo {pages} {261102} (\bibinfo {year} {2020})},\ \Eprint
  {http://arxiv.org/abs/2008.03305} {arXiv:2008.03305 [hep-ph]} \BibitemShut
  {NoStop}%
\bibitem [{\citenamefont {Payez}\ \emph {et~al.}(2015)\citenamefont {Payez},
  \citenamefont {Evoli}, \citenamefont {Fischer}, \citenamefont {Giannotti},
  \citenamefont {Mirizzi},\ and\ \citenamefont {Ringwald}}]{Payez:2014xsa}%
  \BibitemOpen
  \bibfield  {author} {\bibinfo {author} {\bibfnamefont {Alexandre}\
  \bibnamefont {Payez}}, \bibinfo {author} {\bibfnamefont {Carmelo}\
  \bibnamefont {Evoli}}, \bibinfo {author} {\bibfnamefont {Tobias}\
  \bibnamefont {Fischer}}, \bibinfo {author} {\bibfnamefont {Maurizio}\
  \bibnamefont {Giannotti}}, \bibinfo {author} {\bibfnamefont {Alessandro}\
  \bibnamefont {Mirizzi}}, \ and\ \bibinfo {author} {\bibfnamefont {Andreas}\
  \bibnamefont {Ringwald}},\ }\bibfield  {title} {\enquote {\bibinfo {title}
  {{Revisiting the SN1987A gamma-ray limit on ultralight axion-like
  particles}},}\ }\href {\doibase 10.1088/1475-7516/2015/02/006} {\bibfield
  {journal} {\bibinfo  {journal} {JCAP}\ }\textbf {\bibinfo {volume} {02}},\
  \bibinfo {pages} {006} (\bibinfo {year} {2015})},\ \Eprint
  {http://arxiv.org/abs/1410.3747} {arXiv:1410.3747 [astro-ph.HE]} \BibitemShut
  {NoStop}%
\bibitem [{\citenamefont {Workman}\ \emph {et~al.}(2022)\citenamefont {Workman}
  \emph {et~al.}}]{ParticleDataGroup:2022pth}%
  \BibitemOpen
  \bibfield  {author} {\bibinfo {author} {\bibfnamefont {R.~L.}\ \bibnamefont
  {Workman}} \emph {et~al.} (\bibinfo {collaboration} {Particle Data Group}),\
  }\bibfield  {title} {\enquote {\bibinfo {title} {{Review of Particle
  Physics}},}\ }\href {\doibase 10.1093/ptep/ptac097} {\bibfield  {journal}
  {\bibinfo  {journal} {PTEP}\ }\textbf {\bibinfo {volume} {2022}},\ \bibinfo
  {pages} {083C01} (\bibinfo {year} {2022})}\BibitemShut {NoStop}%
\bibitem [{\citenamefont {Heisenberg}\ and\ \citenamefont
  {Euler}(1936)}]{Heisenberg:1936nmg}%
  \BibitemOpen
  \bibfield  {author} {\bibinfo {author} {\bibfnamefont {W.}~\bibnamefont
  {Heisenberg}}\ and\ \bibinfo {author} {\bibfnamefont {H.}~\bibnamefont
  {Euler}},\ }\bibfield  {title} {\enquote {\bibinfo {title} {{Consequences of
  Dirac's theory of positrons}},}\ }\href {\doibase 10.1007/BF01343663}
  {\bibfield  {journal} {\bibinfo  {journal} {Z. Phys.}\ }\textbf {\bibinfo
  {volume} {98}},\ \bibinfo {pages} {714--732} (\bibinfo {year} {1936})},\
  \Eprint {http://arxiv.org/abs/physics/0605038} {arXiv:physics/0605038}
  \BibitemShut {NoStop}%
\bibitem [{\citenamefont {Weisskopf}(1936)}]{Weisskopf:1936hya}%
  \BibitemOpen
  \bibfield  {author} {\bibinfo {author} {\bibfnamefont {V.}~\bibnamefont
  {Weisskopf}},\ }\bibfield  {title} {\enquote {\bibinfo {title} {{The
  electrodynamics of the vacuum based on the quantum theory of the
  electron}},}\ }\href@noop {} {\bibfield  {journal} {\bibinfo  {journal}
  {Kong. Dan. Vid. Sel. Mat. Fys. Med.}\ }\textbf {\bibinfo {volume} {14N6}},\
  \bibinfo {pages} {1--39} (\bibinfo {year} {1936})}\BibitemShut {NoStop}%
\bibitem [{\citenamefont {Adler}(1971)}]{Adler:1971wn}%
  \BibitemOpen
  \bibfield  {author} {\bibinfo {author} {\bibfnamefont {Stephen~L.}\
  \bibnamefont {Adler}},\ }\bibfield  {title} {\enquote {\bibinfo {title}
  {{Photon splitting and photon dispersion in a strong magnetic field}},}\
  }\href {\doibase 10.1016/0003-4916(71)90154-0} {\bibfield  {journal}
  {\bibinfo  {journal} {Annals Phys.}\ }\textbf {\bibinfo {volume} {67}},\
  \bibinfo {pages} {599--647} (\bibinfo {year} {1971})}\BibitemShut {NoStop}%
\bibitem [{\citenamefont {Dunne}(2004)}]{Dunne:2004nc}%
  \BibitemOpen
  \bibfield  {author} {\bibinfo {author} {\bibfnamefont {Gerald~V.}\
  \bibnamefont {Dunne}},\ }\enquote {\bibinfo {title} {{Heisenberg-Euler
  effective Lagrangians: Basics and extensions}},}\ in\ \href {\doibase
  10.1142/9789812775344_0014} {\emph {\bibinfo {booktitle} {{From fields to
  strings: Circumnavigating theoretical physics. Ian Kogan memorial collection
  (3 volume set)}}}},\ \bibinfo {editor} {edited by\ \bibinfo {editor}
  {\bibfnamefont {M.}~\bibnamefont {Shifman}}, \bibinfo {editor} {\bibfnamefont
  {A.}~\bibnamefont {Vainshtein}}, \ and\ \bibinfo {editor} {\bibfnamefont
  {J.}~\bibnamefont {Wheater}}}\ (\bibinfo {year} {2004})\ pp.\ \bibinfo
  {pages} {445--522},\ \Eprint {http://arxiv.org/abs/hep-th/0406216}
  {arXiv:hep-th/0406216} \BibitemShut {NoStop}%
\bibitem [{\citenamefont {Gurevich}\ \emph {et~al.}(1993)\citenamefont
  {Gurevich}, \citenamefont {Beskin},\ and\ \citenamefont
  {Istomin}}]{Beskin:1993xx}%
  \BibitemOpen
  \bibfield  {author} {\bibinfo {author} {\bibfnamefont {A.~V.}\ \bibnamefont
  {Gurevich}}, \bibinfo {author} {\bibfnamefont {V.~S.}\ \bibnamefont
  {Beskin}}, \ and\ \bibinfo {author} {\bibfnamefont {Ya.~N}\ \bibnamefont
  {Istomin}},\ }\href@noop {} {\emph {\bibinfo {title} {{Physics of the pulsar
  magnetosphere}}}}\ (\bibinfo  {publisher} {Cambridge University Press},\
  \bibinfo {year} {1993})\BibitemShut {NoStop}%
\bibitem [{\citenamefont {P\'etri}(2016)}]{Petri:2016tqe}%
  \BibitemOpen
  \bibfield  {author} {\bibinfo {author} {\bibfnamefont {J.}~\bibnamefont
  {P\'etri}},\ }\bibfield  {title} {\enquote {\bibinfo {title} {{Theory of
  pulsar magnetosphere and wind}},}\ }\href {\doibase
  10.1017/S0022377816000763} {\bibfield  {journal} {\bibinfo  {journal} {J.
  Plasma Phys.}\ }\textbf {\bibinfo {volume} {82}},\ \bibinfo {pages}
  {635820502} (\bibinfo {year} {2016})},\ \Eprint
  {http://arxiv.org/abs/1608.04895} {arXiv:1608.04895 [astro-ph.HE]}
  \BibitemShut {NoStop}%
\bibitem [{\citenamefont {Morris}(1986)}]{Morris:1984iz}%
  \BibitemOpen
  \bibfield  {author} {\bibinfo {author} {\bibfnamefont {Donald~E.}\
  \bibnamefont {Morris}},\ }\bibfield  {title} {\enquote {\bibinfo {title}
  {{Axion Mass Limits From Pulsar X-rays}},}\ }\href {\doibase
  10.1103/PhysRevD.34.843} {\bibfield  {journal} {\bibinfo  {journal} {Phys.
  Rev. D}\ }\textbf {\bibinfo {volume} {34}},\ \bibinfo {pages} {843} (\bibinfo
  {year} {1986})}\BibitemShut {NoStop}%
\bibitem [{\citenamefont {Lai}\ and\ \citenamefont {Ho}(2002)}]{Lai:2001di}%
  \BibitemOpen
  \bibfield  {author} {\bibinfo {author} {\bibfnamefont {Dong}\ \bibnamefont
  {Lai}}\ and\ \bibinfo {author} {\bibfnamefont {Wynn C.~G.}\ \bibnamefont
  {Ho}},\ }\bibfield  {title} {\enquote {\bibinfo {title} {{Resonant conversion
  of photon modes due to vacuum polarization in a magnetized plasma:
  implications for x-ray emission from magnetars}},}\ }\href {\doibase
  10.1086/338074} {\bibfield  {journal} {\bibinfo  {journal} {Astrophys. J.}\
  }\textbf {\bibinfo {volume} {566}},\ \bibinfo {pages} {373} (\bibinfo {year}
  {2002})},\ \Eprint {http://arxiv.org/abs/astro-ph/0108127}
  {arXiv:astro-ph/0108127} \BibitemShut {NoStop}%
\bibitem [{\citenamefont {Gill}\ and\ \citenamefont
  {Heyl}(2011)}]{Gill:2011yp}%
  \BibitemOpen
  \bibfield  {author} {\bibinfo {author} {\bibfnamefont {Ramandeep}\
  \bibnamefont {Gill}}\ and\ \bibinfo {author} {\bibfnamefont {Jeremy~S.}\
  \bibnamefont {Heyl}},\ }\bibfield  {title} {\enquote {\bibinfo {title}
  {{Constraining the photon-axion coupling constant with magnetic white
  dwarfs}},}\ }\href {\doibase 10.1103/PhysRevD.84.085001} {\bibfield
  {journal} {\bibinfo  {journal} {Phys. Rev. D}\ }\textbf {\bibinfo {volume}
  {84}},\ \bibinfo {pages} {085001} (\bibinfo {year} {2011})},\ \Eprint
  {http://arxiv.org/abs/1105.2083} {arXiv:1105.2083 [astro-ph.HE]} \BibitemShut
  {NoStop}%
\bibitem [{\citenamefont {Fortin}\ and\ \citenamefont
  {Sinha}(2019)}]{Fortin:2018aom}%
  \BibitemOpen
  \bibfield  {author} {\bibinfo {author} {\bibfnamefont {Jean-Fran\c{c}ois}\
  \bibnamefont {Fortin}}\ and\ \bibinfo {author} {\bibfnamefont {Kuver}\
  \bibnamefont {Sinha}},\ }\bibfield  {title} {\enquote {\bibinfo {title}
  {{X-Ray Polarization Signals from Magnetars with Axion-Like-Particles}},}\
  }\href {\doibase 10.1007/JHEP01(2019)163} {\bibfield  {journal} {\bibinfo
  {journal} {JHEP}\ }\textbf {\bibinfo {volume} {01}},\ \bibinfo {pages} {163}
  (\bibinfo {year} {2019})},\ \Eprint {http://arxiv.org/abs/1807.10773}
  {arXiv:1807.10773 [hep-ph]} \BibitemShut {NoStop}%
\bibitem [{\citenamefont {Dessert}\ \emph {et~al.}(2019)\citenamefont
  {Dessert}, \citenamefont {Long},\ and\ \citenamefont
  {Safdi}}]{Dessert:2019sgw}%
  \BibitemOpen
  \bibfield  {author} {\bibinfo {author} {\bibfnamefont {Christopher}\
  \bibnamefont {Dessert}}, \bibinfo {author} {\bibfnamefont {Andrew~J.}\
  \bibnamefont {Long}}, \ and\ \bibinfo {author} {\bibfnamefont {Benjamin~R.}\
  \bibnamefont {Safdi}},\ }\bibfield  {title} {\enquote {\bibinfo {title}
  {{X-ray Signatures of Axion Conversion in Magnetic White Dwarf Stars}},}\
  }\href {\doibase 10.1103/PhysRevLett.123.061104} {\bibfield  {journal}
  {\bibinfo  {journal} {Phys. Rev. Lett.}\ }\textbf {\bibinfo {volume} {123}},\
  \bibinfo {pages} {061104} (\bibinfo {year} {2019})},\ \Eprint
  {http://arxiv.org/abs/1903.05088} {arXiv:1903.05088 [hep-ph]} \BibitemShut
  {NoStop}%
\bibitem [{\citenamefont {Buschmann}\ \emph {et~al.}(2021)\citenamefont
  {Buschmann}, \citenamefont {Co}, \citenamefont {Dessert},\ and\ \citenamefont
  {Safdi}}]{Buschmann:2019pfp}%
  \BibitemOpen
  \bibfield  {author} {\bibinfo {author} {\bibfnamefont {Malte}\ \bibnamefont
  {Buschmann}}, \bibinfo {author} {\bibfnamefont {Raymond~T.}\ \bibnamefont
  {Co}}, \bibinfo {author} {\bibfnamefont {Christopher}\ \bibnamefont
  {Dessert}}, \ and\ \bibinfo {author} {\bibfnamefont {Benjamin~R.}\
  \bibnamefont {Safdi}},\ }\bibfield  {title} {\enquote {\bibinfo {title}
  {{Axion Emission Can Explain a New Hard X-Ray Excess from Nearby Isolated
  Neutron Stars}},}\ }\href {\doibase 10.1103/PhysRevLett.126.021102}
  {\bibfield  {journal} {\bibinfo  {journal} {Phys. Rev. Lett.}\ }\textbf
  {\bibinfo {volume} {126}},\ \bibinfo {pages} {021102} (\bibinfo {year}
  {2021})},\ \Eprint {http://arxiv.org/abs/1910.04164} {arXiv:1910.04164
  [hep-ph]} \BibitemShut {NoStop}%
\bibitem [{\citenamefont {Dessert}\ \emph
  {et~al.}(2022{\natexlab{a}})\citenamefont {Dessert}, \citenamefont {Long},\
  and\ \citenamefont {Safdi}}]{Dessert:2021bkv}%
  \BibitemOpen
  \bibfield  {author} {\bibinfo {author} {\bibfnamefont {Christopher}\
  \bibnamefont {Dessert}}, \bibinfo {author} {\bibfnamefont {Andrew~J.}\
  \bibnamefont {Long}}, \ and\ \bibinfo {author} {\bibfnamefont {Benjamin~R.}\
  \bibnamefont {Safdi}},\ }\bibfield  {title} {\enquote {\bibinfo {title} {{No
  Evidence for Axions from Chandra Observation of the Magnetic White Dwarf RE
  J0317-853}},}\ }\href {\doibase 10.1103/PhysRevLett.128.071102} {\bibfield
  {journal} {\bibinfo  {journal} {Phys. Rev. Lett.}\ }\textbf {\bibinfo
  {volume} {128}},\ \bibinfo {pages} {071102} (\bibinfo {year}
  {2022}{\natexlab{a}})},\ \Eprint {http://arxiv.org/abs/2104.12772}
  {arXiv:2104.12772 [hep-ph]} \BibitemShut {NoStop}%
\bibitem [{\citenamefont {Fortin}\ \emph
  {et~al.}(2021{\natexlab{b}})\citenamefont {Fortin}, \citenamefont {Guo},
  \citenamefont {Harris}, \citenamefont {Sheridan},\ and\ \citenamefont
  {Sinha}}]{Fortin:2021sst}%
  \BibitemOpen
  \bibfield  {author} {\bibinfo {author} {\bibfnamefont {Jean-Fran\c{c}ois}\
  \bibnamefont {Fortin}}, \bibinfo {author} {\bibfnamefont {Huai-Ke}\
  \bibnamefont {Guo}}, \bibinfo {author} {\bibfnamefont {Steven~P.}\
  \bibnamefont {Harris}}, \bibinfo {author} {\bibfnamefont {Elijah}\
  \bibnamefont {Sheridan}}, \ and\ \bibinfo {author} {\bibfnamefont {Kuver}\
  \bibnamefont {Sinha}},\ }\bibfield  {title} {\enquote {\bibinfo {title}
  {{Magnetars and axion-like particles: probes with the hard X-ray
  spectrum}},}\ }\href {\doibase 10.1088/1475-7516/2021/06/036} {\bibfield
  {journal} {\bibinfo  {journal} {JCAP}\ }\textbf {\bibinfo {volume} {06}},\
  \bibinfo {pages} {036} (\bibinfo {year} {2021}{\natexlab{b}})},\ \Eprint
  {http://arxiv.org/abs/2101.05302} {arXiv:2101.05302 [hep-ph]} \BibitemShut
  {NoStop}%
\bibitem [{\citenamefont {Dessert}\ \emph
  {et~al.}(2022{\natexlab{b}})\citenamefont {Dessert}, \citenamefont {Dunsky},\
  and\ \citenamefont {Safdi}}]{Dessert:2022yqq}%
  \BibitemOpen
  \bibfield  {author} {\bibinfo {author} {\bibfnamefont {Christopher}\
  \bibnamefont {Dessert}}, \bibinfo {author} {\bibfnamefont {David}\
  \bibnamefont {Dunsky}}, \ and\ \bibinfo {author} {\bibfnamefont
  {Benjamin~R.}\ \bibnamefont {Safdi}},\ }\bibfield  {title} {\enquote
  {\bibinfo {title} {{Upper limit on the axion-photon coupling from magnetic
  white dwarf polarization}},}\ }\href {\doibase 10.1103/PhysRevD.105.103034}
  {\bibfield  {journal} {\bibinfo  {journal} {Phys. Rev. D}\ }\textbf {\bibinfo
  {volume} {105}},\ \bibinfo {pages} {103034} (\bibinfo {year}
  {2022}{\natexlab{b}})},\ \Eprint {http://arxiv.org/abs/2203.04319}
  {arXiv:2203.04319 [hep-ph]} \BibitemShut {NoStop}%
\bibitem [{\citenamefont {Dobrynina}\ \emph {et~al.}(2015)\citenamefont
  {Dobrynina}, \citenamefont {Kartavtsev},\ and\ \citenamefont
  {Raffelt}}]{Dobrynina:2014qba}%
  \BibitemOpen
  \bibfield  {author} {\bibinfo {author} {\bibfnamefont {Alexandra}\
  \bibnamefont {Dobrynina}}, \bibinfo {author} {\bibfnamefont {Alexander}\
  \bibnamefont {Kartavtsev}}, \ and\ \bibinfo {author} {\bibfnamefont {Georg}\
  \bibnamefont {Raffelt}},\ }\bibfield  {title} {\enquote {\bibinfo {title}
  {{Photon-photon dispersion of TeV gamma rays and its role for photon-ALP
  conversion}},}\ }\href {\doibase 10.1103/PhysRevD.91.083003} {\bibfield
  {journal} {\bibinfo  {journal} {Phys. Rev. D}\ }\textbf {\bibinfo {volume}
  {91}},\ \bibinfo {pages} {083003} (\bibinfo {year} {2015})},\ \bibinfo {note}
  {[Erratum: Phys.Rev.D 95, 109905 (2017)]},\ \Eprint
  {http://arxiv.org/abs/1412.4777} {arXiv:1412.4777 [astro-ph.HE]} \BibitemShut
  {NoStop}%
\bibitem [{\citenamefont {{Walter}}\ \emph {et~al.}(2010)\citenamefont
  {{Walter}}, \citenamefont {{Eisenbei{\ss}}}, \citenamefont {{Lattimer}},
  \citenamefont {{Kim}}, \citenamefont {{Hambaryan}},\ and\ \citenamefont
  {{Neuh{\"a}user}}}]{Walter:2010xxx}%
  \BibitemOpen
  \bibfield  {author} {\bibinfo {author} {\bibfnamefont {F.~M.}\ \bibnamefont
  {{Walter}}}, \bibinfo {author} {\bibfnamefont {T.}~\bibnamefont
  {{Eisenbei{\ss}}}}, \bibinfo {author} {\bibfnamefont {J.~M.}\ \bibnamefont
  {{Lattimer}}}, \bibinfo {author} {\bibfnamefont {B.}~\bibnamefont {{Kim}}},
  \bibinfo {author} {\bibfnamefont {V.}~\bibnamefont {{Hambaryan}}}, \ and\
  \bibinfo {author} {\bibfnamefont {R.}~\bibnamefont {{Neuh{\"a}user}}},\
  }\bibfield  {title} {\enquote {\bibinfo {title} {{Revisiting the Parallax of
  the Isolated Neutron Star RX J185635-3754 Using HST/ACS Imaging}},}\ }\href
  {\doibase 10.1088/0004-637X/724/1/669} {\bibfield  {journal} {\bibinfo
  {journal} {\apj}\ }\textbf {\bibinfo {volume} {724}},\ \bibinfo {pages}
  {669--677} (\bibinfo {year} {2010})},\ \Eprint
  {http://arxiv.org/abs/1008.1709} {arXiv:1008.1709 [astro-ph.SR]} \BibitemShut
  {NoStop}%
\bibitem [{\citenamefont {Mignani}\ \emph {et~al.}(2017)\citenamefont
  {Mignani}, \citenamefont {Testa}, \citenamefont {Caniulef}, \citenamefont
  {Taverna}, \citenamefont {Turolla}, \citenamefont {Zane},\ and\ \citenamefont
  {Wu}}]{Mignani:2016fwz}%
  \BibitemOpen
  \bibfield  {author} {\bibinfo {author} {\bibfnamefont {R.~P.}\ \bibnamefont
  {Mignani}}, \bibinfo {author} {\bibfnamefont {V.}~\bibnamefont {Testa}},
  \bibinfo {author} {\bibfnamefont {D.~Gonzalez}\ \bibnamefont {Caniulef}},
  \bibinfo {author} {\bibfnamefont {R.}~\bibnamefont {Taverna}}, \bibinfo
  {author} {\bibfnamefont {R.}~\bibnamefont {Turolla}}, \bibinfo {author}
  {\bibfnamefont {S.}~\bibnamefont {Zane}}, \ and\ \bibinfo {author}
  {\bibfnamefont {K.}~\bibnamefont {Wu}},\ }\bibfield  {title} {\enquote
  {\bibinfo {title} {{Evidence for vacuum birefringence from the first
  optical-polarimetry measurement of the isolated neutron star RX
  J1856.5\ensuremath{-}3754}},}\ }\href {\doibase 10.1093/mnras/stw2798}
  {\bibfield  {journal} {\bibinfo  {journal} {Mon. Not. Roy. Astron. Soc.}\
  }\textbf {\bibinfo {volume} {465}},\ \bibinfo {pages} {492--500} (\bibinfo
  {year} {2017})},\ \Eprint {http://arxiv.org/abs/1610.08323} {arXiv:1610.08323
  [astro-ph.HE]} \BibitemShut {NoStop}%
\bibitem [{\citenamefont {Hauser}\ and\ \citenamefont
  {Dwek}(2001)}]{Hauser:2001xs}%
  \BibitemOpen
  \bibfield  {author} {\bibinfo {author} {\bibfnamefont {Michael~G.}\
  \bibnamefont {Hauser}}\ and\ \bibinfo {author} {\bibfnamefont {Eli}\
  \bibnamefont {Dwek}},\ }\bibfield  {title} {\enquote {\bibinfo {title} {{The
  cosmic infrared background: measurements and implications}},}\ }\href
  {\doibase 10.1146/annurev.astro.39.1.249} {\bibfield  {journal} {\bibinfo
  {journal} {Ann. Rev. Astron. Astrophys.}\ }\textbf {\bibinfo {volume} {39}},\
  \bibinfo {pages} {249--307} (\bibinfo {year} {2001})},\ \Eprint
  {http://arxiv.org/abs/astro-ph/0105539} {arXiv:astro-ph/0105539} \BibitemShut
  {NoStop}%
\bibitem [{\citenamefont {{Hirashita}}(2016)}]{2016PASJ...68R...1H}%
  \BibitemOpen
  \bibfield  {author} {\bibinfo {author} {\bibfnamefont {Hiroyuki et~al.}\
  \bibnamefont {{Hirashita}}},\ }\bibfield  {title} {\enquote {\bibinfo {title}
  {{First-generation science cases for ground-based terahertz telescopes}},}\
  }\href {\doibase 10.1093/pasj/psv115} {\bibfield  {journal} {\bibinfo
  {journal} {Publications of the Astronomical Society of Japan}\ }\textbf
  {\bibinfo {volume} {68}},\ \bibinfo {eid} {R1} (\bibinfo {year} {2016})},\
  \Eprint {http://arxiv.org/abs/1511.00839} {arXiv:1511.00839 [astro-ph.GA]}
  \BibitemShut {NoStop}%
\bibitem [{\citenamefont {Berezhiani}\ \emph {et~al.}(1992)\citenamefont
  {Berezhiani}, \citenamefont {Sakharov},\ and\ \citenamefont
  {Khlopov}}]{Berezhiani:1992rk}%
  \BibitemOpen
  \bibfield  {author} {\bibinfo {author} {\bibfnamefont {Z.~G.}\ \bibnamefont
  {Berezhiani}}, \bibinfo {author} {\bibfnamefont {A.~S.}\ \bibnamefont
  {Sakharov}}, \ and\ \bibinfo {author} {\bibfnamefont {M.~Yu.}\ \bibnamefont
  {Khlopov}},\ }\bibfield  {title} {\enquote {\bibinfo {title} {{Primordial
  background of cosmological axions}},}\ }\href@noop {} {\bibfield  {journal}
  {\bibinfo  {journal} {Sov. J. Nucl. Phys.}\ }\textbf {\bibinfo {volume}
  {55}},\ \bibinfo {pages} {1063--1071} (\bibinfo {year} {1992})}\BibitemShut
  {NoStop}%
\bibitem [{\citenamefont {Salvio}\ \emph {et~al.}(2014)\citenamefont {Salvio},
  \citenamefont {Strumia},\ and\ \citenamefont {Xue}}]{Salvio:2013iaa}%
  \BibitemOpen
  \bibfield  {author} {\bibinfo {author} {\bibfnamefont {Alberto}\ \bibnamefont
  {Salvio}}, \bibinfo {author} {\bibfnamefont {Alessandro}\ \bibnamefont
  {Strumia}}, \ and\ \bibinfo {author} {\bibfnamefont {Wei}\ \bibnamefont
  {Xue}},\ }\bibfield  {title} {\enquote {\bibinfo {title} {{Thermal axion
  production}},}\ }\href {\doibase 10.1088/1475-7516/2014/01/011} {\bibfield
  {journal} {\bibinfo  {journal} {JCAP}\ }\textbf {\bibinfo {volume} {01}},\
  \bibinfo {pages} {011} (\bibinfo {year} {2014})},\ \Eprint
  {http://arxiv.org/abs/1310.6982} {arXiv:1310.6982 [hep-ph]} \BibitemShut
  {NoStop}%
\bibitem [{\citenamefont {Chacko}\ \emph {et~al.}(2015)\citenamefont {Chacko},
  \citenamefont {Cui}, \citenamefont {Hong},\ and\ \citenamefont
  {Okui}}]{Chacko:2015noa}%
  \BibitemOpen
  \bibfield  {author} {\bibinfo {author} {\bibfnamefont {Zackaria}\
  \bibnamefont {Chacko}}, \bibinfo {author} {\bibfnamefont {Yanou}\
  \bibnamefont {Cui}}, \bibinfo {author} {\bibfnamefont {Sungwoo}\ \bibnamefont
  {Hong}}, \ and\ \bibinfo {author} {\bibfnamefont {Takemichi}\ \bibnamefont
  {Okui}},\ }\bibfield  {title} {\enquote {\bibinfo {title} {{Hidden dark
  matter sector, dark radiation, and the CMB}},}\ }\href {\doibase
  10.1103/PhysRevD.92.055033} {\bibfield  {journal} {\bibinfo  {journal} {Phys.
  Rev. D}\ }\textbf {\bibinfo {volume} {92}},\ \bibinfo {pages} {055033}
  (\bibinfo {year} {2015})},\ \Eprint {http://arxiv.org/abs/1505.04192}
  {arXiv:1505.04192 [hep-ph]} \BibitemShut {NoStop}%
\bibitem [{\citenamefont {Baumann}\ \emph {et~al.}(2016)\citenamefont
  {Baumann}, \citenamefont {Green},\ and\ \citenamefont
  {Wallisch}}]{Baumann:2016wac}%
  \BibitemOpen
  \bibfield  {author} {\bibinfo {author} {\bibfnamefont {Daniel}\ \bibnamefont
  {Baumann}}, \bibinfo {author} {\bibfnamefont {Daniel}\ \bibnamefont {Green}},
  \ and\ \bibinfo {author} {\bibfnamefont {Benjamin}\ \bibnamefont
  {Wallisch}},\ }\bibfield  {title} {\enquote {\bibinfo {title} {{New Target
  for Cosmic Axion Searches}},}\ }\href {\doibase
  10.1103/PhysRevLett.117.171301} {\bibfield  {journal} {\bibinfo  {journal}
  {Phys. Rev. Lett.}\ }\textbf {\bibinfo {volume} {117}},\ \bibinfo {pages}
  {171301} (\bibinfo {year} {2016})},\ \Eprint
  {http://arxiv.org/abs/1604.08614} {arXiv:1604.08614 [astro-ph.CO]}
  \BibitemShut {NoStop}%
\bibitem [{\citenamefont {Arias-Arag\'on}\ \emph {et~al.}(2021)\citenamefont
  {Arias-Arag\'on}, \citenamefont {D'Eramo}, \citenamefont {Ferreira},
  \citenamefont {Merlo},\ and\ \citenamefont {Notari}}]{Arias-Aragon:2020shv}%
  \BibitemOpen
  \bibfield  {author} {\bibinfo {author} {\bibfnamefont {Fernando}\
  \bibnamefont {Arias-Arag\'on}}, \bibinfo {author} {\bibfnamefont {Francesco}\
  \bibnamefont {D'Eramo}}, \bibinfo {author} {\bibfnamefont {Ricardo~Z.}\
  \bibnamefont {Ferreira}}, \bibinfo {author} {\bibfnamefont {Luca}\
  \bibnamefont {Merlo}}, \ and\ \bibinfo {author} {\bibfnamefont {Alessio}\
  \bibnamefont {Notari}},\ }\bibfield  {title} {\enquote {\bibinfo {title}
  {{Production of Thermal Axions across the ElectroWeak Phase Transition}},}\
  }\href {\doibase 10.1088/1475-7516/2021/03/090} {\bibfield  {journal}
  {\bibinfo  {journal} {JCAP}\ }\textbf {\bibinfo {volume} {03}},\ \bibinfo
  {pages} {090} (\bibinfo {year} {2021})},\ \Eprint
  {http://arxiv.org/abs/2012.04736} {arXiv:2012.04736 [hep-ph]} \BibitemShut
  {NoStop}%
\bibitem [{\citenamefont {Arias}\ \emph {et~al.}(2023)\citenamefont {Arias},
  \citenamefont {Bernal}, \citenamefont {Osi\'nski}, \citenamefont
  {Roszkowski},\ and\ \citenamefont {Venegas}}]{Arias:2023wyg}%
  \BibitemOpen
  \bibfield  {author} {\bibinfo {author} {\bibfnamefont {Paola}\ \bibnamefont
  {Arias}}, \bibinfo {author} {\bibfnamefont {Nicol\'as}\ \bibnamefont
  {Bernal}}, \bibinfo {author} {\bibfnamefont {Jacek~K.}\ \bibnamefont
  {Osi\'nski}}, \bibinfo {author} {\bibfnamefont {Leszek}\ \bibnamefont
  {Roszkowski}}, \ and\ \bibinfo {author} {\bibfnamefont {Moira}\ \bibnamefont
  {Venegas}},\ }\bibfield  {title} {\enquote {\bibinfo {title} {{Revisiting
  signatures of thermal axions in nonstandard cosmologies}},}\ }\href@noop {}
  {\  (\bibinfo {year} {2023})},\ \Eprint {http://arxiv.org/abs/2308.01352}
  {arXiv:2308.01352 [hep-ph]} \BibitemShut {NoStop}%
\bibitem [{\citenamefont {Conlon}\ and\ \citenamefont
  {Marsh}(2013{\natexlab{a}})}]{Conlon:2013isa}%
  \BibitemOpen
  \bibfield  {author} {\bibinfo {author} {\bibfnamefont {Joseph~P.}\
  \bibnamefont {Conlon}}\ and\ \bibinfo {author} {\bibfnamefont {M.~C.~David}\
  \bibnamefont {Marsh}},\ }\bibfield  {title} {\enquote {\bibinfo {title} {{The
  Cosmophenomenology of Axionic Dark Radiation}},}\ }\href {\doibase
  10.1007/JHEP10(2013)214} {\bibfield  {journal} {\bibinfo  {journal} {JHEP}\
  }\textbf {\bibinfo {volume} {10}},\ \bibinfo {pages} {214} (\bibinfo {year}
  {2013}{\natexlab{a}})},\ \Eprint {http://arxiv.org/abs/1304.1804}
  {arXiv:1304.1804 [hep-ph]} \BibitemShut {NoStop}%
\bibitem [{\citenamefont {Dror}\ \emph {et~al.}(2021)\citenamefont {Dror},
  \citenamefont {Murayama},\ and\ \citenamefont {Rodd}}]{Dror:2021nyr}%
  \BibitemOpen
  \bibfield  {author} {\bibinfo {author} {\bibfnamefont {Jeff~A.}\ \bibnamefont
  {Dror}}, \bibinfo {author} {\bibfnamefont {Hitoshi}\ \bibnamefont
  {Murayama}}, \ and\ \bibinfo {author} {\bibfnamefont {Nicholas~L.}\
  \bibnamefont {Rodd}},\ }\bibfield  {title} {\enquote {\bibinfo {title}
  {{Cosmic axion background}},}\ }\href {\doibase 10.1103/PhysRevD.103.115004}
  {\bibfield  {journal} {\bibinfo  {journal} {Phys. Rev. D}\ }\textbf {\bibinfo
  {volume} {103}},\ \bibinfo {pages} {115004} (\bibinfo {year} {2021})},\
  \bibinfo {note} {[Erratum: Phys.Rev.D 106, 119902 (2022)]},\ \Eprint
  {http://arxiv.org/abs/2101.09287} {arXiv:2101.09287 [hep-ph]} \BibitemShut
  {NoStop}%
\bibitem [{\citenamefont {Buschmann}\ \emph {et~al.}(2022)\citenamefont
  {Buschmann}, \citenamefont {Dessert}, \citenamefont {Foster}, \citenamefont
  {Long},\ and\ \citenamefont {Safdi}}]{Buschmann:2021juv}%
  \BibitemOpen
  \bibfield  {author} {\bibinfo {author} {\bibfnamefont {Malte}\ \bibnamefont
  {Buschmann}}, \bibinfo {author} {\bibfnamefont {Christopher}\ \bibnamefont
  {Dessert}}, \bibinfo {author} {\bibfnamefont {Joshua~W.}\ \bibnamefont
  {Foster}}, \bibinfo {author} {\bibfnamefont {Andrew~J.}\ \bibnamefont
  {Long}}, \ and\ \bibinfo {author} {\bibfnamefont {Benjamin~R.}\ \bibnamefont
  {Safdi}},\ }\bibfield  {title} {\enquote {\bibinfo {title} {{Upper Limit on
  the QCD Axion Mass from Isolated Neutron Star Cooling}},}\ }\href {\doibase
  10.1103/PhysRevLett.128.091102} {\bibfield  {journal} {\bibinfo  {journal}
  {Phys. Rev. Lett.}\ }\textbf {\bibinfo {volume} {128}},\ \bibinfo {pages}
  {091102} (\bibinfo {year} {2022})},\ \Eprint
  {http://arxiv.org/abs/2111.09892} {arXiv:2111.09892 [hep-ph]} \BibitemShut
  {NoStop}%
\bibitem [{\citenamefont {Prabhu}(2021)}]{Prabhu:2021zve}%
  \BibitemOpen
  \bibfield  {author} {\bibinfo {author} {\bibfnamefont {Anirudh}\ \bibnamefont
  {Prabhu}},\ }\bibfield  {title} {\enquote {\bibinfo {title} {{Axion
  production in pulsar magnetosphere gaps}},}\ }\href {\doibase
  10.1103/PhysRevD.104.055038} {\bibfield  {journal} {\bibinfo  {journal}
  {Phys. Rev. D}\ }\textbf {\bibinfo {volume} {104}},\ \bibinfo {pages}
  {055038} (\bibinfo {year} {2021})},\ \Eprint
  {http://arxiv.org/abs/2104.14569} {arXiv:2104.14569 [hep-ph]} \BibitemShut
  {NoStop}%
\bibitem [{\citenamefont {Noordhuis}\ \emph
  {et~al.}(2023{\natexlab{a}})\citenamefont {Noordhuis}, \citenamefont
  {Prabhu}, \citenamefont {Witte}, \citenamefont {Chen}, \citenamefont {Cruz},\
  and\ \citenamefont {Weniger}}]{Noordhuis:2022ljw}%
  \BibitemOpen
  \bibfield  {author} {\bibinfo {author} {\bibfnamefont {Dion}\ \bibnamefont
  {Noordhuis}}, \bibinfo {author} {\bibfnamefont {Anirudh}\ \bibnamefont
  {Prabhu}}, \bibinfo {author} {\bibfnamefont {Samuel~J.}\ \bibnamefont
  {Witte}}, \bibinfo {author} {\bibfnamefont {Alexander~Y.}\ \bibnamefont
  {Chen}}, \bibinfo {author} {\bibfnamefont {F\'abio}\ \bibnamefont {Cruz}}, \
  and\ \bibinfo {author} {\bibfnamefont {Christoph}\ \bibnamefont {Weniger}},\
  }\bibfield  {title} {\enquote {\bibinfo {title} {{Novel Constraints on Axions
  Produced in Pulsar Polar-Cap Cascades}},}\ }\href {\doibase
  10.1103/PhysRevLett.131.111004} {\bibfield  {journal} {\bibinfo  {journal}
  {Phys. Rev. Lett.}\ }\textbf {\bibinfo {volume} {131}},\ \bibinfo {pages}
  {111004} (\bibinfo {year} {2023}{\natexlab{a}})},\ \Eprint
  {http://arxiv.org/abs/2209.09917} {arXiv:2209.09917 [hep-ph]} \BibitemShut
  {NoStop}%
\bibitem [{\citenamefont {Prabhu}(2023)}]{Prabhu:2023cgb}%
  \BibitemOpen
  \bibfield  {author} {\bibinfo {author} {\bibfnamefont {Anirudh}\ \bibnamefont
  {Prabhu}},\ }\bibfield  {title} {\enquote {\bibinfo {title} {{Axion-mediated
  Transport of Fast Radio Bursts Originating in Inner Magnetospheres of
  Magnetars}},}\ }\href {\doibase 10.3847/2041-8213/acc7a7} {\bibfield
  {journal} {\bibinfo  {journal} {Astrophys. J. Lett.}\ }\textbf {\bibinfo
  {volume} {946}},\ \bibinfo {pages} {L52} (\bibinfo {year} {2023})},\ \Eprint
  {http://arxiv.org/abs/2302.11645} {arXiv:2302.11645 [astro-ph.HE]}
  \BibitemShut {NoStop}%
\bibitem [{\citenamefont {Bai}\ and\ \citenamefont
  {de~Lima}(2024)}]{Bai:2023bbg}%
  \BibitemOpen
  \bibfield  {author} {\bibinfo {author} {\bibfnamefont {Yang}\ \bibnamefont
  {Bai}}\ and\ \bibinfo {author} {\bibfnamefont {Carlos~Henrique}\ \bibnamefont
  {de~Lima}},\ }\bibfield  {title} {\enquote {\bibinfo {title}
  {{Electrobaryonic axion: hair of neutron stars}},}\ }\href {\doibase
  10.1007/JHEP05(2024)312} {\bibfield  {journal} {\bibinfo  {journal} {JHEP}\
  }\textbf {\bibinfo {volume} {05}},\ \bibinfo {pages} {312} (\bibinfo {year}
  {2024})},\ \Eprint {http://arxiv.org/abs/2311.18794} {arXiv:2311.18794
  [hep-ph]} \BibitemShut {NoStop}%
\bibitem [{\citenamefont {Van~Tilburg}(2021)}]{VanTilburg:2020jvl}%
  \BibitemOpen
  \bibfield  {author} {\bibinfo {author} {\bibfnamefont {Ken}\ \bibnamefont
  {Van~Tilburg}},\ }\bibfield  {title} {\enquote {\bibinfo {title} {{Stellar
  basins of gravitationally bound particles}},}\ }\href {\doibase
  10.1103/PhysRevD.104.023019} {\bibfield  {journal} {\bibinfo  {journal}
  {Phys. Rev. D}\ }\textbf {\bibinfo {volume} {104}},\ \bibinfo {pages}
  {023019} (\bibinfo {year} {2021})},\ \Eprint
  {http://arxiv.org/abs/2006.12431} {arXiv:2006.12431 [hep-ph]} \BibitemShut
  {NoStop}%
\bibitem [{\citenamefont {Cicoli}\ \emph {et~al.}(2013)\citenamefont {Cicoli},
  \citenamefont {Conlon},\ and\ \citenamefont {Quevedo}}]{Cicoli:2012aq}%
  \BibitemOpen
  \bibfield  {author} {\bibinfo {author} {\bibfnamefont {Michele}\ \bibnamefont
  {Cicoli}}, \bibinfo {author} {\bibfnamefont {Joseph~P.}\ \bibnamefont
  {Conlon}}, \ and\ \bibinfo {author} {\bibfnamefont {Fernando}\ \bibnamefont
  {Quevedo}},\ }\bibfield  {title} {\enquote {\bibinfo {title} {{Dark radiation
  in LARGE volume models}},}\ }\href {\doibase 10.1103/PhysRevD.87.043520}
  {\bibfield  {journal} {\bibinfo  {journal} {Phys. Rev. D}\ }\textbf {\bibinfo
  {volume} {87}},\ \bibinfo {pages} {043520} (\bibinfo {year} {2013})},\
  \Eprint {http://arxiv.org/abs/1208.3562} {arXiv:1208.3562 [hep-ph]}
  \BibitemShut {NoStop}%
\bibitem [{\citenamefont {Higaki}\ and\ \citenamefont
  {Takahashi}(2012)}]{Higaki:2012ar}%
  \BibitemOpen
  \bibfield  {author} {\bibinfo {author} {\bibfnamefont {Tetsutaro}\
  \bibnamefont {Higaki}}\ and\ \bibinfo {author} {\bibfnamefont {Fuminobu}\
  \bibnamefont {Takahashi}},\ }\bibfield  {title} {\enquote {\bibinfo {title}
  {{Dark Radiation and Dark Matter in Large Volume Compactifications}},}\
  }\href {\doibase 10.1007/JHEP11(2012)125} {\bibfield  {journal} {\bibinfo
  {journal} {JHEP}\ }\textbf {\bibinfo {volume} {11}},\ \bibinfo {pages} {125}
  (\bibinfo {year} {2012})},\ \Eprint {http://arxiv.org/abs/1208.3563}
  {arXiv:1208.3563 [hep-ph]} \BibitemShut {NoStop}%
\bibitem [{\citenamefont {Conlon}\ and\ \citenamefont
  {Marsh}(2013{\natexlab{b}})}]{Conlon:2013txa}%
  \BibitemOpen
  \bibfield  {author} {\bibinfo {author} {\bibfnamefont {Joseph~P.}\
  \bibnamefont {Conlon}}\ and\ \bibinfo {author} {\bibfnamefont {M.~C.~David}\
  \bibnamefont {Marsh}},\ }\bibfield  {title} {\enquote {\bibinfo {title}
  {{Excess Astrophysical Photons from a 0.1\textendash{}1 keV Cosmic Axion
  Background}},}\ }\href {\doibase 10.1103/PhysRevLett.111.151301} {\bibfield
  {journal} {\bibinfo  {journal} {Phys. Rev. Lett.}\ }\textbf {\bibinfo
  {volume} {111}},\ \bibinfo {pages} {151301} (\bibinfo {year}
  {2013}{\natexlab{b}})},\ \Eprint {http://arxiv.org/abs/1305.3603}
  {arXiv:1305.3603 [astro-ph.CO]} \BibitemShut {NoStop}%
\bibitem [{\citenamefont {Higaki}\ \emph
  {et~al.}(2013{\natexlab{b}})\citenamefont {Higaki}, \citenamefont
  {Nakayama},\ and\ \citenamefont {Takahashi}}]{Higaki:2013lra}%
  \BibitemOpen
  \bibfield  {author} {\bibinfo {author} {\bibfnamefont {Tetsutaro}\
  \bibnamefont {Higaki}}, \bibinfo {author} {\bibfnamefont {Kazunori}\
  \bibnamefont {Nakayama}}, \ and\ \bibinfo {author} {\bibfnamefont {Fuminobu}\
  \bibnamefont {Takahashi}},\ }\bibfield  {title} {\enquote {\bibinfo {title}
  {{Moduli-Induced Axion Problem}},}\ }\href {\doibase 10.1007/JHEP07(2013)005}
  {\bibfield  {journal} {\bibinfo  {journal} {JHEP}\ }\textbf {\bibinfo
  {volume} {07}},\ \bibinfo {pages} {005} (\bibinfo {year}
  {2013}{\natexlab{b}})},\ \Eprint {http://arxiv.org/abs/1304.7987}
  {arXiv:1304.7987 [hep-ph]} \BibitemShut {NoStop}%
\bibitem [{\citenamefont {Hebecker}\ \emph {et~al.}(2014)\citenamefont
  {Hebecker}, \citenamefont {Mangat}, \citenamefont {Rompineve},\ and\
  \citenamefont {Witkowski}}]{Hebecker:2014gka}%
  \BibitemOpen
  \bibfield  {author} {\bibinfo {author} {\bibfnamefont {Arthur}\ \bibnamefont
  {Hebecker}}, \bibinfo {author} {\bibfnamefont {Patrick}\ \bibnamefont
  {Mangat}}, \bibinfo {author} {\bibfnamefont {Fabrizio}\ \bibnamefont
  {Rompineve}}, \ and\ \bibinfo {author} {\bibfnamefont {Lukas~T.}\
  \bibnamefont {Witkowski}},\ }\bibfield  {title} {\enquote {\bibinfo {title}
  {{Dark Radiation predictions from general Large Volume Scenarios}},}\ }\href
  {\doibase 10.1007/JHEP09(2014)140} {\bibfield  {journal} {\bibinfo  {journal}
  {JHEP}\ }\textbf {\bibinfo {volume} {09}},\ \bibinfo {pages} {140} (\bibinfo
  {year} {2014})},\ \Eprint {http://arxiv.org/abs/1403.6810} {arXiv:1403.6810
  [hep-ph]} \BibitemShut {NoStop}%
\bibitem [{\citenamefont {Cicoli}\ \emph {et~al.}(2014)\citenamefont {Cicoli},
  \citenamefont {Conlon}, \citenamefont {Marsh},\ and\ \citenamefont
  {Rummel}}]{Cicoli:2014bfa}%
  \BibitemOpen
  \bibfield  {author} {\bibinfo {author} {\bibfnamefont {Michele}\ \bibnamefont
  {Cicoli}}, \bibinfo {author} {\bibfnamefont {Joseph~P.}\ \bibnamefont
  {Conlon}}, \bibinfo {author} {\bibfnamefont {M.~C.~David}\ \bibnamefont
  {Marsh}}, \ and\ \bibinfo {author} {\bibfnamefont {Markus}\ \bibnamefont
  {Rummel}},\ }\bibfield  {title} {\enquote {\bibinfo {title} {{3.55 keV photon
  line and its morphology from a 3.55 keV axionlike particle line}},}\ }\href
  {\doibase 10.1103/PhysRevD.90.023540} {\bibfield  {journal} {\bibinfo
  {journal} {Phys. Rev. D}\ }\textbf {\bibinfo {volume} {90}},\ \bibinfo
  {pages} {023540} (\bibinfo {year} {2014})},\ \Eprint
  {http://arxiv.org/abs/1403.2370} {arXiv:1403.2370 [hep-ph]} \BibitemShut
  {NoStop}%
\bibitem [{\citenamefont {Cui}\ \emph {et~al.}(2018)\citenamefont {Cui},
  \citenamefont {Pospelov},\ and\ \citenamefont {Pradler}}]{Cui:2017ytb}%
  \BibitemOpen
  \bibfield  {author} {\bibinfo {author} {\bibfnamefont {Yanou}\ \bibnamefont
  {Cui}}, \bibinfo {author} {\bibfnamefont {Maxim}\ \bibnamefont {Pospelov}}, \
  and\ \bibinfo {author} {\bibfnamefont {Josef}\ \bibnamefont {Pradler}},\
  }\bibfield  {title} {\enquote {\bibinfo {title} {{Signatures of Dark
  Radiation in Neutrino and Dark Matter Detectors}},}\ }\href {\doibase
  10.1103/PhysRevD.97.103004} {\bibfield  {journal} {\bibinfo  {journal} {Phys.
  Rev. D}\ }\textbf {\bibinfo {volume} {97}},\ \bibinfo {pages} {103004}
  (\bibinfo {year} {2018})},\ \Eprint {http://arxiv.org/abs/1711.04531}
  {arXiv:1711.04531 [hep-ph]} \BibitemShut {NoStop}%
\bibitem [{\citenamefont {Moroi}\ and\ \citenamefont
  {Yin}(2021)}]{Moroi:2020has}%
  \BibitemOpen
  \bibfield  {author} {\bibinfo {author} {\bibfnamefont {Takeo}\ \bibnamefont
  {Moroi}}\ and\ \bibinfo {author} {\bibfnamefont {Wen}\ \bibnamefont {Yin}},\
  }\bibfield  {title} {\enquote {\bibinfo {title} {{Light Dark Matter from
  Inflaton Decay}},}\ }\href {\doibase 10.1007/JHEP03(2021)301} {\bibfield
  {journal} {\bibinfo  {journal} {JHEP}\ }\textbf {\bibinfo {volume} {03}},\
  \bibinfo {pages} {301} (\bibinfo {year} {2021})},\ \Eprint
  {http://arxiv.org/abs/2011.09475} {arXiv:2011.09475 [hep-ph]} \BibitemShut
  {NoStop}%
\bibitem [{\citenamefont {Jaeckel}\ and\ \citenamefont
  {Yin}(2022)}]{Jaeckel:2021ert}%
  \BibitemOpen
  \bibfield  {author} {\bibinfo {author} {\bibfnamefont {Joerg}\ \bibnamefont
  {Jaeckel}}\ and\ \bibinfo {author} {\bibfnamefont {Wen}\ \bibnamefont
  {Yin}},\ }\bibfield  {title} {\enquote {\bibinfo {title} {{Shining ALP dark
  radiation}},}\ }\href {\doibase 10.1103/PhysRevD.105.115003} {\bibfield
  {journal} {\bibinfo  {journal} {Phys. Rev. D}\ }\textbf {\bibinfo {volume}
  {105}},\ \bibinfo {pages} {115003} (\bibinfo {year} {2022})},\ \Eprint
  {http://arxiv.org/abs/2110.03692} {arXiv:2110.03692 [hep-ph]} \BibitemShut
  {NoStop}%
\bibitem [{\citenamefont {Bell}\ \emph {et~al.}(2024)\citenamefont {Bell},
  \citenamefont {Busoni}, \citenamefont {Robles},\ and\ \citenamefont
  {Virgato}}]{Bell:2023ysh}%
  \BibitemOpen
  \bibfield  {author} {\bibinfo {author} {\bibfnamefont {Nicole~F.}\
  \bibnamefont {Bell}}, \bibinfo {author} {\bibfnamefont {Giorgio}\
  \bibnamefont {Busoni}}, \bibinfo {author} {\bibfnamefont {Sandra}\
  \bibnamefont {Robles}}, \ and\ \bibinfo {author} {\bibfnamefont {Michael}\
  \bibnamefont {Virgato}},\ }\bibfield  {title} {\enquote {\bibinfo {title}
  {{Thermalization and annihilation of dark matter in neutron stars}},}\ }\href
  {\doibase 10.1088/1475-7516/2024/04/006} {\bibfield  {journal} {\bibinfo
  {journal} {JCAP}\ }\textbf {\bibinfo {volume} {04}},\ \bibinfo {pages} {006}
  (\bibinfo {year} {2024})},\ \Eprint {http://arxiv.org/abs/2312.11892}
  {arXiv:2312.11892 [hep-ph]} \BibitemShut {NoStop}%
\bibitem [{\citenamefont {Kibble}(1976)}]{Kibble:1976sj}%
  \BibitemOpen
  \bibfield  {author} {\bibinfo {author} {\bibfnamefont {T.~W.~B.}\
  \bibnamefont {Kibble}},\ }\bibfield  {title} {\enquote {\bibinfo {title}
  {{Topology of Cosmic Domains and Strings}},}\ }\href {\doibase
  10.1088/0305-4470/9/8/029} {\bibfield  {journal} {\bibinfo  {journal} {J.
  Phys. A}\ }\textbf {\bibinfo {volume} {9}},\ \bibinfo {pages} {1387--1398}
  (\bibinfo {year} {1976})}\BibitemShut {NoStop}%
\bibitem [{\citenamefont {Yamaguchi}\ \emph {et~al.}(1999)\citenamefont
  {Yamaguchi}, \citenamefont {Kawasaki},\ and\ \citenamefont
  {Yokoyama}}]{Yamaguchi:1998gx}%
  \BibitemOpen
  \bibfield  {author} {\bibinfo {author} {\bibfnamefont {Masahide}\
  \bibnamefont {Yamaguchi}}, \bibinfo {author} {\bibfnamefont {M.}~\bibnamefont
  {Kawasaki}}, \ and\ \bibinfo {author} {\bibfnamefont {Jun'ichi}\ \bibnamefont
  {Yokoyama}},\ }\bibfield  {title} {\enquote {\bibinfo {title} {{Evolution of
  axionic strings and spectrum of axions radiated from them}},}\ }\href
  {\doibase 10.1103/PhysRevLett.82.4578} {\bibfield  {journal} {\bibinfo
  {journal} {Phys. Rev. Lett.}\ }\textbf {\bibinfo {volume} {82}},\ \bibinfo
  {pages} {4578--4581} (\bibinfo {year} {1999})},\ \Eprint
  {http://arxiv.org/abs/hep-ph/9811311} {arXiv:hep-ph/9811311} \BibitemShut
  {NoStop}%
\bibitem [{\citenamefont {Hagmann}\ \emph {et~al.}(1999)\citenamefont
  {Hagmann}, \citenamefont {Chang},\ and\ \citenamefont
  {Sikivie}}]{Hagmann:1998me}%
  \BibitemOpen
  \bibfield  {author} {\bibinfo {author} {\bibfnamefont {C.}~\bibnamefont
  {Hagmann}}, \bibinfo {author} {\bibfnamefont {Sanghyeon}\ \bibnamefont
  {Chang}}, \ and\ \bibinfo {author} {\bibfnamefont {P.}~\bibnamefont
  {Sikivie}},\ }\bibfield  {title} {\enquote {\bibinfo {title} {{Axions from
  string decay}},}\ }\href {\doibase 10.1016/S0920-5632(98)00506-4} {\bibfield
  {journal} {\bibinfo  {journal} {Nucl. Phys. B Proc. Suppl.}\ }\textbf
  {\bibinfo {volume} {72}},\ \bibinfo {pages} {81--86} (\bibinfo {year}
  {1999})},\ \Eprint {http://arxiv.org/abs/hep-ph/9807428}
  {arXiv:hep-ph/9807428} \BibitemShut {NoStop}%
\bibitem [{\citenamefont {Yamaguchi}\ and\ \citenamefont
  {Yokoyama}(2003)}]{Yamaguchi:2002sh}%
  \BibitemOpen
  \bibfield  {author} {\bibinfo {author} {\bibfnamefont {Masahide}\
  \bibnamefont {Yamaguchi}}\ and\ \bibinfo {author} {\bibfnamefont {Jun'ichi}\
  \bibnamefont {Yokoyama}},\ }\bibfield  {title} {\enquote {\bibinfo {title}
  {{Quantitative evolution of global strings from the Lagrangian view
  point}},}\ }\href {\doibase 10.1103/PhysRevD.67.103514} {\bibfield  {journal}
  {\bibinfo  {journal} {Phys. Rev. D}\ }\textbf {\bibinfo {volume} {67}},\
  \bibinfo {pages} {103514} (\bibinfo {year} {2003})},\ \Eprint
  {http://arxiv.org/abs/hep-ph/0210343} {arXiv:hep-ph/0210343} \BibitemShut
  {NoStop}%
\bibitem [{\citenamefont {Hiramatsu}\ \emph {et~al.}(2011)\citenamefont
  {Hiramatsu}, \citenamefont {Kawasaki}, \citenamefont {Sekiguchi},
  \citenamefont {Yamaguchi},\ and\ \citenamefont
  {Yokoyama}}]{Hiramatsu:2010yu}%
  \BibitemOpen
  \bibfield  {author} {\bibinfo {author} {\bibfnamefont {Takashi}\ \bibnamefont
  {Hiramatsu}}, \bibinfo {author} {\bibfnamefont {Masahiro}\ \bibnamefont
  {Kawasaki}}, \bibinfo {author} {\bibfnamefont {Toyokazu}\ \bibnamefont
  {Sekiguchi}}, \bibinfo {author} {\bibfnamefont {Masahide}\ \bibnamefont
  {Yamaguchi}}, \ and\ \bibinfo {author} {\bibfnamefont {Jun'ichi}\
  \bibnamefont {Yokoyama}},\ }\bibfield  {title} {\enquote {\bibinfo {title}
  {{Improved estimation of radiated axions from cosmological axionic
  strings}},}\ }\href {\doibase 10.1103/PhysRevD.83.123531} {\bibfield
  {journal} {\bibinfo  {journal} {Phys. Rev. D}\ }\textbf {\bibinfo {volume}
  {83}},\ \bibinfo {pages} {123531} (\bibinfo {year} {2011})},\ \Eprint
  {http://arxiv.org/abs/1012.5502} {arXiv:1012.5502 [hep-ph]} \BibitemShut
  {NoStop}%
\bibitem [{\citenamefont {Hiramatsu}\ \emph {et~al.}(2012)\citenamefont
  {Hiramatsu}, \citenamefont {Kawasaki}, \citenamefont {Saikawa},\ and\
  \citenamefont {Sekiguchi}}]{Hiramatsu:2012gg}%
  \BibitemOpen
  \bibfield  {author} {\bibinfo {author} {\bibfnamefont {Takashi}\ \bibnamefont
  {Hiramatsu}}, \bibinfo {author} {\bibfnamefont {Masahiro}\ \bibnamefont
  {Kawasaki}}, \bibinfo {author} {\bibfnamefont {Ken'ichi}\ \bibnamefont
  {Saikawa}}, \ and\ \bibinfo {author} {\bibfnamefont {Toyokazu}\ \bibnamefont
  {Sekiguchi}},\ }\bibfield  {title} {\enquote {\bibinfo {title} {{Production
  of dark matter axions from collapse of string-wall systems}},}\ }\href
  {\doibase 10.1103/PhysRevD.85.105020} {\bibfield  {journal} {\bibinfo
  {journal} {Phys. Rev. D}\ }\textbf {\bibinfo {volume} {85}},\ \bibinfo
  {pages} {105020} (\bibinfo {year} {2012})},\ \bibinfo {note} {[Erratum:
  Phys.Rev.D 86, 089902 (2012)]},\ \Eprint {http://arxiv.org/abs/1202.5851}
  {arXiv:1202.5851 [hep-ph]} \BibitemShut {NoStop}%
\bibitem [{\citenamefont {Kawasaki}\ \emph {et~al.}(2015)\citenamefont
  {Kawasaki}, \citenamefont {Saikawa},\ and\ \citenamefont
  {Sekiguchi}}]{Kawasaki:2014sqa}%
  \BibitemOpen
  \bibfield  {author} {\bibinfo {author} {\bibfnamefont {Masahiro}\
  \bibnamefont {Kawasaki}}, \bibinfo {author} {\bibfnamefont {Ken'ichi}\
  \bibnamefont {Saikawa}}, \ and\ \bibinfo {author} {\bibfnamefont {Toyokazu}\
  \bibnamefont {Sekiguchi}},\ }\bibfield  {title} {\enquote {\bibinfo {title}
  {{Axion dark matter from topological defects}},}\ }\href {\doibase
  10.1103/PhysRevD.91.065014} {\bibfield  {journal} {\bibinfo  {journal} {Phys.
  Rev. D}\ }\textbf {\bibinfo {volume} {91}},\ \bibinfo {pages} {065014}
  (\bibinfo {year} {2015})},\ \Eprint {http://arxiv.org/abs/1412.0789}
  {arXiv:1412.0789 [hep-ph]} \BibitemShut {NoStop}%
\bibitem [{\citenamefont {Lopez-Eiguren}\ \emph {et~al.}(2017)\citenamefont
  {Lopez-Eiguren}, \citenamefont {Lizarraga}, \citenamefont {Hindmarsh},\ and\
  \citenamefont {Urrestilla}}]{Lopez-Eiguren:2017dmc}%
  \BibitemOpen
  \bibfield  {author} {\bibinfo {author} {\bibfnamefont {Asier}\ \bibnamefont
  {Lopez-Eiguren}}, \bibinfo {author} {\bibfnamefont {Joanes}\ \bibnamefont
  {Lizarraga}}, \bibinfo {author} {\bibfnamefont {Mark}\ \bibnamefont
  {Hindmarsh}}, \ and\ \bibinfo {author} {\bibfnamefont {Jon}\ \bibnamefont
  {Urrestilla}},\ }\bibfield  {title} {\enquote {\bibinfo {title} {{Cosmic
  Microwave Background constraints for global strings and global monopoles}},}\
  }\href {\doibase 10.1088/1475-7516/2017/07/026} {\bibfield  {journal}
  {\bibinfo  {journal} {JCAP}\ }\textbf {\bibinfo {volume} {07}},\ \bibinfo
  {pages} {026} (\bibinfo {year} {2017})},\ \Eprint
  {http://arxiv.org/abs/1705.04154} {arXiv:1705.04154 [astro-ph.CO]}
  \BibitemShut {NoStop}%
\bibitem [{\citenamefont {Gorghetto}\ \emph {et~al.}(2018)\citenamefont
  {Gorghetto}, \citenamefont {Hardy},\ and\ \citenamefont
  {Villadoro}}]{Gorghetto:2018myk}%
  \BibitemOpen
  \bibfield  {author} {\bibinfo {author} {\bibfnamefont {Marco}\ \bibnamefont
  {Gorghetto}}, \bibinfo {author} {\bibfnamefont {Edward}\ \bibnamefont
  {Hardy}}, \ and\ \bibinfo {author} {\bibfnamefont {Giovanni}\ \bibnamefont
  {Villadoro}},\ }\bibfield  {title} {\enquote {\bibinfo {title} {{Axions from
  Strings: the Attractive Solution}},}\ }\href {\doibase
  10.1007/JHEP07(2018)151} {\bibfield  {journal} {\bibinfo  {journal} {JHEP}\
  }\textbf {\bibinfo {volume} {07}},\ \bibinfo {pages} {151} (\bibinfo {year}
  {2018})},\ \Eprint {http://arxiv.org/abs/1806.04677} {arXiv:1806.04677
  [hep-ph]} \BibitemShut {NoStop}%
\bibitem [{\citenamefont {Kawasaki}\ \emph {et~al.}(2018)\citenamefont
  {Kawasaki}, \citenamefont {Sekiguchi}, \citenamefont {Yamaguchi},\ and\
  \citenamefont {Yokoyama}}]{Kawasaki:2018bzv}%
  \BibitemOpen
  \bibfield  {author} {\bibinfo {author} {\bibfnamefont {Masahiro}\
  \bibnamefont {Kawasaki}}, \bibinfo {author} {\bibfnamefont {Toyokazu}\
  \bibnamefont {Sekiguchi}}, \bibinfo {author} {\bibfnamefont {Masahide}\
  \bibnamefont {Yamaguchi}}, \ and\ \bibinfo {author} {\bibfnamefont
  {Jun'ichi}\ \bibnamefont {Yokoyama}},\ }\bibfield  {title} {\enquote
  {\bibinfo {title} {{Long-term dynamics of cosmological axion strings}},}\
  }\href {\doibase 10.1093/ptep/pty098} {\bibfield  {journal} {\bibinfo
  {journal} {PTEP}\ }\textbf {\bibinfo {volume} {2018}},\ \bibinfo {pages}
  {091E01} (\bibinfo {year} {2018})},\ \Eprint
  {http://arxiv.org/abs/1806.05566} {arXiv:1806.05566 [hep-ph]} \BibitemShut
  {NoStop}%
\bibitem [{\citenamefont {Vaquero}\ \emph {et~al.}(2019)\citenamefont
  {Vaquero}, \citenamefont {Redondo},\ and\ \citenamefont
  {Stadler}}]{Vaquero:2018tib}%
  \BibitemOpen
  \bibfield  {author} {\bibinfo {author} {\bibfnamefont {Alejandro}\
  \bibnamefont {Vaquero}}, \bibinfo {author} {\bibfnamefont {Javier}\
  \bibnamefont {Redondo}}, \ and\ \bibinfo {author} {\bibfnamefont {Julia}\
  \bibnamefont {Stadler}},\ }\bibfield  {title} {\enquote {\bibinfo {title}
  {{Early seeds of axion miniclusters}},}\ }\href {\doibase
  10.1088/1475-7516/2019/04/012} {\bibfield  {journal} {\bibinfo  {journal}
  {JCAP}\ }\textbf {\bibinfo {volume} {04}},\ \bibinfo {pages} {012} (\bibinfo
  {year} {2019})},\ \Eprint {http://arxiv.org/abs/1809.09241} {arXiv:1809.09241
  [astro-ph.CO]} \BibitemShut {NoStop}%
\bibitem [{\citenamefont {Martins}(2019)}]{Martins:2018dqg}%
  \BibitemOpen
  \bibfield  {author} {\bibinfo {author} {\bibfnamefont {C.~J. A.~P.}\
  \bibnamefont {Martins}},\ }\bibfield  {title} {\enquote {\bibinfo {title}
  {{Scaling properties of cosmological axion strings}},}\ }\href {\doibase
  10.1016/j.physletb.2018.11.031} {\bibfield  {journal} {\bibinfo  {journal}
  {Phys. Lett. B}\ }\textbf {\bibinfo {volume} {788}},\ \bibinfo {pages}
  {147--151} (\bibinfo {year} {2019})},\ \Eprint
  {http://arxiv.org/abs/1811.12678} {arXiv:1811.12678 [astro-ph.CO]}
  \BibitemShut {NoStop}%
\bibitem [{\citenamefont {Buschmann}\ \emph {et~al.}(2020)\citenamefont
  {Buschmann}, \citenamefont {Foster},\ and\ \citenamefont
  {Safdi}}]{Buschmann:2019icd}%
  \BibitemOpen
  \bibfield  {author} {\bibinfo {author} {\bibfnamefont {Malte}\ \bibnamefont
  {Buschmann}}, \bibinfo {author} {\bibfnamefont {Joshua~W.}\ \bibnamefont
  {Foster}}, \ and\ \bibinfo {author} {\bibfnamefont {Benjamin~R.}\
  \bibnamefont {Safdi}},\ }\bibfield  {title} {\enquote {\bibinfo {title}
  {{Early-Universe Simulations of the Cosmological Axion}},}\ }\href {\doibase
  10.1103/PhysRevLett.124.161103} {\bibfield  {journal} {\bibinfo  {journal}
  {Phys. Rev. Lett.}\ }\textbf {\bibinfo {volume} {124}},\ \bibinfo {pages}
  {161103} (\bibinfo {year} {2020})},\ \Eprint
  {http://arxiv.org/abs/1906.00967} {arXiv:1906.00967 [astro-ph.CO]}
  \BibitemShut {NoStop}%
\bibitem [{\citenamefont {Hindmarsh}\ \emph {et~al.}(2020)\citenamefont
  {Hindmarsh}, \citenamefont {Lizarraga}, \citenamefont {Lopez-Eiguren},\ and\
  \citenamefont {Urrestilla}}]{Hindmarsh:2019csc}%
  \BibitemOpen
  \bibfield  {author} {\bibinfo {author} {\bibfnamefont {Mark}\ \bibnamefont
  {Hindmarsh}}, \bibinfo {author} {\bibfnamefont {Joanes}\ \bibnamefont
  {Lizarraga}}, \bibinfo {author} {\bibfnamefont {Asier}\ \bibnamefont
  {Lopez-Eiguren}}, \ and\ \bibinfo {author} {\bibfnamefont {Jon}\ \bibnamefont
  {Urrestilla}},\ }\bibfield  {title} {\enquote {\bibinfo {title} {{Scaling
  Density of Axion Strings}},}\ }\href {\doibase
  10.1103/PhysRevLett.124.021301} {\bibfield  {journal} {\bibinfo  {journal}
  {Phys. Rev. Lett.}\ }\textbf {\bibinfo {volume} {124}},\ \bibinfo {pages}
  {021301} (\bibinfo {year} {2020})},\ \Eprint
  {http://arxiv.org/abs/1908.03522} {arXiv:1908.03522 [astro-ph.CO]}
  \BibitemShut {NoStop}%
\bibitem [{\citenamefont {Klaer}\ and\ \citenamefont
  {Moore}(2020)}]{Klaer:2019fxc}%
  \BibitemOpen
  \bibfield  {author} {\bibinfo {author} {\bibfnamefont {Vincent~B.}\
  \bibnamefont {Klaer}}\ and\ \bibinfo {author} {\bibfnamefont {Guy~D.}\
  \bibnamefont {Moore}},\ }\bibfield  {title} {\enquote {\bibinfo {title}
  {{Global cosmic string networks as a function of tension}},}\ }\href
  {\doibase 10.1088/1475-7516/2020/06/021} {\bibfield  {journal} {\bibinfo
  {journal} {JCAP}\ }\textbf {\bibinfo {volume} {06}},\ \bibinfo {pages} {021}
  (\bibinfo {year} {2020})},\ \Eprint {http://arxiv.org/abs/1912.08058}
  {arXiv:1912.08058 [hep-ph]} \BibitemShut {NoStop}%
\bibitem [{\citenamefont {Gorghetto}\ \emph {et~al.}(2021)\citenamefont
  {Gorghetto}, \citenamefont {Hardy},\ and\ \citenamefont
  {Villadoro}}]{Gorghetto:2020qws}%
  \BibitemOpen
  \bibfield  {author} {\bibinfo {author} {\bibfnamefont {Marco}\ \bibnamefont
  {Gorghetto}}, \bibinfo {author} {\bibfnamefont {Edward}\ \bibnamefont
  {Hardy}}, \ and\ \bibinfo {author} {\bibfnamefont {Giovanni}\ \bibnamefont
  {Villadoro}},\ }\bibfield  {title} {\enquote {\bibinfo {title} {{More axions
  from strings}},}\ }\href {\doibase 10.21468/SciPostPhys.10.2.050} {\bibfield
  {journal} {\bibinfo  {journal} {SciPost Phys.}\ }\textbf {\bibinfo {volume}
  {10}},\ \bibinfo {pages} {050} (\bibinfo {year} {2021})},\ \Eprint
  {http://arxiv.org/abs/2007.04990} {arXiv:2007.04990 [hep-ph]} \BibitemShut
  {NoStop}%
\bibitem [{\citenamefont {Hindmarsh}\ \emph {et~al.}(2021)\citenamefont
  {Hindmarsh}, \citenamefont {Lizarraga}, \citenamefont {Lopez-Eiguren},\ and\
  \citenamefont {Urrestilla}}]{Hindmarsh:2021vih}%
  \BibitemOpen
  \bibfield  {author} {\bibinfo {author} {\bibfnamefont {Mark}\ \bibnamefont
  {Hindmarsh}}, \bibinfo {author} {\bibfnamefont {Joanes}\ \bibnamefont
  {Lizarraga}}, \bibinfo {author} {\bibfnamefont {Asier}\ \bibnamefont
  {Lopez-Eiguren}}, \ and\ \bibinfo {author} {\bibfnamefont {Jon}\ \bibnamefont
  {Urrestilla}},\ }\bibfield  {title} {\enquote {\bibinfo {title} {{Approach to
  scaling in axion string networks}},}\ }\href {\doibase
  10.1103/PhysRevD.103.103534} {\bibfield  {journal} {\bibinfo  {journal}
  {Phys. Rev. D}\ }\textbf {\bibinfo {volume} {103}},\ \bibinfo {pages}
  {103534} (\bibinfo {year} {2021})},\ \Eprint
  {http://arxiv.org/abs/2102.07723} {arXiv:2102.07723 [astro-ph.CO]}
  \BibitemShut {NoStop}%
\bibitem [{\citenamefont {Harari}\ and\ \citenamefont
  {Sikivie}(1987)}]{Harari:1987ht}%
  \BibitemOpen
  \bibfield  {author} {\bibinfo {author} {\bibfnamefont {Diego}\ \bibnamefont
  {Harari}}\ and\ \bibinfo {author} {\bibfnamefont {P.}~\bibnamefont
  {Sikivie}},\ }\bibfield  {title} {\enquote {\bibinfo {title} {{On the
  Evolution of Global Strings in the Early Universe}},}\ }\href {\doibase
  10.1016/0370-2693(87)90032-3} {\bibfield  {journal} {\bibinfo  {journal}
  {Phys. Lett. B}\ }\textbf {\bibinfo {volume} {195}},\ \bibinfo {pages}
  {361--365} (\bibinfo {year} {1987})}\BibitemShut {NoStop}%
\bibitem [{\citenamefont {Long}(2018)}]{Long:2018nsl}%
  \BibitemOpen
  \bibfield  {author} {\bibinfo {author} {\bibfnamefont {Andrew~J.}\
  \bibnamefont {Long}},\ }\bibfield  {title} {\enquote {\bibinfo {title}
  {{Cosmological Aspects of the Clockwork Axion}},}\ }\href {\doibase
  10.1007/JHEP07(2018)066} {\bibfield  {journal} {\bibinfo  {journal} {JHEP}\
  }\textbf {\bibinfo {volume} {07}},\ \bibinfo {pages} {066} (\bibinfo {year}
  {2018})},\ \Eprint {http://arxiv.org/abs/1803.07086} {arXiv:1803.07086
  [hep-ph]} \BibitemShut {NoStop}%
\bibitem [{\citenamefont {{Barry, Pete and Chang, Clarence and Dona, Kristin
  and Khatiwada, Rakshya and Knirck, Stefan and Kurinsky, Noah and Liu, Jesse
  and Marsh, David and Miller, David and Noroozian, Omid and Sonnenschein,
  Andrew}}(2021)}]{Barry:2021}%
  \BibitemOpen
  \bibfield  {author} {\bibinfo {author} {\bibnamefont {{Barry, Pete and Chang,
  Clarence and Dona, Kristin and Khatiwada, Rakshya and Knirck, Stefan and
  Kurinsky, Noah and Liu, Jesse and Marsh, David and Miller, David and
  Noroozian, Omid and Sonnenschein, Andrew}}},\ }\href@noop {} {\enquote
  {\bibinfo {title} {{Opening the terahertz axion window}},}\ }\bibinfo
  {howpublished} {Available at
  \url{https://www.snowmass21.org/docs/files/summaries/CF/SNOWMASS21-CF2_CF0-AF7_AF0-IF1_IF2-UF2_UF0_Jesse_Liu-179.pdf}
  (2024/06/19)} (\bibinfo {year} {2021})\BibitemShut {NoStop}%
\bibitem [{\citenamefont {Liu}\ \emph {et~al.}(2022)\citenamefont {Liu} \emph
  {et~al.}}]{BREAD:2021tpx}%
  \BibitemOpen
  \bibfield  {author} {\bibinfo {author} {\bibfnamefont {Jesse}\ \bibnamefont
  {Liu}} \emph {et~al.} (\bibinfo {collaboration} {BREAD}),\ }\bibfield
  {title} {\enquote {\bibinfo {title} {{Broadband Solenoidal Haloscope for
  Terahertz Axion Detection}},}\ }\href {\doibase
  10.1103/PhysRevLett.128.131801} {\bibfield  {journal} {\bibinfo  {journal}
  {Phys. Rev. Lett.}\ }\textbf {\bibinfo {volume} {128}},\ \bibinfo {pages}
  {131801} (\bibinfo {year} {2022})},\ \Eprint
  {http://arxiv.org/abs/2111.12103} {arXiv:2111.12103 [physics.ins-det]}
  \BibitemShut {NoStop}%
\bibitem [{\citenamefont {Wilczek}(1987)}]{Wilczek:1987mv}%
  \BibitemOpen
  \bibfield  {author} {\bibinfo {author} {\bibfnamefont {Frank}\ \bibnamefont
  {Wilczek}},\ }\bibfield  {title} {\enquote {\bibinfo {title} {{Two
  Applications of Axion Electrodynamics}},}\ }\href {\doibase
  10.1103/PhysRevLett.58.1799} {\bibfield  {journal} {\bibinfo  {journal}
  {Phys. Rev. Lett.}\ }\textbf {\bibinfo {volume} {58}},\ \bibinfo {pages}
  {1799} (\bibinfo {year} {1987})}\BibitemShut {NoStop}%
\bibitem [{\citenamefont {Sitenko}(1967)}]{Sitenko:1967}%
  \BibitemOpen
  \bibfield  {author} {\bibinfo {author} {\bibfnamefont {A.~G.}\ \bibnamefont
  {Sitenko}},\ }\href@noop {} {\emph {\bibinfo {title} {{Electromagnetic
  Fluctuations in Plasma}}}}\ (\bibinfo  {publisher} {Academic Press},\
  \bibinfo {address} {{New York, New York}},\ \bibinfo {year}
  {1967})\BibitemShut {NoStop}%
\bibitem [{\citenamefont {Caputo}\ \emph {et~al.}(2023)\citenamefont {Caputo},
  \citenamefont {Witte}, \citenamefont {Philippov},\ and\ \citenamefont
  {Jacobson}}]{Caputo:2023cpv}%
  \BibitemOpen
  \bibfield  {author} {\bibinfo {author} {\bibfnamefont {Andrea}\ \bibnamefont
  {Caputo}}, \bibinfo {author} {\bibfnamefont {Samuel~J.}\ \bibnamefont
  {Witte}}, \bibinfo {author} {\bibfnamefont {Alexander~A.}\ \bibnamefont
  {Philippov}}, \ and\ \bibinfo {author} {\bibfnamefont {Ted}\ \bibnamefont
  {Jacobson}},\ }\bibfield  {title} {\enquote {\bibinfo {title} {{Pulsar
  Nulling and Vacuum Radio Emission from Axion Clouds}},}\ }\href@noop {} {\
  (\bibinfo {year} {2023})},\ \Eprint {http://arxiv.org/abs/2311.14795}
  {arXiv:2311.14795 [hep-ph]} \BibitemShut {NoStop}%
\bibitem [{\citenamefont {Noordhuis}\ \emph
  {et~al.}(2023{\natexlab{b}})\citenamefont {Noordhuis}, \citenamefont
  {Prabhu}, \citenamefont {Weniger},\ and\ \citenamefont
  {Witte}}]{Noordhuis:2023wid}%
  \BibitemOpen
  \bibfield  {author} {\bibinfo {author} {\bibfnamefont {Dion}\ \bibnamefont
  {Noordhuis}}, \bibinfo {author} {\bibfnamefont {Anirudh}\ \bibnamefont
  {Prabhu}}, \bibinfo {author} {\bibfnamefont {Christoph}\ \bibnamefont
  {Weniger}}, \ and\ \bibinfo {author} {\bibfnamefont {Samuel~J.}\ \bibnamefont
  {Witte}},\ }\bibfield  {title} {\enquote {\bibinfo {title} {{Axion Clouds
  around Neutron Stars}},}\ }\href@noop {} {\  (\bibinfo {year}
  {2023}{\natexlab{b}})},\ \Eprint {http://arxiv.org/abs/2307.11811}
  {arXiv:2307.11811 [hep-ph]} \BibitemShut {NoStop}%
\bibitem [{\citenamefont {Lorimer}\ \emph {et~al.}(2006)\citenamefont {Lorimer}
  \emph {et~al.}}]{Lorimer:2006qs}%
  \BibitemOpen
  \bibfield  {author} {\bibinfo {author} {\bibfnamefont {D.~R.}\ \bibnamefont
  {Lorimer}} \emph {et~al.},\ }\bibfield  {title} {\enquote {\bibinfo {title}
  {{The Parkes multibeam pulsar survey: VI. Discovery and timing of 142 pulsars
  and a Galactic population analysis}},}\ }\href {\doibase
  10.1111/j.1365-2966.2006.10887.x} {\bibfield  {journal} {\bibinfo  {journal}
  {Mon. Not. Roy. Astron. Soc.}\ }\textbf {\bibinfo {volume} {372}},\ \bibinfo
  {pages} {777--800} (\bibinfo {year} {2006})},\ \Eprint
  {http://arxiv.org/abs/astro-ph/0607640} {arXiv:astro-ph/0607640} \BibitemShut
  {NoStop}%
\bibitem [{\citenamefont {N\"attil\"a}\ and\ \citenamefont
  {Kajava}(2022)}]{Nattila:2022evn}%
  \BibitemOpen
  \bibfield  {author} {\bibinfo {author} {\bibfnamefont {Joonas}\ \bibnamefont
  {N\"attil\"a}}\ and\ \bibinfo {author} {\bibfnamefont {Jari J.~E.}\
  \bibnamefont {Kajava}},\ }\enquote {\bibinfo {title} {{Fundamental physics
  with neutron stars}},}\ in\ \href {\doibase 10.1007/978-981-16-4544-0_105-1}
  {\emph {\bibinfo {booktitle} {{Handbook of X-ray and Gamma-ray
  Astrophysics}}}},\ \bibinfo {editor} {edited by\ \bibinfo {editor}
  {\bibfnamefont {Cosimo}\ \bibnamefont {Bambi}}\ and\ \bibinfo {editor}
  {\bibfnamefont {Andrea}\ \bibnamefont {Santangelo}}}\ (\bibinfo {year}
  {2022})\ \Eprint {http://arxiv.org/abs/2211.15721} {arXiv:2211.15721
  [astro-ph.HE]} \BibitemShut {NoStop}%
\end{thebibliography}%

\end{document}